\documentclass[aps,prb,reprint,showpacs,intlimits,preprintnumbers,longbibliography,amsmath,amssymb]{revtex4-1}
\usepackage{graphicx}
\usepackage{bm}
\usepackage{hyperref}
\usepackage{color}
\newcommand\tpp{t^{++}}
\newcommand\tpm{t^{+-}}
\newcommand\tmm{t^{--}}
\DeclareMathOperator{\Tr}{Tr}
\DeclareMathOperator{\Real}{Re}
\DeclareMathOperator{\Img}{Im}
\DeclareMathOperator{\adj}{adj}

\begin{document}

\title{Resonant enhancement of thermoelectric properties by correlated hopping for the Falicov-Kimball model on Bethe lattice}
\author{D.~A.~Dobushovskyi, A.~M.~Shvaika}
\affiliation{Institute for Condensed Matter Physics of the National Academy of Sciences of Ukraine,
	Lviv, 79011 Ukraine}
\author{V. Zlati\'c}
\affiliation{Institute of Physics, Zagreb POB 304, Croatia}
\affiliation{Department of Physics, Faculty of Science, University of Split, HR-21000 Split, Croatia}

\begin{abstract}
The effect of correlated hopping on the charge and heat transport of strongly correlated particles 
is studied for the Falicov-Kimball model on the Bethe lattice. 
Exact solutions for the one particle density of states (DOS) and two particle transport function 
(the ``quasiparticle'' scattering time) are derived using dynamical mean field theory. 
For a wide range of the correlated hopping,  the transport function exhibits singularities due to the 
resonant  two-particle contribution, whereas the one particle DOS does not show any anomalous features. 
By tuning the number of itinerant electrons, so as to bring the Fermi level close to the resonant frequency,  
we get a large increase of the electrical and thermal conductivities, and the thermoelectric power. 
When the hopping amplitude between the occupied sites is reduced sufficiently, the itinerant electrons localize 
in the clusters of sites occupied by $f$ electrons. This gives rise to an additional narrow band in the DOS 
between the lower and upper Hubbard bands,  but has only a minor effect on the thermoelectric properties. 
\end{abstract}	
\pacs{72.15.Jf, 72.20.Pa, 71.27.+a, 71.10.Fd}

\maketitle

\section{Introduction}

The effects of electron correlations on various phenomena in different materials, from the one- and two-dimensional organic conductors, 
through three-dimensional solids, up to the optical lattices, have been attracting considerable interest  for more than half a century.  
The majority of publication dealing with these problems considered only the local Coulomb correlations of the Hubbard or Anderson type, 
$U\sum_i \hat{n}_{i\uparrow}\hat{n}_{i\downarrow}$. 
However, as noticed by Hubbard in his seminal article,\cite{hubbard:238}  
the second quantized representation of the inter-electron Coulomb interaction should also take into account, 
besides the local term, the nonlocal contributions.  
He pointed to the inter-site Coulomb interaction $\sum_{ij} V_{ij} \hat{n}_{i}\hat{n}_{j}$ and the so-called correlated hopping,
\begin{equation}
\sum_{ij\sigma} t_{ij}^{(2)} (\hat{n}_{i\bar{\sigma}}+\hat{n}_{j\bar{\sigma}})c_{i\sigma}^{\dagger}c_{j\sigma},
\end{equation}
which reflects the fact that the overlap of different many-body states is not the same, 
so that the value of inter-site hopping depends on the occupation of these states. 
The origin of the correlated hopping can be either a direct inter-site interaction or an indirect effective one,\cite{foglio:4554,simon:7471} 
which can produce the multi-particle interactions, e.g., the three-particle one $\sum_{ij\sigma} t_{ij}^{(3)} \hat{n}_{i\bar{\sigma}}c_{i\sigma}^{\dagger}c_{j\sigma}\hat{n}_{j\bar{\sigma}}$.

While the effects of local Coulomb interaction described by the Hubbard and Anderson models have widely been investigated in the theory of strongly correlated electron systems, correlated hopping has attracted much less attention.%
\footnote{The name used for such contributions is not well established yet. Apart from the term ``correlated hopping,'' many other terms circulate, 
such as ``assisted hopping,'' ``bond-charge interaction (repulsion),'' ``occupation-dependent hopping,'' ``correlated hybridization,'' etc.
Similar contributions in the theory of disordered systems are also known as an ``off-diagonal disorder''\cite{blackman:2412}.}
It was considered in connection with the new mechanisms for high temperature superconductivity,\cite{hirsch:326,bulka:10303} 
properties of organic compounds\cite{arrachea:1173} and molecular crystals,\cite{tsuchiizu:044519} electron-hole asymmetry,\cite{didukh:7893} 
and enhancement of magnetic properties.\cite{kollar:045107} 
Recently, correlated hopping has been examined in relation to quantum 
dots\cite{meir:196802,hubsch:196401,tooski:055001} and fermionic\cite{jurgensen:043623,liberto:013624} and 
bosonic\cite{eckholt:093028,luhmann:033021,jurgensen:193003} atoms on optical lattices.
However, due to its nonlocal character, the theoretical treatment of correlated hopping is difficult  and, in most cases, the solutions can only be obtained by rather drastic approximations.

The exact results, that one can obtain in some special cases, are of great importance, 
as they can be used for benchmarking various  approximations. 
Here, we study the effects of correlated hopping using the Falicov-Kimball model, the simplest model  of strongly correlated electrons.\cite{falicov:997}  
In its canonical form, it considers the local interaction between the itinerant $d$ electrons and localized $f$ electrons.
It is a binary alloy type model and its ground state phase diagram for the one-dimensional $(D=1)$ 
and two-dimensional $(D=2)$ cases displays a variety of modulated 
phases.\cite{gajek:3473,wojtkiewicz:233103,wojtkiewicz:3467,cencarikova:42701}
The dynamical mean field theory (DMFT)\cite{metzner:324,brandt:365,georges:13} yields 
an exact solution of the Falicov-Kimball model in infinite dimensions, 
based on which the phase diagram and phase transitions in ordered phase were investigated 
and different spectral functions and responses were calculated.\cite{freericks:1333} 
An extension of the model, so as to include the  correlated hopping, was also considered 
and the DMFT solution with a nonlocal self-energy was obtained.\cite{schiller:15660,shvaika:075101,shvaika:119901}

In this article we study the effect of correlated hopping on the charge and heat transport of the Falicov-Kimball model on the Bethe lattice. 
The correlated hopping makes the Falicov-Kimball model similar to the binary alloy model with off-diagonal disorder,\cite{blackman:2412}  
studied long ago by Hoshino and Niizeki.\cite{hoshino:1320}
Using the Mott's relation and considering the special case of the hopping matrix with zero determinant\cite{shiba:77} (see below) they 
obtained the thermoelectric power of the model. 
More recently,\cite{shvaika:43704} the general case of the hopping matrix was addressed on 
a $D\to\infty$ hypercubic lattice with the Gaussian density of states (DOS). 
Unfortunately, in this case one encounters several difficulties related  to an infinite bandwidth 
and a finite ``quasiparticle'' scattering time at $\omega\to\pm\infty$.\cite{freericks:195120} 
In the case of  a Bethe lattice with a semi elliptic density of states, the bandwidth is finite 
and  the ``quasiparticle'' scattering time vanishes, whenever the DOS is zero, 
so one is closer to the three-dimensional systems.\cite{joura:165105}
Here, we calculate the single particle and two particle properties of an infinite-dimensional Falicov-Kimball model with correlated hopping and show that, unlike the single particle DOS, the transport properties exhibit a number of surprising features.  
In infinite dimensions,  the canonical model can have, at very low temperatures, a phase transition to an ordered phase. 
(For a Bethe lattice there are only two possibilities: two-sublattice chessboard charge-density-wave ordering 
or segregation into two phases with different particle densities.\cite{freericks:13438})
However,  since the phase diagram for the model with correlated hopping is not known, we present the results 
for the homogeneous phase all the way down to $T=0$. 

The paper is organized as follows. 
In Sec.~\ref{sec:formalism}, we present the DMFT solution for the Falicov-Kimball model with correlated hopping on a Bethe lattice. 
Section~\ref{sec:transport} provides the derivation of the charge and energy transport coefficients in a homogeneous phase. 
In Sec.~\ref{sec:results}, we consider peculiarities of the charge and heat transport for different values of the 
correlated hopping and doping. The results are summarized in Sec.~\ref{sec:conclusions}. 

\section{DMFT formalism for correlated hopping on Bethe lattice}\label{sec:formalism}

\subsection{The model Hamiltonian}
The Hamiltonian of the Falicov-Kimball model\cite{falicov:997} with correlated hopping has the form
\begin{align}
	H &= H_{\textrm{loc}} + H_t,
	\nonumber\\
	H_{\textrm{loc}} &= \sum_i \left[Un_{id}n_{if} - \mu_f n_{if} - \mu_d n_{id}\right],
	\nonumber\\
	H_t&=\sum_{\langle ij\rangle} \frac{t_{ij}^{*}}{\sqrt{Z}} \Bigl[t_1d_i^{\dag}d_j +
	t_2d_i^{\dag}d_j\left(n_{if}+n_{jf}\right) 
	\nonumber\\
	&+ t_3d_i^{\dag}d_jn_{if}n_{jf}\Bigr],
\end{align}
where $H_{\textrm{loc}}$ describes local correlations between the itinerant $d$ electrons and localized $f$ electrons 
and $H_{t}$ describes nonlocal terms on the Bethe lattice with infinite coordination number, $Z\rightarrow\infty$, 
including the nearest-neighbor inter-site hopping with amplitude $t_1$ and nonlocal correlations with amplitudes $t_2$ and $t_3$ --- 
the so-called correlated hopping.
Because the number of localized particles is conserved, $[n_{if},H]=0$, one can introduce the projection operators $P_i^+=n_{if}$ and $P_i^-=1-n_{if}$, and define the 
projected $d$-electron operators
\begin{align}
\bm{d}_i&=\begin{pmatrix} d_i P_i^+ \\ d_i P_i^- \end{pmatrix}, 
\label{eq:proj_d}
\end{align}
such that the nonlocal term assumes a matrix form,\cite{shvaika:075101}
\begin{align}
H_t&=\sum_{\langle ij\rangle} \frac{t_{ij}^{*}}{\sqrt{Z}} \Bigl[
t^{++}P_i^+d_i^{\dag}d_jP_j^+ + t^{--}P_i^-d_i^{\dag}d_jP_j^-
\nonumber\\
&+t^{+-}P_i^+d_i^{\dag}d_jP_j^- +
t^{-+}P_i^-d_i^{\dag}d_jP_j^+\Bigr]
\nonumber\\
&= \sum_{\langle ij\rangle} \frac{t_{ij}^{*}}{\sqrt{Z}} \bm{d}_i^{\dag}\mathbf{t}\bm{d}_j~.
\end{align}
The hopping matrix is 
\begin{align}
\mathbf{t}&=\begin{bmatrix}
t^{++} & t^{+-} \\
t^{-+} & t^{--}
\end{bmatrix} 
\label{eq:t_mtrx}
\end{align}
and the connection between its matrix elements and initial hopping amplitudes reads
\begin{align}
t^{--}&=t_1,             &t_1&=t^{--},\nonumber \\
t^{+-}&=t^{-+}=t_1+t_2,  &t_2&=t^{+-(-+)}-t^{--},
\\
t^{++}&=t_1+2t_2+t_3,    &t_3&=t^{++}+t^{--}-t^{+-}-t^{-+}.
\nonumber
\end{align}
Our aim is to find the transport properties of the model which are related, 
for $Z\rightarrow\infty$, to the single particle Green's function. 

\subsection{The single particle Green's function}

The off-diagonal Green's function for projected $d$ electrons is defined by the matrix 
$\mathbf{G}_{ij}=[ {G}_{ij}^{\alpha\beta}]$, where ${\alpha,\beta=\pm} $. 
On the imaginary time-axis we have 
\begin{equation}
\mathbf{G}_{ij}(\tau-\tau') = -\left\langle \mathcal{T} \bm{d}_i(\tau) \otimes \bm{d}_j^{\dagger}(\tau')\right\rangle \,, 
\end{equation}
where $\mathcal{T} $ is the imaginary-time ordering operator, $ \otimes $ denotes the direct (Cartesian) product of two vectors,
and the angular bracket denotes the quantum statistical 
averaging with respect to $H$. 
The Green's function is calculated by treating $H_t$ as perturbation,  i.e.,  by expanding around the atomic limit. This leads 
to the Dyson-type equation which can be written in the matrix form as 
\begin{align}
\mathbf{G}_{ij}(\omega) = \bm{\Xi}_{ij}(\omega) + \sum_{\langle i'j'\rangle} \bm{\Xi}_{ij'}(\omega)  \cdot \frac{t_{j'i'}^{*}}{\sqrt{Z}} \mathbf{t} \cdot \mathbf{G}_{i'j}(\omega),
\end{align}
where $\bm{\Xi}_{ij}(\omega)$ is the irreducible cumulant,\cite{metzner:8549,georges:13}  
which cannot be split into two disconnected parts by removing a single hopping line.

In the $Z\to\infty$  limit, the irreducible cumulant is local,\cite{metzner:8549}
\begin{equation}
\bm{\Xi}_{ij}(\omega) = \delta_{ij} \bm{\Xi}(\omega),
\end{equation}
and can be computed by the DMFT. In that approach, the local Green's functions of the lattice is equated with the Green's functions of an auxiliary impurity embedded in a self-consistent bath, described 
by the time-dependent mean fields ($\lambda$ fields). 
Introducing the unperturbed  DOS of the Bethe lattice, 
\begin{equation}\label{eq:dos}
\rho(\epsilon) = \frac{2}{\pi W^2}\sqrt{W^2-\epsilon^2} \,, 
\end{equation}
we write the DMFT equation in the matrix form as\cite{shvaika:075101}
\begin{align}   \label{eq:DMFT} 
 \mathbf{G}_{\text{local}}(\omega)
 &\equiv\mathbf{G}_{ii}(\omega)=  \int\limits_{-\infty}^{+\infty}d\epsilon \rho(\epsilon)
  \mathbf{G}_{\epsilon} (\omega) 
  \nonumber\\
 & = 
 \left[\bm{\Xi}^{-1}(\omega) - \bm{\Lambda}(\omega)\right]^{-1} 
  = \mathbf{G}_{\text{imp}}(\omega),
\end{align}
where $\bm{\Lambda}(\omega)=[\lambda^{\alpha\beta}(\omega)]$ is the $\lambda$ matrix,  
$\mathbf{G}_{\text{imp}}(\omega)$ is the Green's function for the auxiliary impurity problem, and 
\begin{align}
\mathbf{G}_{\epsilon} (\omega)&=\left[\bm{\Xi}^{-1}(\omega) - \mathbf{t}\epsilon\right]^{-1},	
\end{align}
is the lattice Green's function matrix with the components
\begin{align}
G_{\epsilon}^{\beta\alpha}(\omega) &= \frac{A_{\beta\alpha}(\omega) - B_{\beta\alpha}\epsilon}{C(\omega) - D(\omega)\epsilon + \epsilon^2 \det\mathbf{t}} .
\label{GkvsE}
\end{align}
Here, we introduced the matrix adjugate to $\bm{\Xi}^{-1}(\omega)$, 
\begin{equation}
\mathbf{A}(\omega) = \adj \bm{\Xi}^{-1}(\omega) = \bm{\Xi}(\omega)/\det\mathbf{\Xi}(\omega)
\end{equation}
and the matrix adjugate of the hopping matrix $\mathbf{t}$, 
\begin{equation}
\mathbf{B} = \adj \mathbf{t} = \mathbf{t}^{-1}\det\mathbf{t} \,.
\end{equation}
For the $2\times2$ matrices,  the scalars $C(\omega)$ and $D(\omega)$ are given by 
\begin{equation}
C(\omega) = \det \mathbf{A}(\omega) = \det \bm{\Xi}^{-1}(\omega) = 1/\det \bm{\Xi}(\omega),
\end{equation} 
and
\begin{align}
D(\omega) = \Tr \left[\mathbf{A}(\omega)\mathbf{t}\right] = \Tr \left[\bm{\Xi}^{-1}(\omega)\mathbf{B}\right].
\end{align}
The DMFT equation \eqref{eq:DMFT} contains two unknowns: the irreducible cumulant $\bm{\Xi}(\omega)$ and 
the dynamical mean field $\bm{\Lambda}(\omega)$,  both of which are related to  $\mathbf{G}_{\text{imp}}(\omega)$. 
The Green's function of the impurity model with correlated hopping
is given by the exact expression
\begin{align}
  G_{\text{imp}}^{++}(\omega) & = w_1 g_1(\omega),
 \nonumber \\
  G_{\text{imp}}^{--}(\omega) & = w_0 g_0(\omega),
  \nonumber\\
  G_{\text{imp}}^{+-}(\omega) & = G_{\text{imp}}^{-+}(\omega) = 0,
  \label{eq:Gimp}
\end{align}
where $w_1=\langle P^+\rangle=\langle n_f\rangle$, $w_0=\langle P^-\rangle=\langle 1- n_f\rangle$, and
\begin{align}
	g_0(\omega) & =	\frac{1}{\omega+\mu_d - \lambda^{--}(\omega)},
	\nonumber \\
	g_1(\omega) & = \frac{1}{\omega+\mu_d - U - \lambda^{++}(\omega)}
	\label{eq:gimp0}
\end{align}
are the impurity Green's functions of a conduction electron 
in the presence of an $f$ state which is either permanently empty or occupied 
(i.e., locators in the CPA theory\cite{blackman:2412}).
The impurity Green's function yields the renormalized DOS of the lattice, 
\begin{align}
	A_d(\omega) &= -\frac{1}{\pi} \sum_{\alpha,\beta=\pm} \Img G_{\text{imp}}^{\alpha\beta}(\omega)
	\nonumber\\
	&= -\frac{1}{\pi} \left[w_0 \Img g_{0}(\omega) + w_1 \Img g_{1}(\omega)\right] \; .
\end{align}
and, for the Bethe lattice, we can write the DMFT equation \eqref{eq:DMFT} as\cite{shvaika:075101}
\begin{equation}\label{eq:DMFT_Bethe}
\bm{\Lambda}(\omega)=\frac{W^2}{4}\mathbf{t}\mathbf{G}_{\text{imp}}(\omega)\mathbf{t}.
\end{equation}
In numerical calculations, we use $W=2$, which defines our energy scale. 
Equation~\eqref{eq:gimp0} allows us to express the diagonal components of the $\bm{\Lambda}$ matrix in Eq.~\eqref{eq:DMFT_Bethe} in terms of $g_0$ and $g_1$, and write the system of equations
\begin{align}
&\omega+\mu_d-U-\frac{1}{g_1(\omega)}
\nonumber\\
&=\frac{W^2}{4}\left[(\tpp)^2 w_1 g_1(\omega)+(\tpm)^2 w_0 g_0(\omega)\right] ,
\nonumber\\
&\omega+\mu_d-\frac{1}{g_0(\omega)}
\nonumber\\
&=\frac{W^2}{4}\left[(\tpm)^2 w_1 g_1(\omega)+(\tmm)^2 w_0 g_0(\omega)\right] ,
\label{eq:g0system}
\end{align}
which, in general, provides the fourth-order  polynomial equations for $g_0(\omega)$ or $g_1(\omega)$.

Previous investigations of the Falicov-Kimball model with correlated hopping\cite{shvaika:075101,shvaika:43704} have shown that 
the renormalized DOS and transport properties depend strongly on the structure of the hopping matrix \eqref{eq:t_mtrx}, 
and that one can distinguish five different cases (in what follows, unless  stated explicitly, we take $\tmm=t_1=1$):
\begin{enumerate}
	\item[(a)] For $\tpp \neq 0$, $\tpm \neq 0$, and $\det\mathbf{t}=0$, the right-hand parts 
	of Eqs.~\eqref{eq:g0system} are proportional to each other, so that 
	\begin{align}
		&\frac{1}{\tpp}\left[\omega+\mu_d-U-\frac{1}{g_1(\omega)}\right]
	\nonumber\\
		&=\frac{1}{\tmm}\left[\omega+\mu_d-\frac{1}{g_0(\omega)}\right]
	\nonumber\\
		&=\frac{W^2}{4}\left[\tpp w_1 g_1(\omega)+\tmm w_0 g_0(\omega)\right] .
	\label{eq:g0system0}
	\end{align}
	This case was considered by Shiba\cite{shiba:77} and Hoshino and Niizeki.\cite{hoshino:1320}
	The regular Falicov-Kimball model, without the correlated hopping, belongs also to this case.
	\item[(b)] For $\tpp = 0$, $\tpm \neq 0$, and $\det\mathbf{t} \neq 0$, one of the diagonal components 
	of the hopping matrix vanishes, so that the direct hopping of $d$ particles between 
	the sites occupied by the $f$ particles is reduced.
	\item[(c)] For $\tpp \neq 0$, $\tpm = 0$, and  $\det\mathbf{t} \neq 0$, the  hopping matrix is diagonal 
	and the hopping of $d$ particles is allowed only between the sites with the same occupation of $f$ states.
	\item[(d)] For $\tpp = 0$, $\tpm = 0$, and $\det\mathbf{t} = 0$ we are dealing with the simultaneous realization 
	of all three previous cases, and the hopping of $d$ particles is allowed only between the sites 
	which are {\it not} occupied by the $f$ particles.
	\item[(e)] The most general case is obtained for $\tpp \neq 0$, $\tpm \neq 0$, $\det\mathbf{t} \neq 0$.
\end{enumerate} 
For the cases (a,b) and (c,d), the impurity Green's functions $g_0(\omega)$ or $g_1(\omega)$ are defined by the 
cubic and quadratic equations, respectively. 
Within the solutions of these equations, we choose the one with negative imaginary parts, which yields the retarded Green's functions. 
For the general case (e), the quartic polynomial equation with real coefficients has either four real roots, or two real and two mutually 
conjugated complex roots, or two pairs of mutually conjugated complex roots. The correct physical solution is the one 
with negative imaginary parts, which gives the retarded Green's functions $g_0(\omega)$ and $g_1(\omega)$.  
It can be shown that there is always just a single set of physical solutions. 
When $\omega$ has only the real roots, the physical solution is obtained by using the spectral relation
\begin{equation}
\Real g_{0,1}(\omega)=-\frac{1}{\pi}\int_{-\infty}^{+\infty} d\omega^{\prime} \frac{\Img g_{0,1}(\omega^{\prime})}{\omega-\omega^{\prime}}, 
\end{equation}
which yields the correct retarded Green's functions and renormalized single-particle density of states.

For the local single-particle Green's function 
\begin{equation}
G_{ii}(\tau-\tau') = -\left\langle \mathcal{T} d_i(\tau) d_i^{\dagger}(\tau')\right\rangle ,
\end{equation}
we have 
\begin{align}
G_{ii}(\omega) &= \sum_{\alpha,\beta=\pm} G_{\text{imp}}^{\alpha\beta}(\omega) = w_0 g_{0}(\omega) + w_1 g_{1}(\omega)
\nonumber\\
&= \left[\omega + \mu_d - \Sigma(\omega) - \lambda_{\text{HF}}(\omega) \right]^{-1},
\label{eq:G_ii}
\end{align}
where $\lambda_{\text{HF}}(\omega)=w_1\lambda^{++}(\omega)+w_0\lambda^{--}(\omega)$ is the Hartree-Fock dynamical mean field 
and
\begin{equation}
\Sigma(\omega) = U w_1 + \frac{\tilde{U}^2(\omega)w_1w_0}{\omega + \mu_d - \tilde{U}(\omega) w_0 - \lambda^{--}(\omega)}
\end{equation}
is local self-energy, which is different from the nonlocal one for the lattice Green's function.\cite{shvaika:075101} 
The parameter
\begin{equation}
\tilde{U}(\omega) = U + \lambda^{++}(\omega) - \lambda^{--}(\omega)
\end{equation}
is an effective retarded Coulomb interaction. These expressions reveal the dual nature of correlated hopping: 
on one hand, it modifies the hopping amplitude in the Hartree-Fock term $\lambda_{\text{HF}}(\omega)$, 
and, on the other hand, it leads to the nonlocal two-particle correlation, which gives rise to the retarded local interaction.

For a given value of the $d$-electron concentration, $n_d=\langle n_d\rangle$, the chemical potential $\mu_d$ is obtained by solving the equation
\begin{equation}
n_d = -\frac{1}{\pi} \int_{-\infty}^{+\infty} d\omega f(\omega) \Img G_{ii}(\omega),
\end{equation}
where $f(\omega)=1/(e^{\omega/T}+1)$ is the Fermi function and $G_{ii}(\omega)$ is given by Eq.~\eqref{eq:G_ii}.

\section{Transport coefficients in the presence of correlated hopping}\label{sec:transport}

We now calculate the transport properties of correlated electrons by linear response theory. 
The Falicov-Kimball model with correlated hopping satisfies the Boltzmann (Jonson-Mahan) 
theorem,\cite{jonson:4223,jonson:9350,freericks:035133,shvaika:43704} so that all  transport 
integrals follow from a single transport function, 
\begin{equation}\label{JM}
\mathbf{L}_{lm} = \frac{\sigma_0}{e^2} \int_{-\infty}^{+\infty} d \omega \left[-\frac{d f(\omega)}{d \omega}\right] \mathbf{I}(\omega) \omega^{l+m-2},
\end{equation}
where $\mathbf{I}(\omega)$ is the transport function  
which is given below. 

The dc charge conductivity $\bm{\sigma}_{\textrm{dc}} $, the Seebeck coefficient (thermoelectric power $\bm{E}=\mathbf{S}\nabla T$), 
and the electronic contribution to thermal conductivity $\bm{\kappa}_{\textrm{e}}$ follow immediately from the transport integrals as,  
\begin{eqnarray}
\bm{\sigma}_{\textrm{dc}} &=& e^2 \mathbf{L}_{11} ,
\\[2ex]
\mathbf{S} &=&  \frac{1}{eT} \mathbf{L}_{11}^{-1} \mathbf{L}_{12},
\label{thermcond}
\\[2ex]
\bm{\kappa}_{\textrm{e}} &=&  \frac1T \left[\mathbf{L}_{22} - \mathbf{L}_{21} \mathbf{L}_{11}^{-1} \mathbf{L}_{12}\right]~.
\end{eqnarray}
 
The DMFT expression for transport function, generalized to the case of correlated hopping, reads
\begin{align}
&I(\omega) = \frac{1}{\pi} \int d \epsilon \rho(\epsilon) \Phi_{xx}(\epsilon) \Tr \left[\mathbf{t}\, \Img \mathbf{G}_{\epsilon} (\omega)\,  \mathbf{t}\, \Img \mathbf{G}_{\epsilon}(\omega)\right]
\nonumber\\
&= \frac{1}{\pi} \sum_{\alpha\beta\alpha'\beta'}\!\! t^{\alpha\beta} t^{\alpha'\beta'} \!\!\!\int\!\! d \epsilon \rho(\epsilon) \Phi_{xx}(\epsilon) \Img G_{\epsilon}^{\beta\alpha'} (\omega) \Img G_{\epsilon}^{\beta'\alpha}(\omega), 
\label{eq:transport}
\end{align}
where $\Phi_{xx}(\epsilon)$  is the so-called  lattice-specific transport DOS.\cite{arsenault:205109} 
For a $D=\infty$ hypercubic lattice with Gaussian DOS, we have 
$\Phi_{xx}(\epsilon)=W^2/2D$, whereas for the $Z=\infty$ Bethe lattice with semielliptic DOS,  
the $f$-sum rule yields\cite{chung:11955}
\begin{equation}
\Phi_{xx}(\epsilon) = \frac{1}{3Z}\left(W^2-\epsilon^2\right) \,.
\label{eq:Phi_Bethe}
\end{equation}
The integral over  $\epsilon$  in Eq.~\eqref{eq:transport} can now be evaluated and we find that 
the final result depends on the value of $\det \mathbf{t}$. 

For  $\det \mathbf{t}=0$, we find
\begin{equation}\label{eq:I0w}
I(\omega)=\frac{1}{2\pi} \left\{ \Real \Psi^{\prime}\left[\frac{C(\omega)}{D(\omega)} \right] - 
\frac{\Img \Psi\left[\frac{C(\omega)}{D(\omega)}\right] }{\Img \left[\frac{C(\omega)}{D(\omega)}\right]} \right\}  ,
\end{equation}
where
\begin{align}
	\Psi(\zeta)&=\int d \epsilon \frac{\rho(\epsilon)}{\zeta - \epsilon} \Phi_{xx}(\epsilon) ,
	\nonumber \\
	\Psi^{\prime}(\zeta)&=\frac{d \Psi(\zeta)}{d \zeta}.
\end{align}
For the semielliptic DOS, we find
\begin{align}
\Psi(\zeta)&=\frac{1}{3}\left[(W^2 -\zeta^2) F(\zeta)+ \zeta \right] ,
\nonumber \\
\Psi^{\prime}(\zeta)&=\frac{1}{3}\left[(W^2 -\zeta^2) F^{\prime}(\zeta)+1-2 \zeta F(\zeta) \right] ,
\end{align}
where 
\begin{align}
F(\zeta) &= \int d \epsilon \frac{\rho(\epsilon)}{\zeta - \epsilon}= \frac{2}{W^2}\left(\zeta - \sqrt{\zeta^2 -W^2}\right) ,
\nonumber\\
F^{\prime}(\zeta) &=  \frac{d F(\zeta)}{d \zeta}=\frac{\zeta F(\zeta)-2 }{\zeta^2 - W^2}.
\end{align}
The transport function \eqref{eq:I0w} is similar to the one for the Falicov-Kimball model without correlated hopping,  
provided we replace the inverse irreducible cumulant $\Xi^{-1}(\omega)=\omega+\mu_d-\Sigma(\omega)$ by 
the expression 
\begin{equation}
\frac{C(\omega)}{D(\omega)}=\frac{1}{\Tr \left[\mathbf{\Xi}(\omega)\mathbf{t}\right]}.
\end{equation}

For  $\det \mathbf{t}\neq0$, the transport function reads
\begin{widetext}
\begin{align}
I(\omega)&=\frac{1}{2\pi} \left[ \Real \left\{\Psi^{\prime}[E_1(\omega)]+\Psi^{\prime}[E_2(\omega)] \right\} - \frac{\Img \Psi\left[E_1(\omega)\right] }{\Img E_1(\omega)}-\frac{\Img \Psi\left[E_2(\omega)\right] }{\Img E_2(\omega)}\right.
\nonumber\\
&-\left. K(\omega)\left\lbrace\frac{1}{\Img E_1(\omega)}\Img\frac{ \Psi[E_1(\omega)] } {\left[E_1(\omega)-E_2(\omega)\right]\left[E_1(\omega)-E^*_2(\omega)\right]} 
+\frac{1}{\Img E_2(\omega)}\Img \frac{ \Psi[E_2(\omega)] } {\left[E_2(\omega)-E_1(\omega)\right]\left[E_2(\omega)-E^*_1(\omega)\right]}\right\rbrace
\right],
\label{eq:Iw}
\end{align}
\end{widetext}
where $E_{1}$ and $E_{2}$ are the roots of the denominator in Eq.~\eqref{GkvsE}, 
$C(\omega) - D(\omega)\epsilon + \epsilon^2 \det\mathbf{t}=0$, given by  
\begin{align}
E_{1}(\omega) &= \frac{D(\omega)}{2\det \mathbf{t}} \left[ 1 + \sqrt{1 - \frac{4C(\omega)}{D^2(\omega)}\det \mathbf{t}}\, \right],
\label{eq:E1}
\\
E_{2}(\omega) &= \frac{2C(\omega)}{D(\omega)} \left[ 1 + \sqrt{1 - \frac{4C(\omega)}{D^2(\omega)}\det \mathbf{t}}\, \right]^{-1},
\label{eq:E2}
\end{align}
and $K(\omega)$ reads
\begin{equation}
K(\omega)=2 \Real [E_1(\omega)E^*_2(\omega)]-\frac{1}{\det \mathbf{t}}\Real \Tr [\mathbf{A}^*(\omega) \bm{\Xi}^{-1}(\omega)].
\end{equation}
In the limit $\det \mathbf{t}\to0$, the eigenvalue $E_{1}$ diverges and its contribution to the transport function vanishes, 
whereas $E_2(\omega)\to C(\omega)/D(\omega)$, so that we recover the transport function given by expression 
\eqref{eq:I0w}.

\section{Results}\label{sec:results}

In the first part of this section we show how the interacting density of states $A_d(\omega)$ and transport function $I(\omega)$ 
depend on frequency $\omega$  and correlated hopping $t_2$. 
The functional form of both functions, changes with the concentration of $f$ electrons $n_f=\langle n_f\rangle$ 
but does not depend on the chemical potential of $d$ particles $\mu_d$. 
Thus, we plot $A_d(\omega)$ and $I(\omega)$ with $\omega+\mu_d$ on the abscissa. 
Furthermore,  since $t_2$ and $t_3$ enter all the expressions only through the matrix elements $t^{\alpha\beta}$
defined in Eq.~\eqref{eq:t_mtrx}, we show the results for $t_3=0$ only. 
Finite values of $t_3$ can change the results in a quantitative but not in a qualitative way.
The data are presented  for various concentrations of $f$ particles, 
in the weak ($U=0.25$) and the strong ($U=2$) coupling regime.  
In the second part of this section, we discuss the effect of correlated hopping on $A_d(\omega)$ and  
$I(\omega)$, for typical values of the parameters, and show the corresponding transport coefficients in the weak- and strong-coupling limit.

\subsection{Overall features of the DOS and transport function}\label{sec:A and I}

\begin{figure*}    
	\centering
	\includegraphics[width=0.4\linewidth]{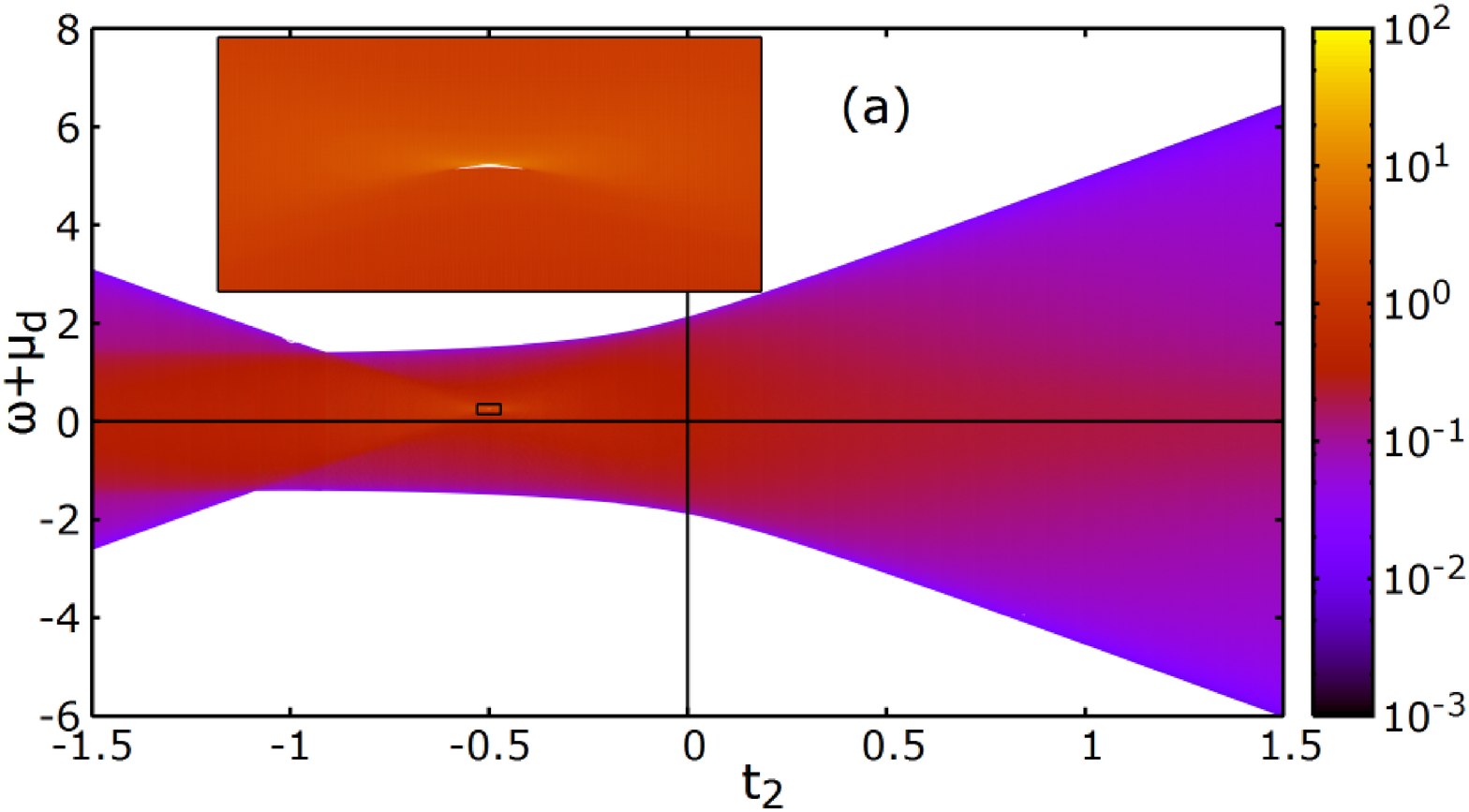}\quad
	\includegraphics[width=0.4\linewidth]{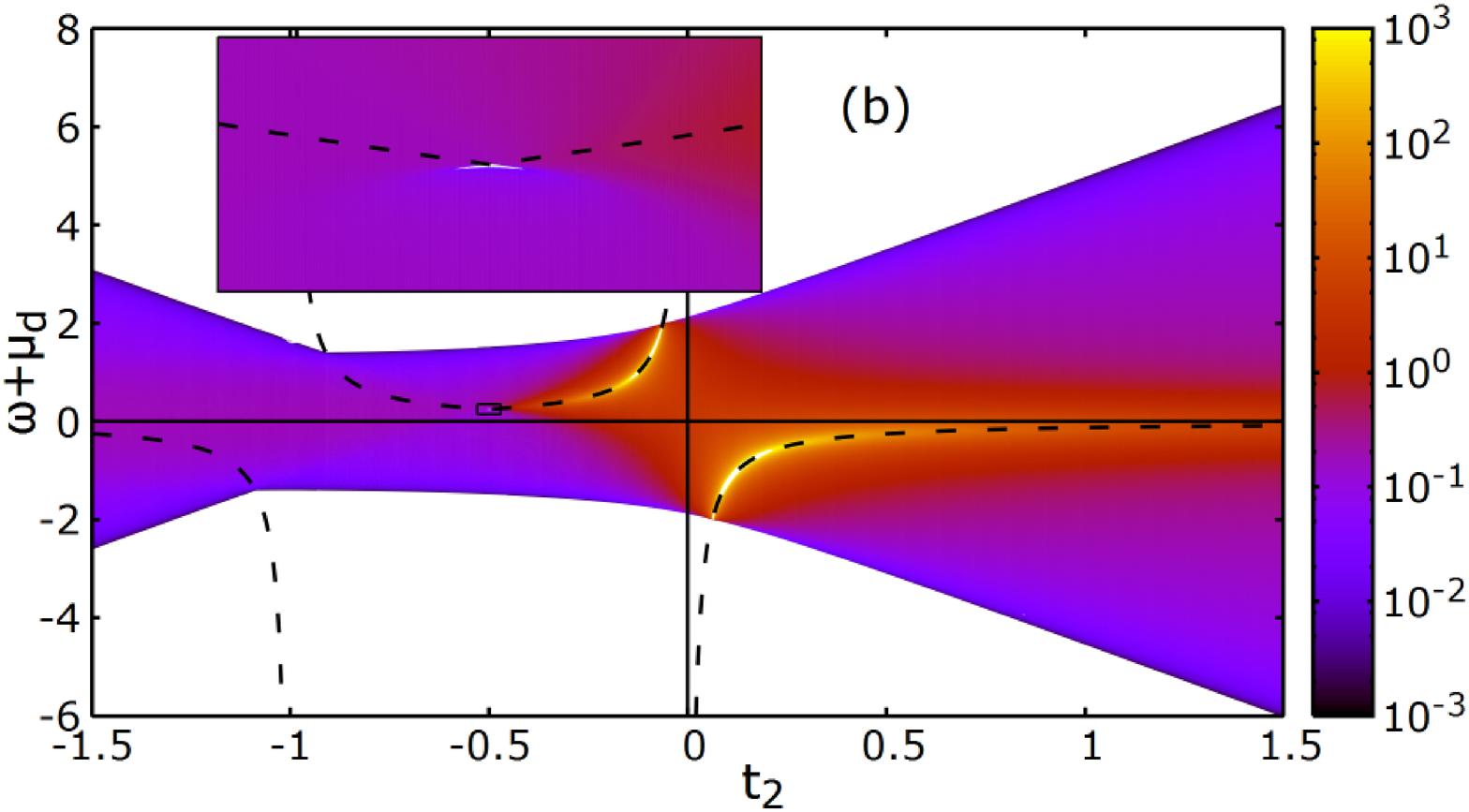}\\
	\includegraphics[width=0.4\linewidth]{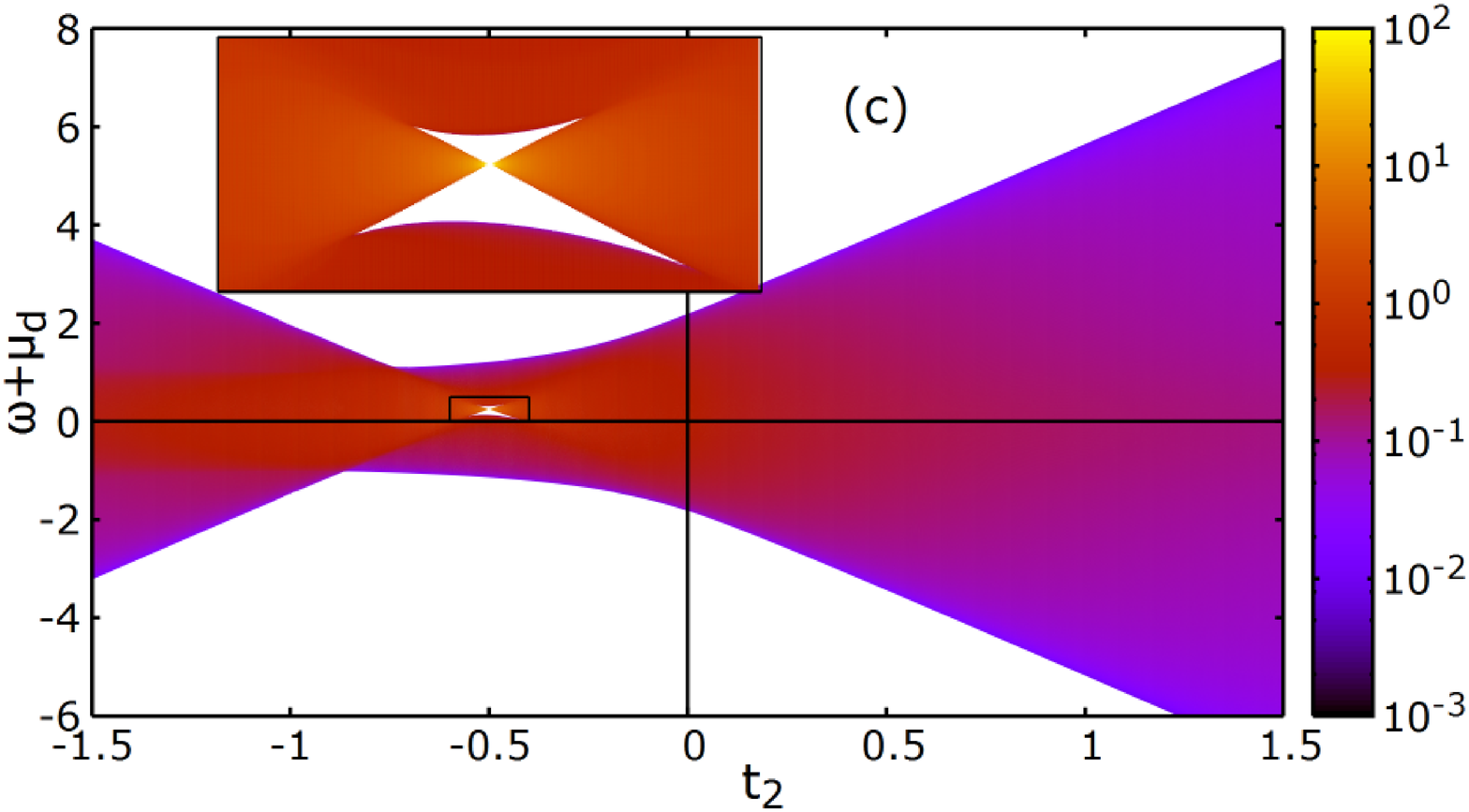}\quad
	\includegraphics[width=0.4\linewidth]{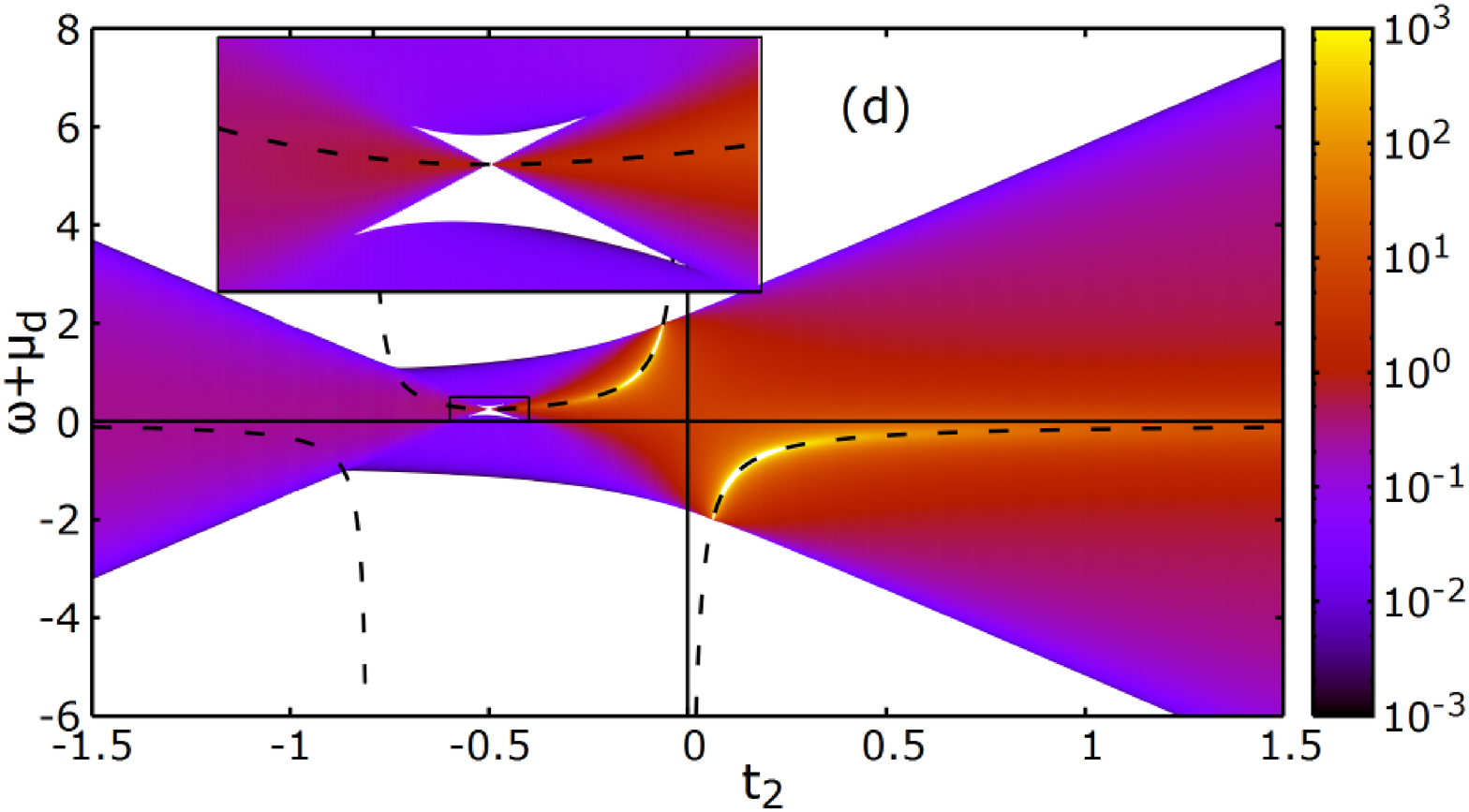}\\
	\includegraphics[width=0.4\linewidth]{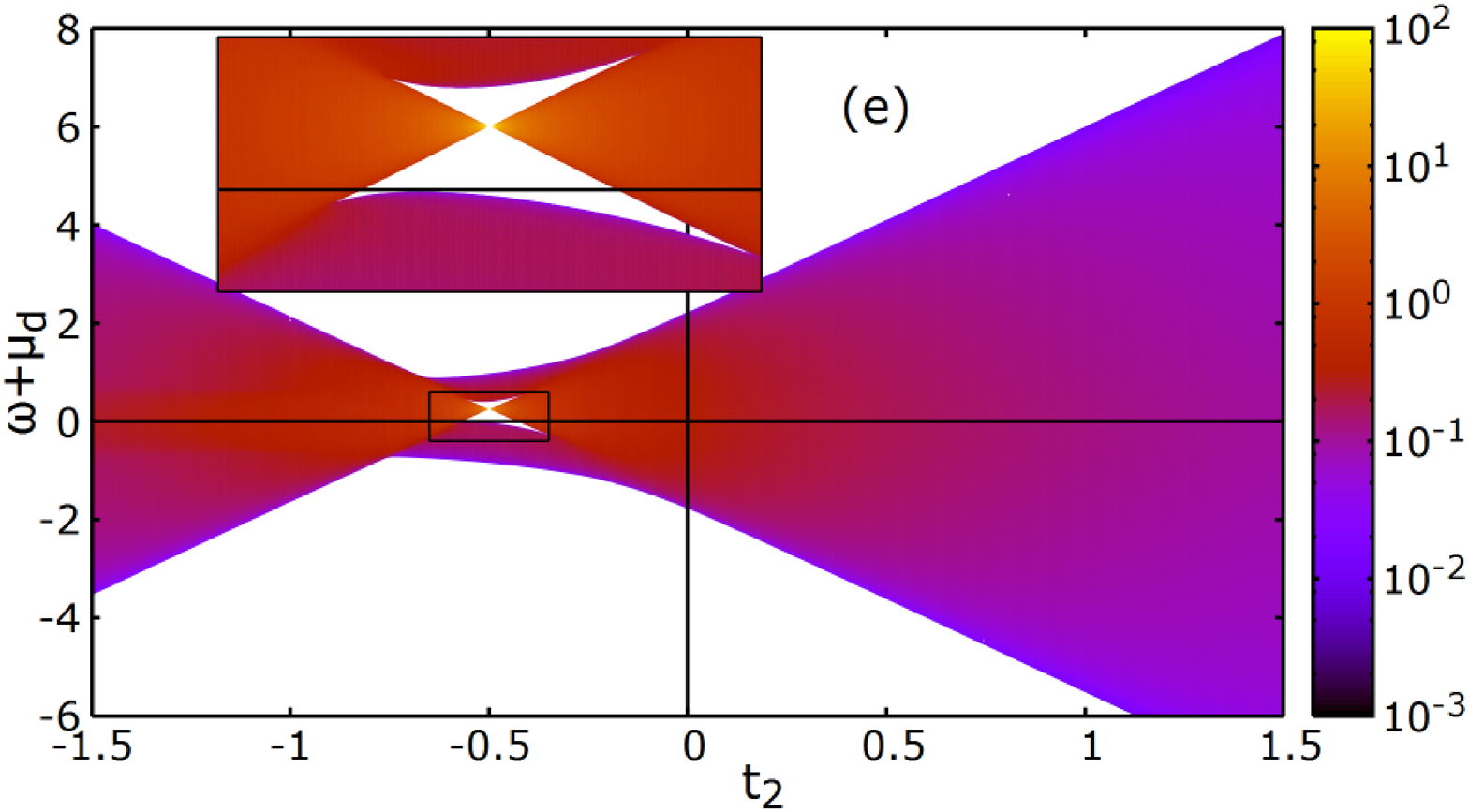}\quad
	\includegraphics[width=0.4\linewidth]{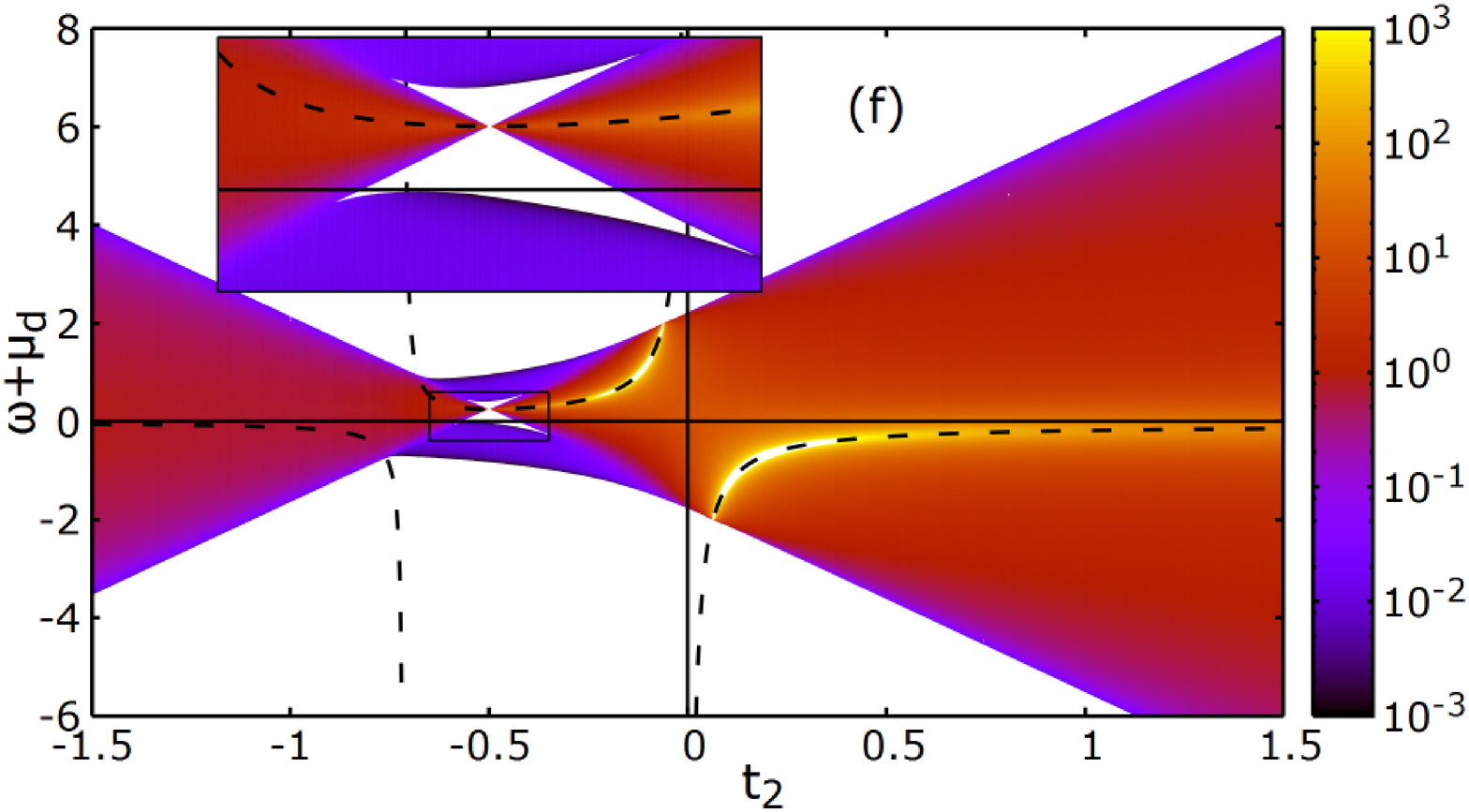}
	\caption{(Color online) 
	The density of states $A_d(\omega)$, left panels (a), (c), (e), and transport function $I(\omega)$, right panels (b), (d), (f), shown as a function of frequency $\omega$ and correlated hopping parameter $t_2$ on a false color plot for $U=0.25$, $t_3=0$, and for $f$-particle concentrations (a,b) $n_f=0.5$, (c,d) $0.75$, and (e,f) $0.9$ (top to bottom). Inserts represent enlarged rectangular areas 
	with (a,b) $t_2\in[-0.53,-0.47]$, $\omega+\mu_d\in[0.2,0.3]$, (c,d) $t_2\in[-0.6,-0.4]$, $\omega+\mu_d\in[0.0,0.5]$, and (e,f) $t_2\in[-0.65,-0.35]$, $\omega+\mu_d\in[-0.4,0.6]$.
	Dashed line indicates the resonant frequency $\omega_{\text{res}}$ given by Eq.~\eqref{eq:w_res}.}
	\label{fig:log_u=025}
\end{figure*}

\begin{figure*}    
	\centering
	\includegraphics[width=0.4\linewidth]{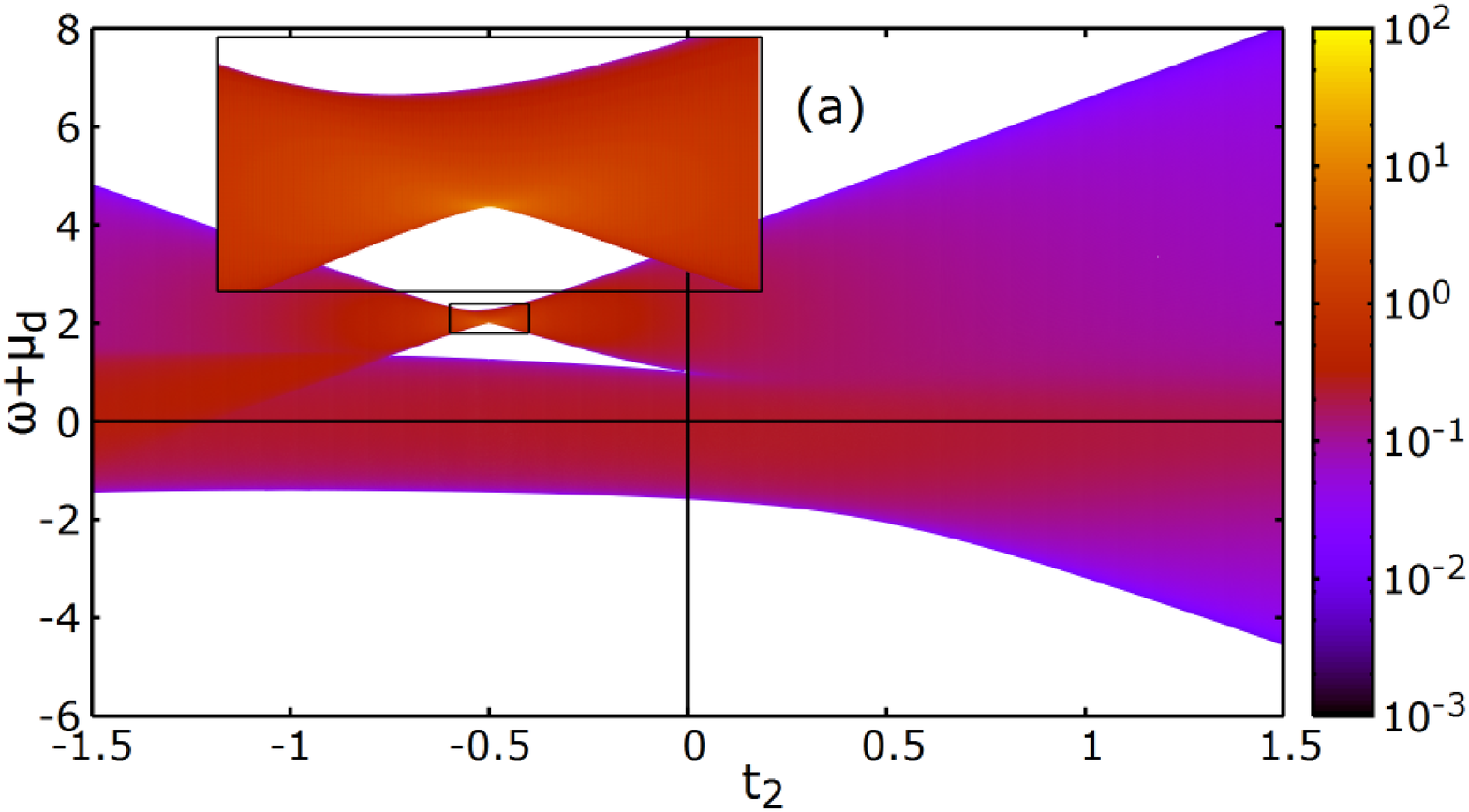}\quad
	\includegraphics[width=0.4\linewidth]{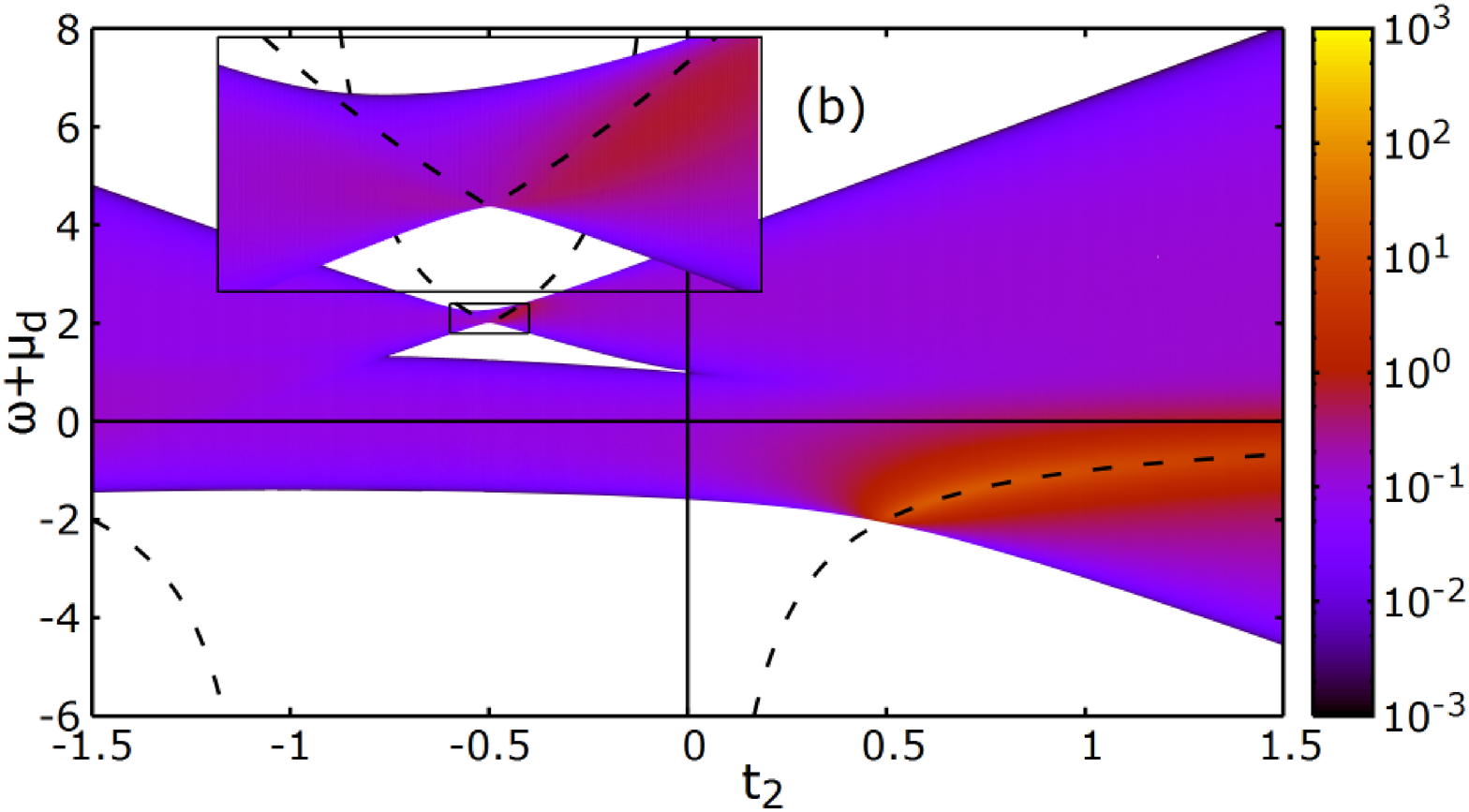}\\
	\includegraphics[width=0.4\linewidth]{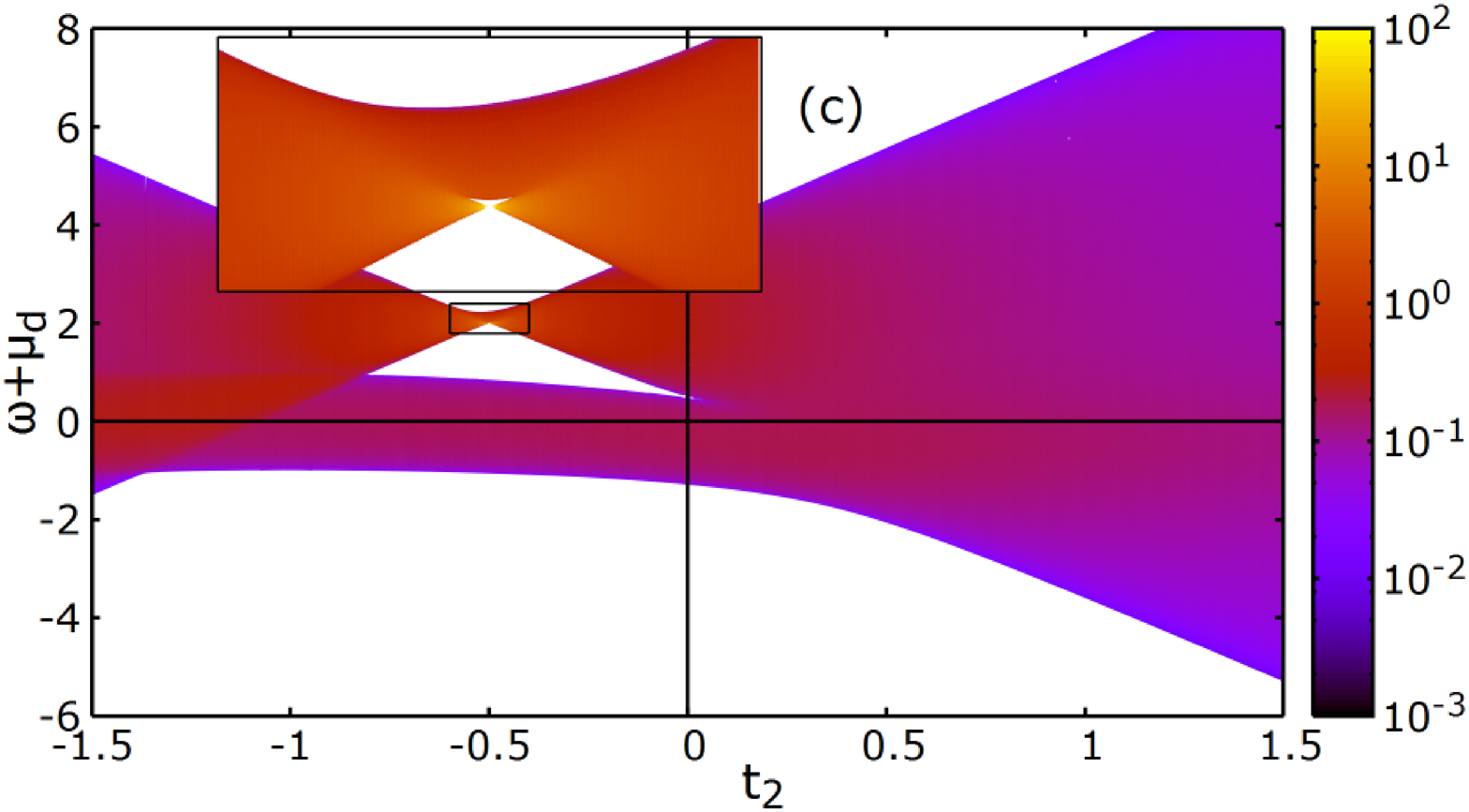}\quad
	\includegraphics[width=0.4\linewidth]{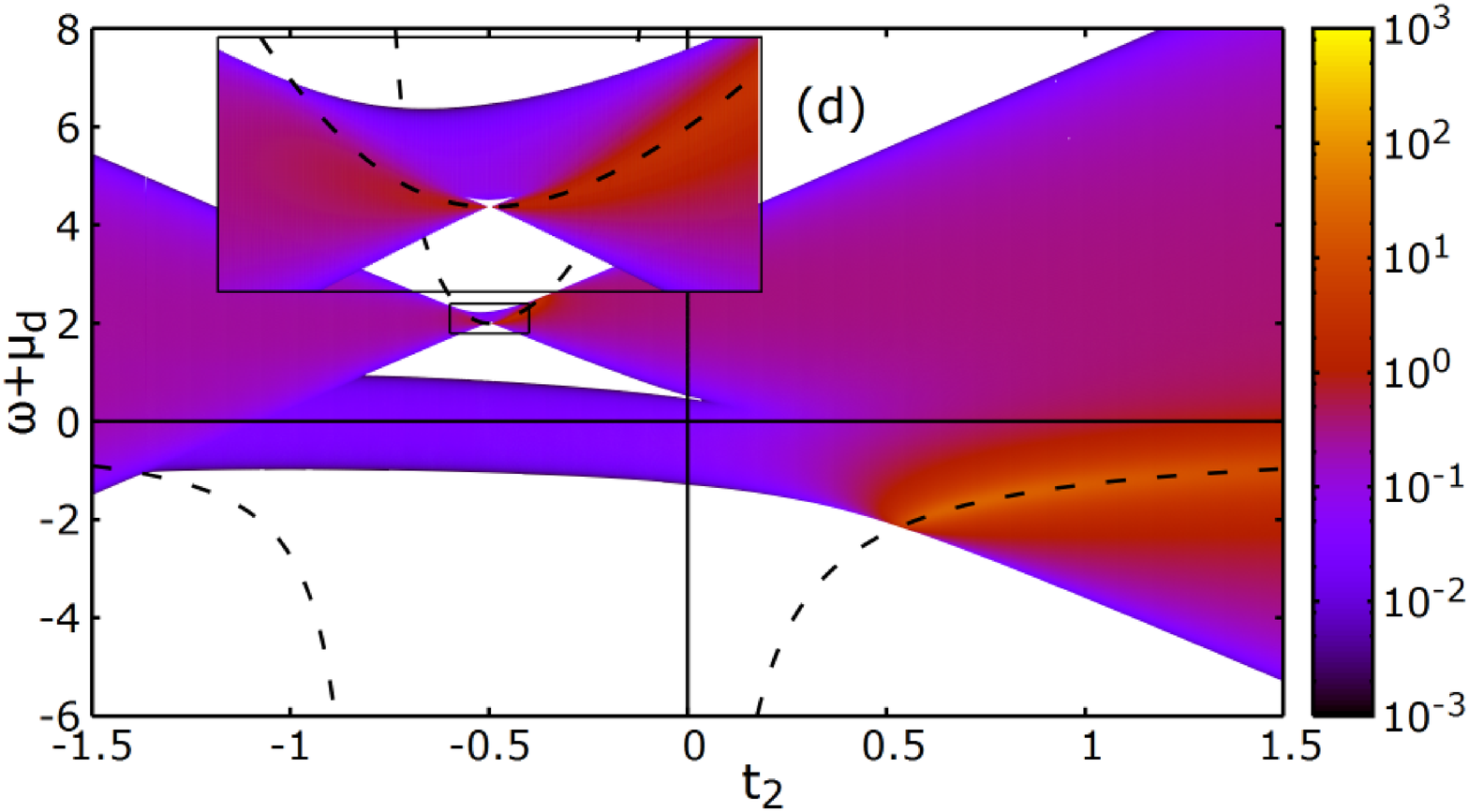}\\
	\includegraphics[width=0.4\linewidth]{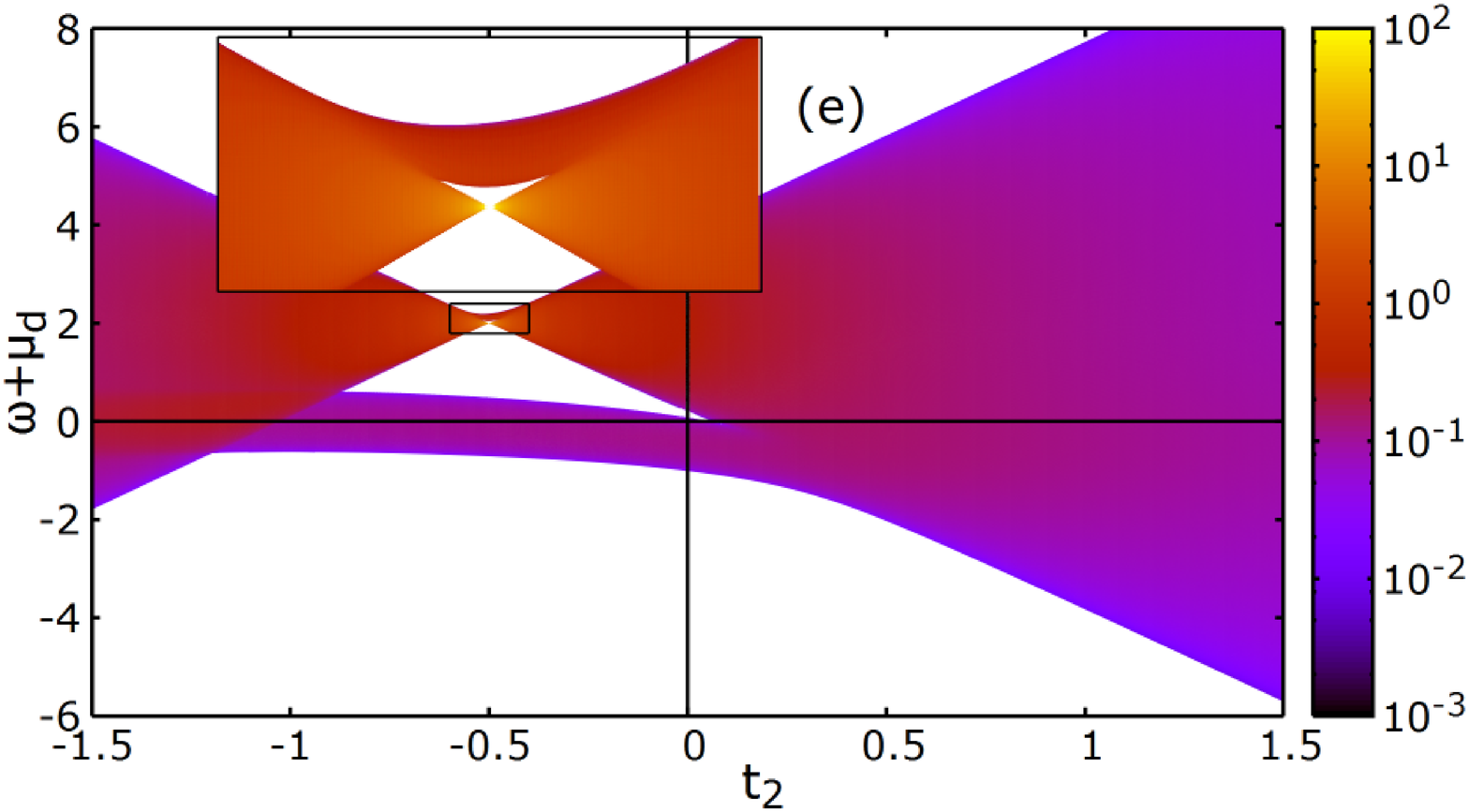}\quad
	\includegraphics[width=0.4\linewidth]{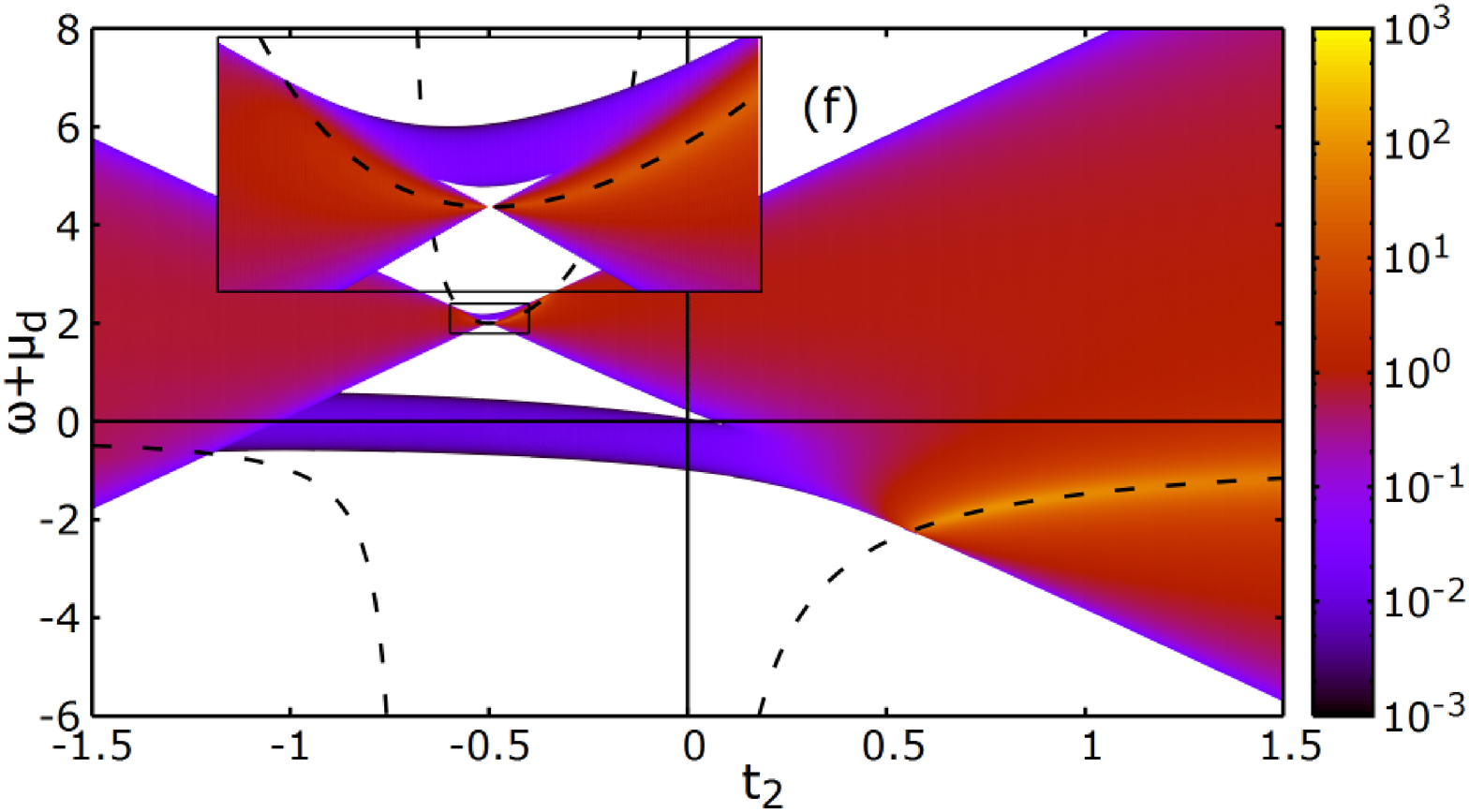}
	\caption{(Color online) 
	Same as in Fig.~\protect\ref{fig:log_u=025} for $U=2$. 
	Inserts represent enlarged rectangular areas with $t_2\in[-0.6,-0.4]$, $\omega+\mu_d\in[1.8,2.4]$.
	}
	\label{fig:log_u=20}
\end{figure*}
The dependence of $A_d(\omega)$ and  $I(\omega)$ on $\omega+\mu_d$ and $t_2$ is shown  in Fig.~\ref{fig:log_u=025} on a a false color plot, 
for $n_f=0.5$, 0.75, and  0.9, in the weak-coupling limit ($U=0.25$, $t_3=0$). 
The data show that  $A_d(\omega)$ and  $I(\omega)$ have the same band width, 
which increases for  $t_2>0$ and $t_2<-1$, while it shrinks for $-1<t_2<0$. 

The DOS is a smooth and almost semi elliptic function for all values of  $t_2$, except for  
$t_2\approx -(t_1+t_3)/2\approx-0.5$, where the matrix element $t^{++}$ is small and changes sign. 
At half filling, $n_f=0.5$, and in the vicinity of $t_2=-0.5$ ($t^{++}=0$), 
the DOS exhibits a tiny gap at $\omega+\mu_d=U$ [see insert in Fig.~\ref{fig:log_u=025}(a)], 
with a sharp peak at the bottom-edge of the upper band [see Fig.~\ref{fig:dos_t2_-05U_025nf05} below]. 
For larger values of $n_f$,  the DOS  develops a more prominent gaps,  generating, first, a two-band  and, then, a three-band structure. 
In the case of the two-band structure, the spectral weights of the lower and upper bands are  $w_0=1-n_f$ and $w_1=n_f$, 
respectively, which is similar to what one finds in the doped Mott-Hubbard insulator phase of the Falicov-Kimball model without 
correlated hopping.\cite{joura:165105,zlatic:266601,zlatic:155101} 
As we approach the $t^{++}=0$ point, an additional gap appears in the upper Hubbard band and the two-peak structure 
is transformed into a three-peak one;  
the spectral weights of the lower and upper bands are the same $w_0=1-n_f$, while the spectral weight of the middle band is $2n_f-1$. 
For $t^{++}\to0$, the  DOS of the middle band narrows to a $\delta$~peak. 

The  mid-band emerges, for  $n_f>1/2$, because the probability of the neighboring sites being occupied by 
the $f$ particles becomes macroscopically large, so that the clusters of sites occupied by the $f$ particles are created. 
Since the value of $t_2$  is such that the hopping matrix element $t^{++}$ is very small, a direct hopping of $d$ particles 
between these sites is suppressed. Thus, the $d$ particles within the cluster are localized and we observe  
a band of localized states which are separated from the upper Hubbard band by the localization gap. 
As $n_f\to1$, almost all the sites are occupied by $f$ particles, so that the cluster covers the whole lattice and 
the weight of the lower and upper band decreases as $1-n_f$ and the DOS is dominated by a narrow peak 
of localized states located at $\omega+\mu_d=U$. 
The presence of the clusters of localized states strongly affects the properties of itinerant $d$ electrons around 
the Fermi level $E_{\textrm{F}}$ (the chemical potential $\mu_d$ at zero temperature). 
For $n_d<1-n_f$, the Fermi level is   in the lower band; 
for $1-n_f<n_d<n_f$, it is fixed in the narrow mid-peak and the localized $d$ states of clusters are filled; 
for $n_d>n_f$, the Fermi level is  in the upper band.

In the strong-coupling limit,  $A_d(\omega)$ and  $I(\omega)$  change in a qualitative way.  
As shown in Fig.~\ref{fig:log_u=20} (left panel), the Mott-Hubbard gap emerges now in an extended range of $t_2$ values, 
including the case without the correlated hopping (Mott transition in the regular Falicov-Kimball model). 
The localization band is pushed into the upper Hubbard band and is very narrow; the corresponding localization gap 
exists only in the immediate vicinity of the $t^{++}=0$ value.

The above results show that the definition of the weak and strong coupling is modified by the dual nature of correlated hopping. 
Figures~\ref{fig:log_u=025} and \ref{fig:log_u=20} indicate three regions of the parameter space with markedly different band structure. 
For large positive and negative $t_2$, a single band, with a bandwidth which depends linearly on $t_2$, is observed. 
Here, the amplitude of the Hartree-Fock hopping is larger than the effective Coulomb interaction, which is typical of a weak coupling regime. 
The intermediate values of $t_2$ are characterized by a large reduction of the band width and an opening of a band gap. 
The reduced amplitude of the Hartree-Fock hopping and the Mott-Hubbard gap are typical features of the strong coupling regime.
In the third case,  for $t^{++}\simeq 0$,  in addition to the Hubbard bands, there emerges a band of localized states. 

The  transport function, plotted in the right panel of Figs.~\ref{fig:log_u=025} and \ref{fig:log_u=20}, shows a completely different 
behavior than the density of states. For  $t_2>-(t_1+t_3)/2$, and $t^{++}>0$, $I(\omega)$ exhibits a large enhancement at 
a resonant frequency, $\omega=\omega_{\text{res}}$, which is indicated in Figs.~\ref{fig:log_u=025} and \ref{fig:log_u=20}  
by dashed lines.
Numerical analysis of Eq.~\eqref{eq:Iw} shows that the resonant contribution to $I(\omega)$ comes from the term 
${\Img \Psi\left[E_2(\omega)\right] }/{\Img E_2(\omega)}$. At the resonance, the $\lambda$ fields and Green's functions 
satisfy the interference condition, 
\begin{equation}\label{eq:res_cond}
\frac{\lambda^{++}(\omega)}{\lambda^{--}(\omega)} = \frac{g_0(\omega)}{g_1(\omega)} = \eta,
\end{equation}
which gives
\begin{equation}
\omega_{\text{res}} + \mu_d = \frac{U}{1-\eta}
\label{eq:w_res}
\end{equation}
with
\begin{align}
\eta &= \frac{(t^{+-})^2}{(t^{--})^2} 
\label{eq:x_res}\\
&- \frac{(t^{+-})^2-\sqrt{(t^{+-})^4+4w_1w_0 \left[(t^{++}t^{--})^2-(t^{+-})^4\right] }}{2(t^{--})^2w_0}.
\nonumber
\end{align}
 For two special values of $t_2$, given by the solutions of the equation $\eta=1$ or
\begin{equation}
(t^{--})^2 w_0 + (t^{+-})^2 (w_1 - w_0) - (t^{++})^2 w_1 = 0,
\end{equation}
the resonant frequency is at infinity, i.e., it is shifted outside the bands. 
The regular Falicov-Kimball model, where $t^{--} = t^{+-} = t^{++} = t_1$, is precisely at one of these special points with 
$\omega_{\text{res}}\to\pm\infty$, so that the   transport function has no resonant contribution. 
On the other hand,  for  $t^{++}\to0$, $\omega_{\textrm{res}}$ is within the localized band and the resonant contributions 
become prominent for $t^{++}>0$, while it is suppressed for $t^{++}<0$.

Since the expressions given by Eqs.~\eqref{eq:w_res} and \eqref{eq:x_res} for $\omega_{\text{res}}$ do not depend on the parameter $W$, characterizing the semi elliptic DOS for the Bethe lattice, we conjecture that  the same expressions also hold for other lattices with different unperturbed DOS. Direct numeric calculations for the hypercubic lattice with the Gaussian DOS confirm this conjecture but, in that case, there are no clear band edges and the resonant peak is within the band  for any value of correlated hopping.\cite{shvaika:43704}

\subsection{Transport coefficients}\label{sec: transport}

\subsubsection{Weakly coupling regime}

\begin{figure}    
\centering
\includegraphics[width=0.8\linewidth]{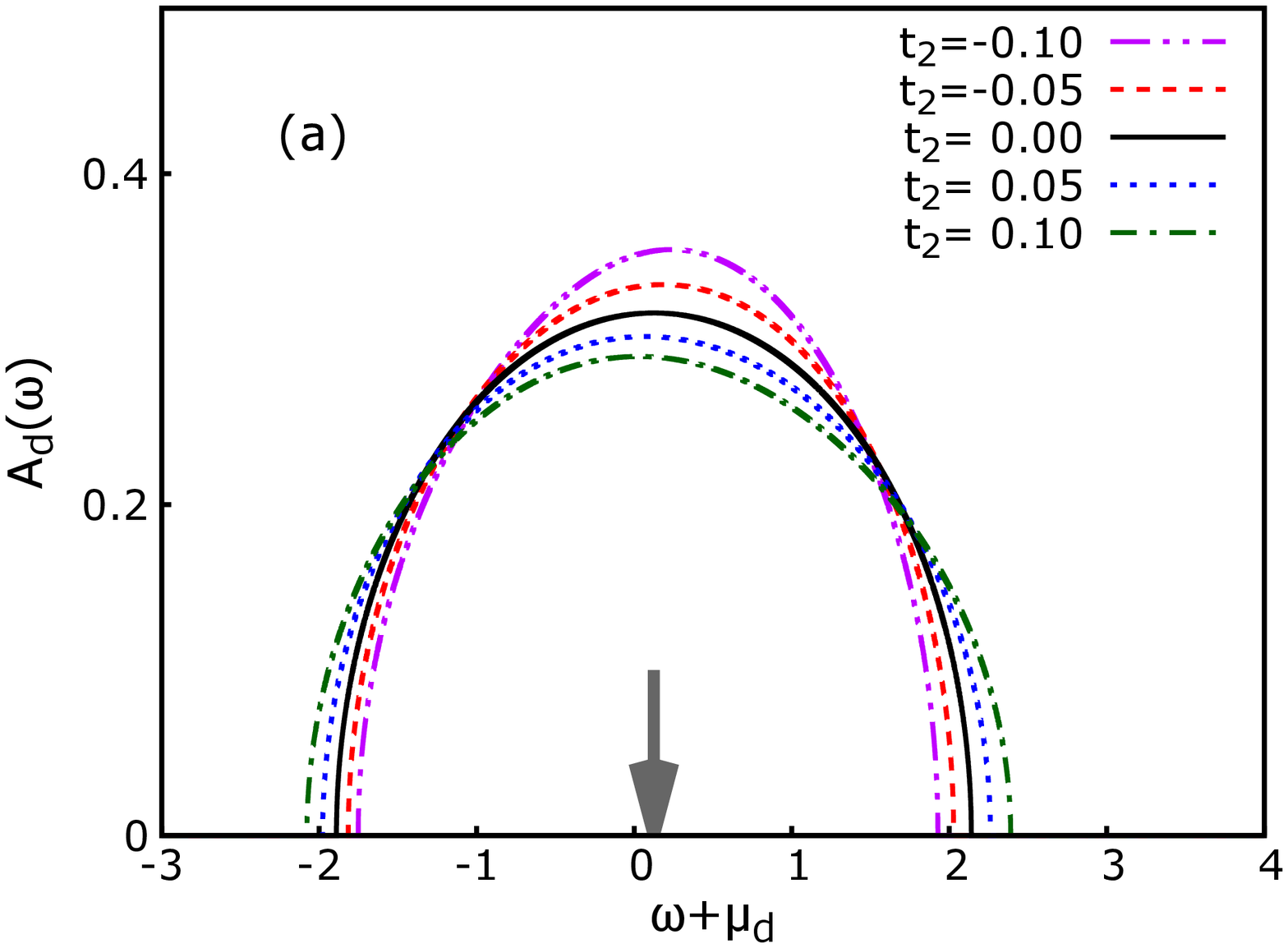}\\
\includegraphics[width=0.8\linewidth]{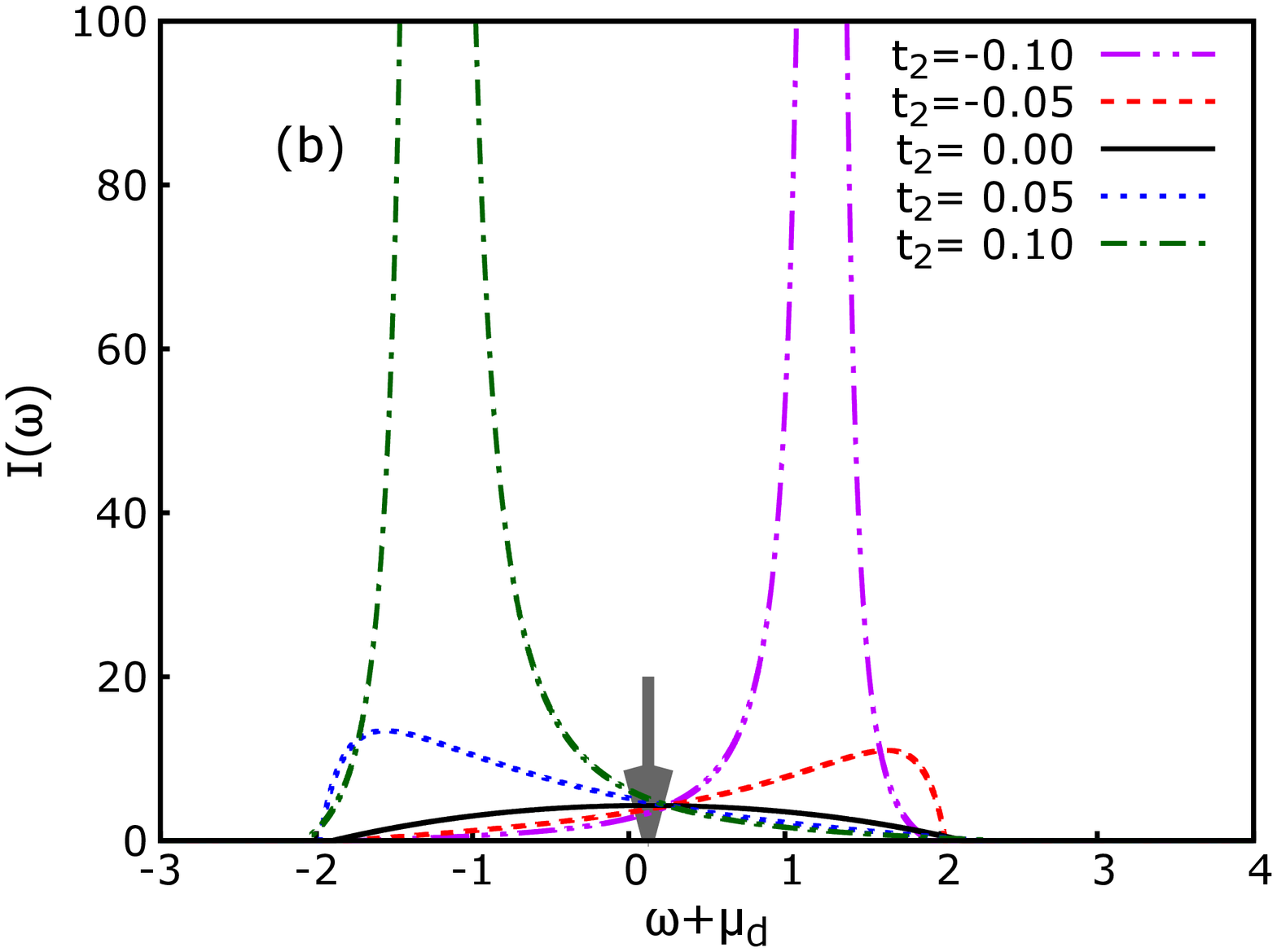}
\caption{(Color online) The interacting DOS (panel a) and transport function (panel b)  
plotted versus frequency for $U=0.25$ 
at half filling ($n_f=n_d=1/2$) and for $t_2=-0.1$, $-0.05$, $0$, $0.05$, $0.1$ ($t_3=0$). 
The Fermi level $E_{\textrm{F}}$ depends on the hopping   $t_2$ and 
the gray arrow indicates a narrow frequency interval in which $E_{\textrm{F}}$  is  located for different values of $t_2$. 
}
\label{fig:dos_t2_00U_025nf05}
\end{figure}

\begin{figure}    
	\centering
	\includegraphics[width=0.8\linewidth]{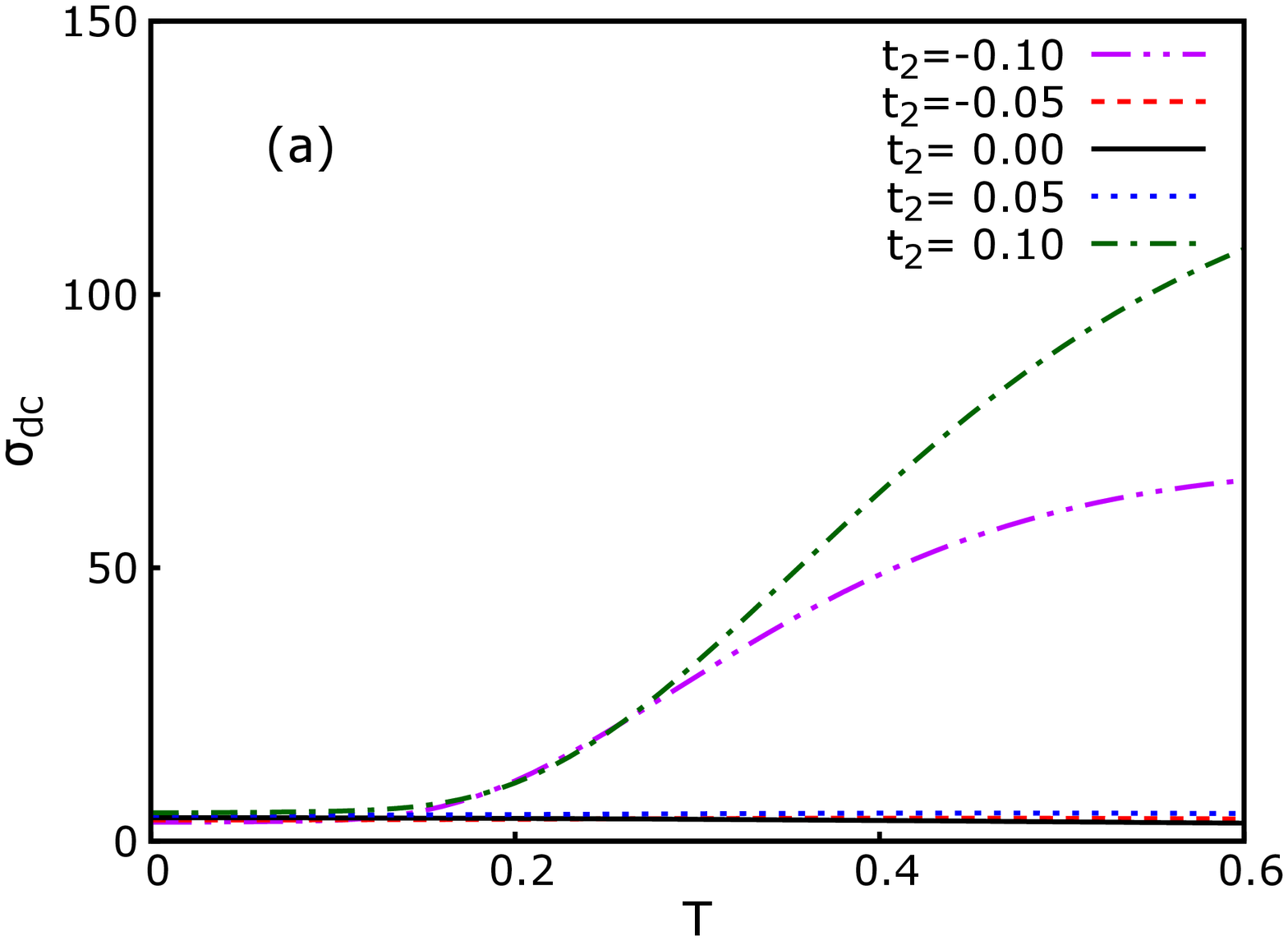}\\
	\includegraphics[width=0.8\linewidth]{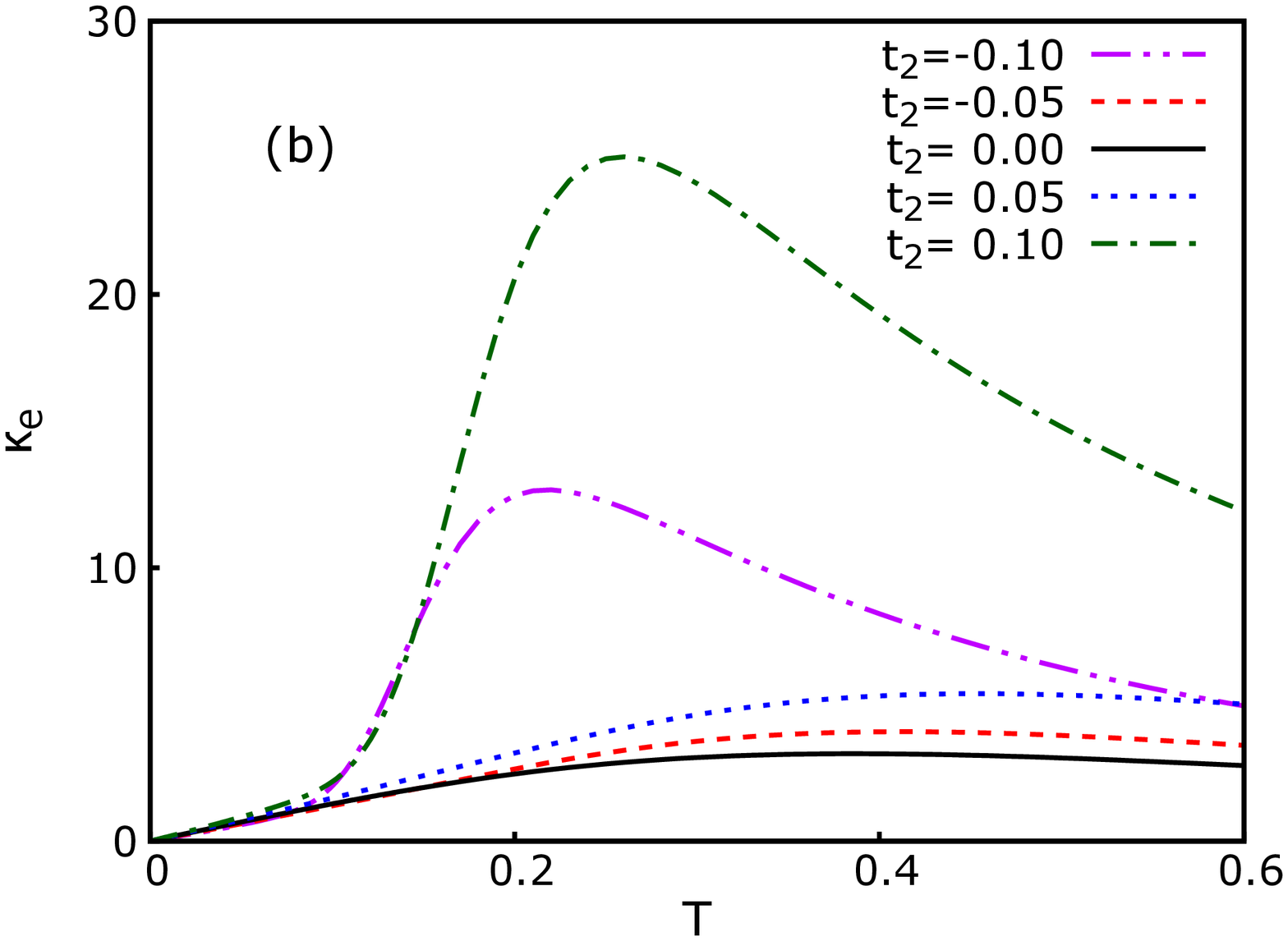}\\
	\includegraphics[width=0.8\linewidth]{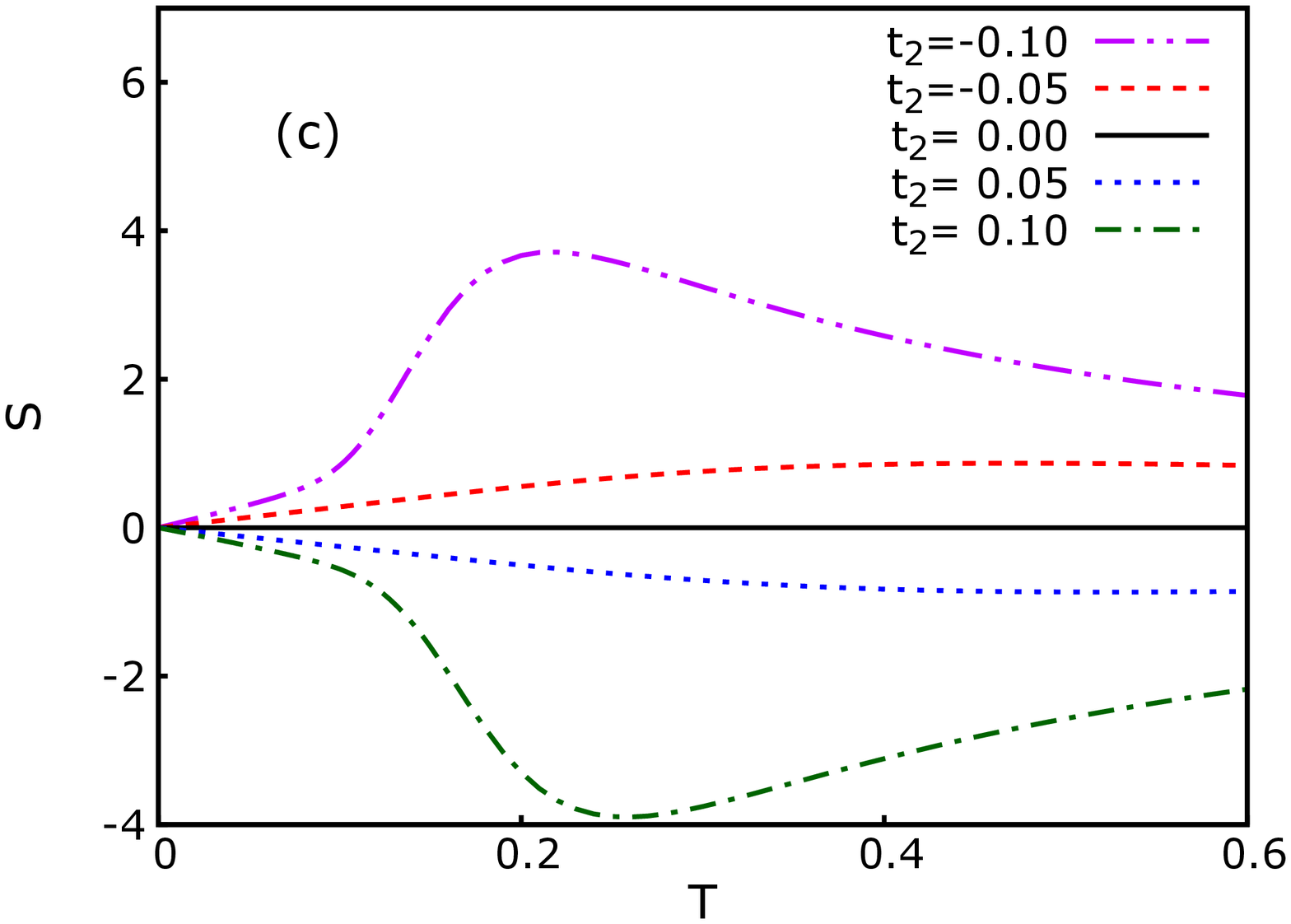}
	\caption{(Color online) Temperature dependences of the dc conductivity $\sigma_{\textrm{dc}}$, 
	thermal conductivity $\kappa_{\textrm{e}}$, and Seebeck coefficient $S$ for 
	the same parameters as in Fig.~\ref{fig:dos_t2_00U_025nf05}. }
	\label{fig:tr_t2_00U_025nf05}
\end{figure}

\begin{figure}    
	\centering
	\includegraphics[width=0.8\linewidth]{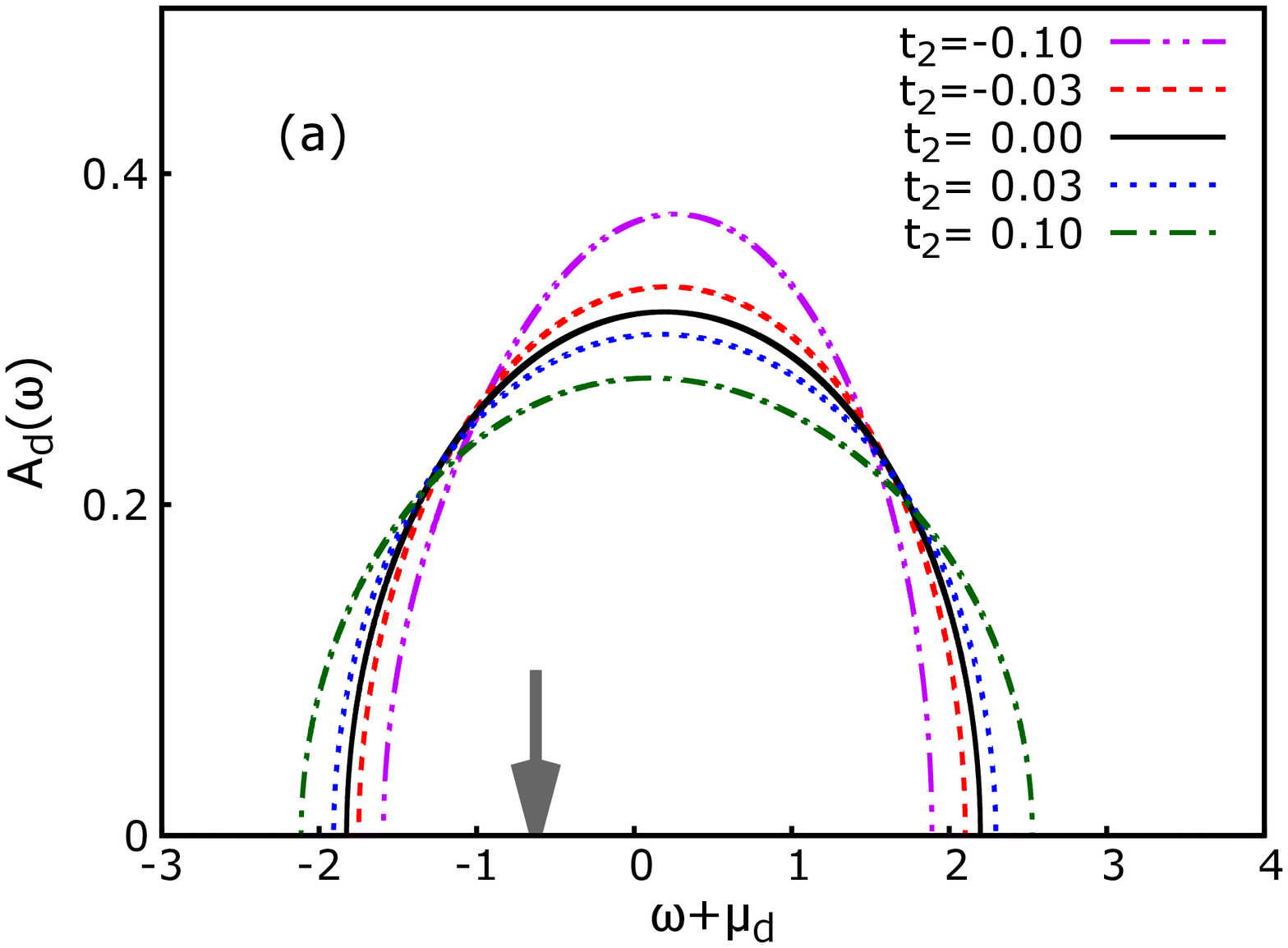}\\
	\includegraphics[width=0.8\linewidth]{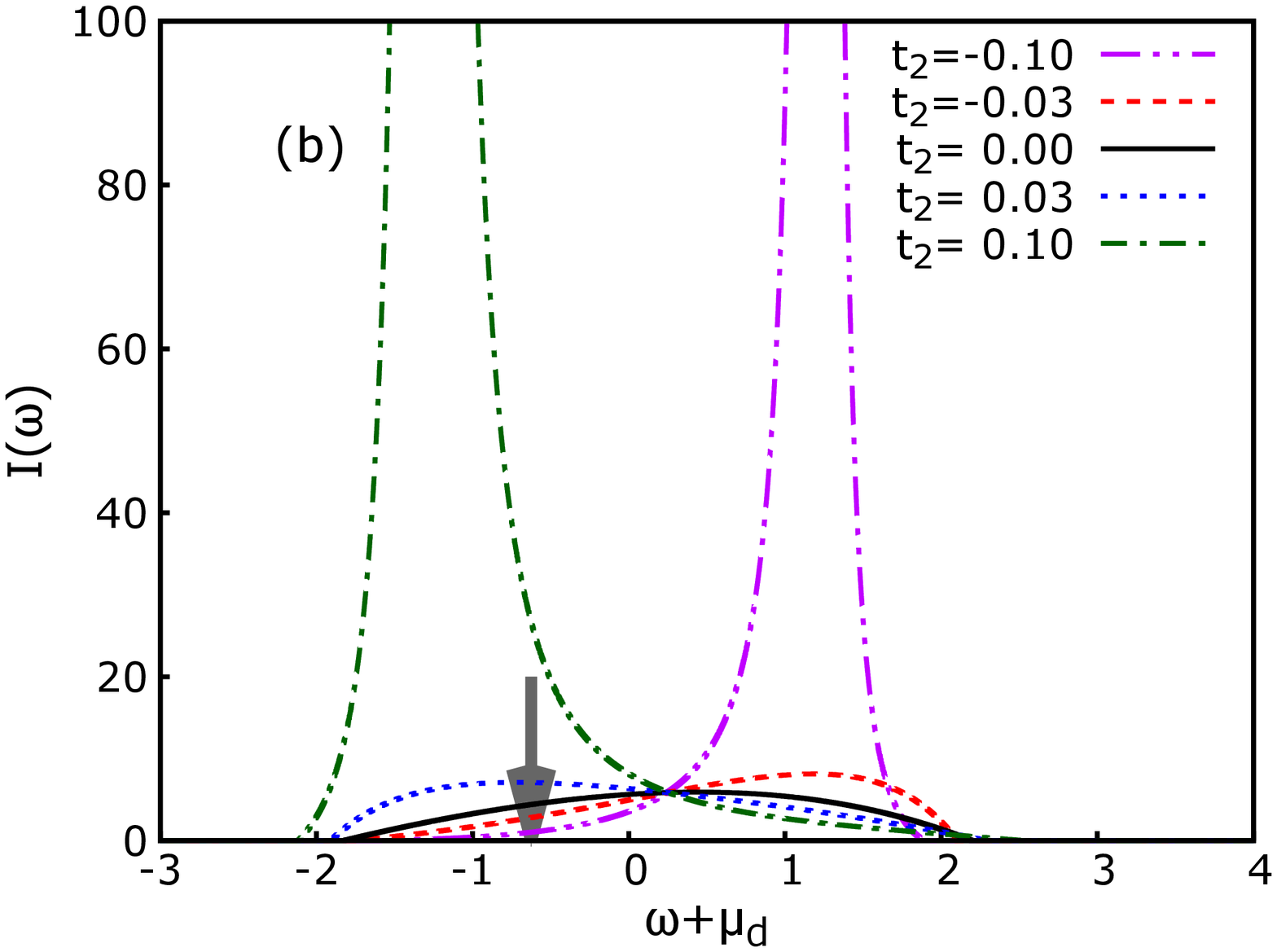}
	\caption{(Color online) 
	The interacting DOS (panel a) and transport function (panel b)  
         plotted versus frequency for $U=0.25$,  $n_f=0.75$, and $n_d=1-n_f=0.25$ 
	and for $t_2=-0.1$, $-0.03$, $0$, $0.03$, $0.1$. 
	The gray arrow indicates a narrow $\omega$-interval in which $E_{\textrm{F}}$  is  located for different values of $t_2$. 
	}
	\label{fig:dos_t2_00U_025nf075}
\end{figure}

\begin{figure}    
	\centering
	\includegraphics[width=0.8\linewidth]{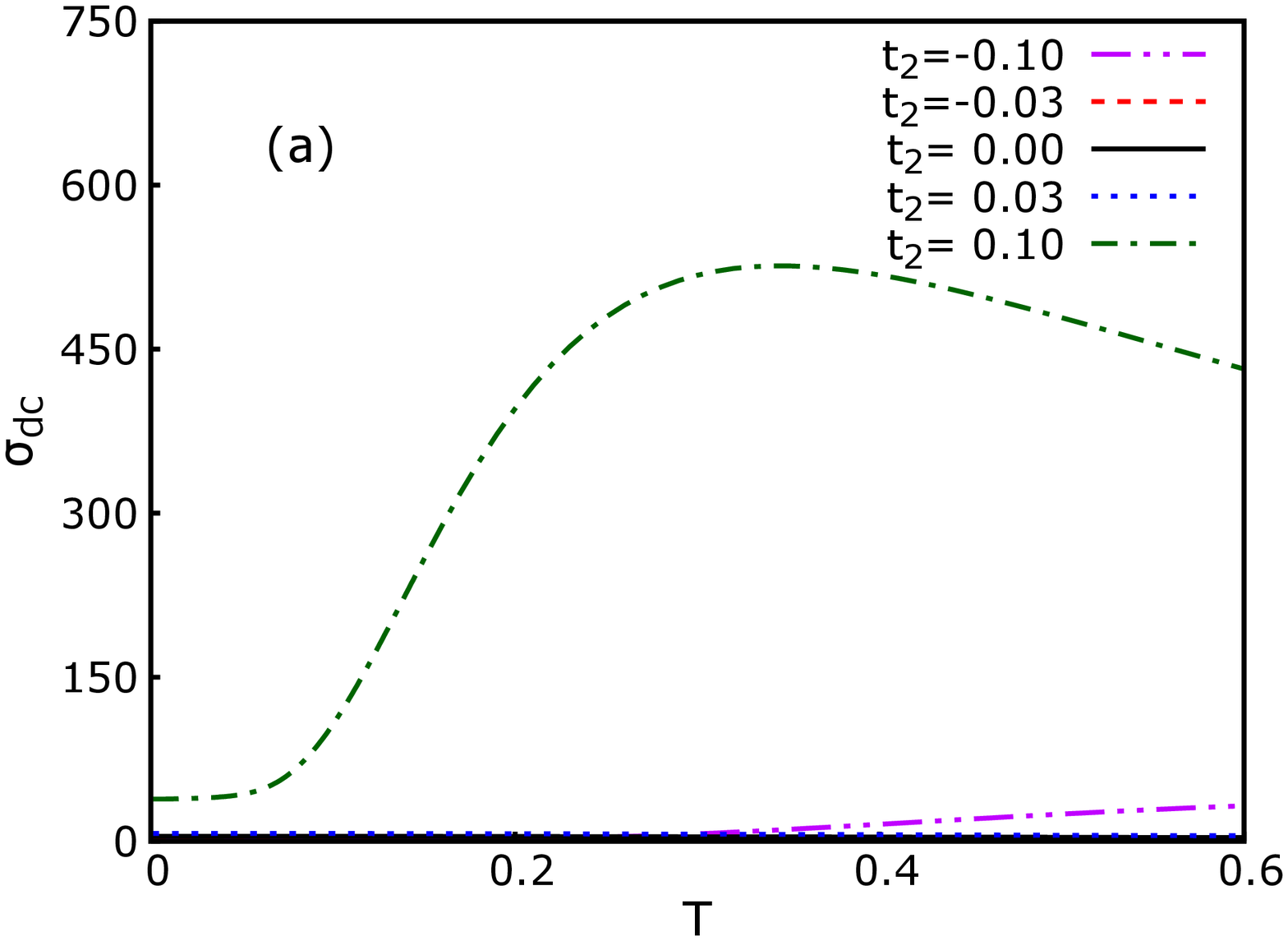}\\
	\includegraphics[width=0.8\linewidth]{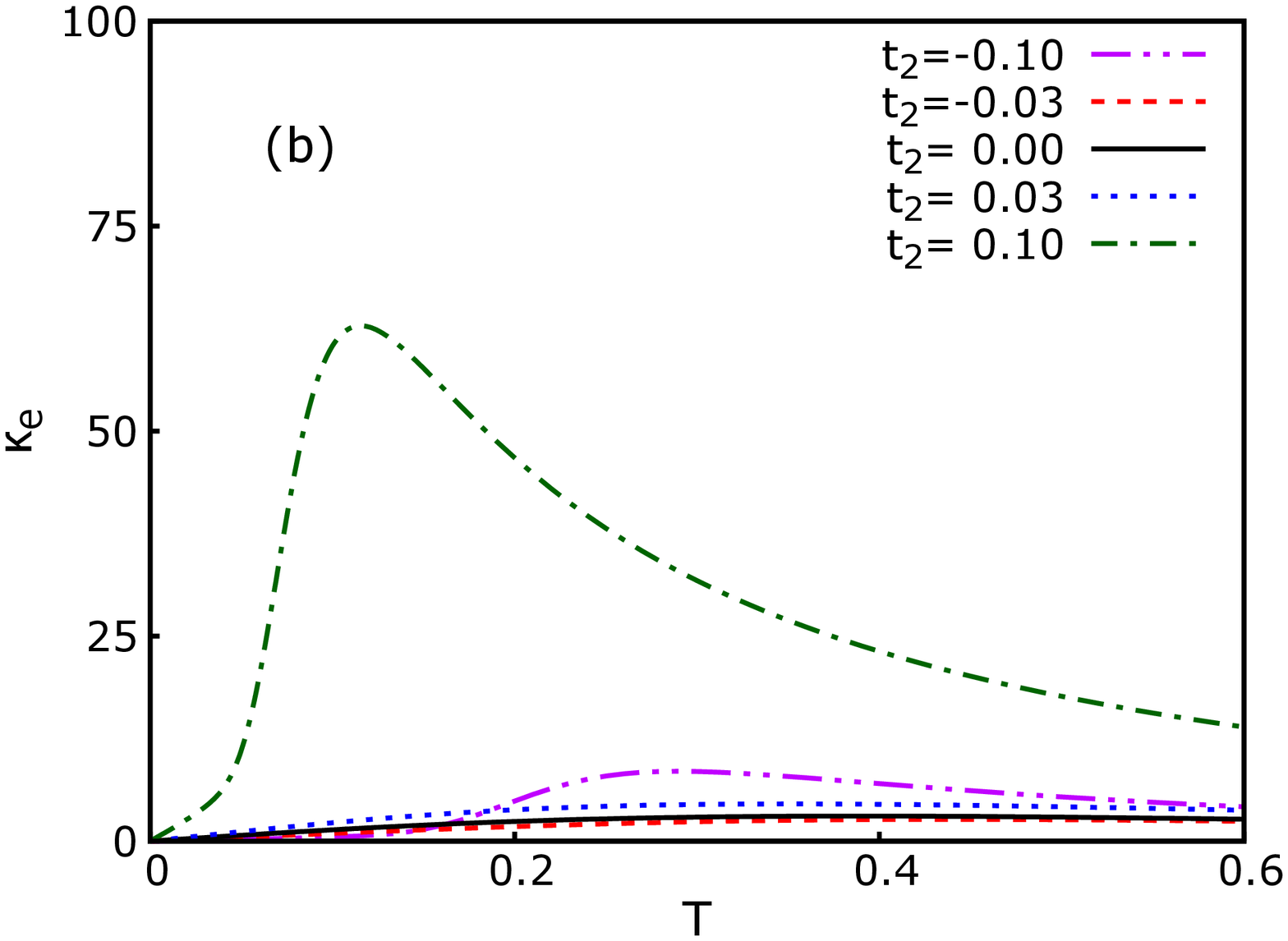}\\
	\includegraphics[width=0.8\linewidth]{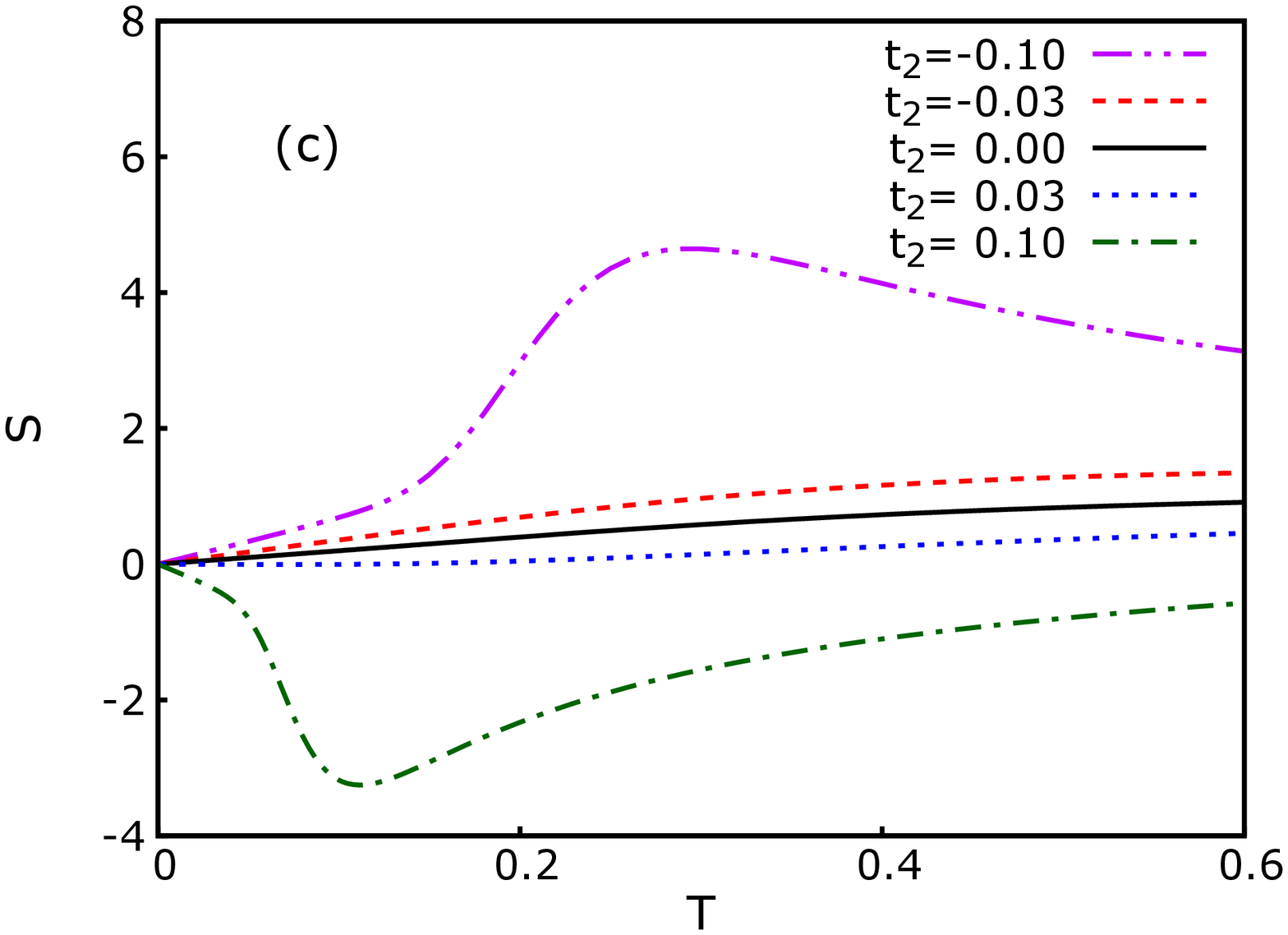}
	\caption{(Color online) 
	 Temperature dependences of the dc conductivity $\sigma_{\textrm{dc}}$, 
	thermal conductivity $\kappa_{\textrm{e}}$, and Seebeck coefficient $S$ is shown for 
	the same parameters as in Fig.~\ref{fig:dos_t2_00U_025nf075}.
	}
	\label{fig:tr_t2_00U_025nf075}
\end{figure}

\begin{figure}    
	\centering
	\includegraphics[width=0.8\linewidth]{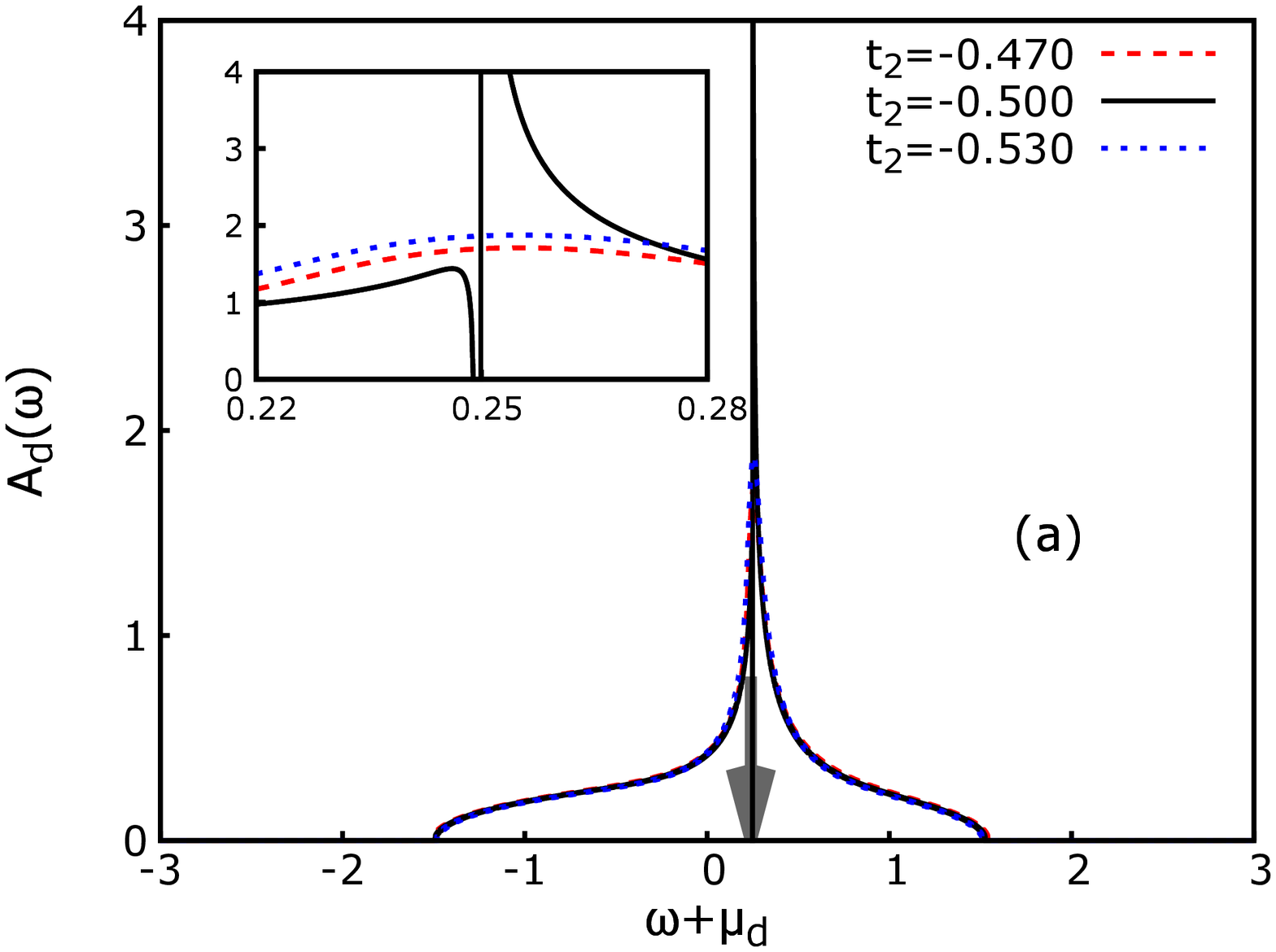}\\
	\includegraphics[width=0.8\linewidth]{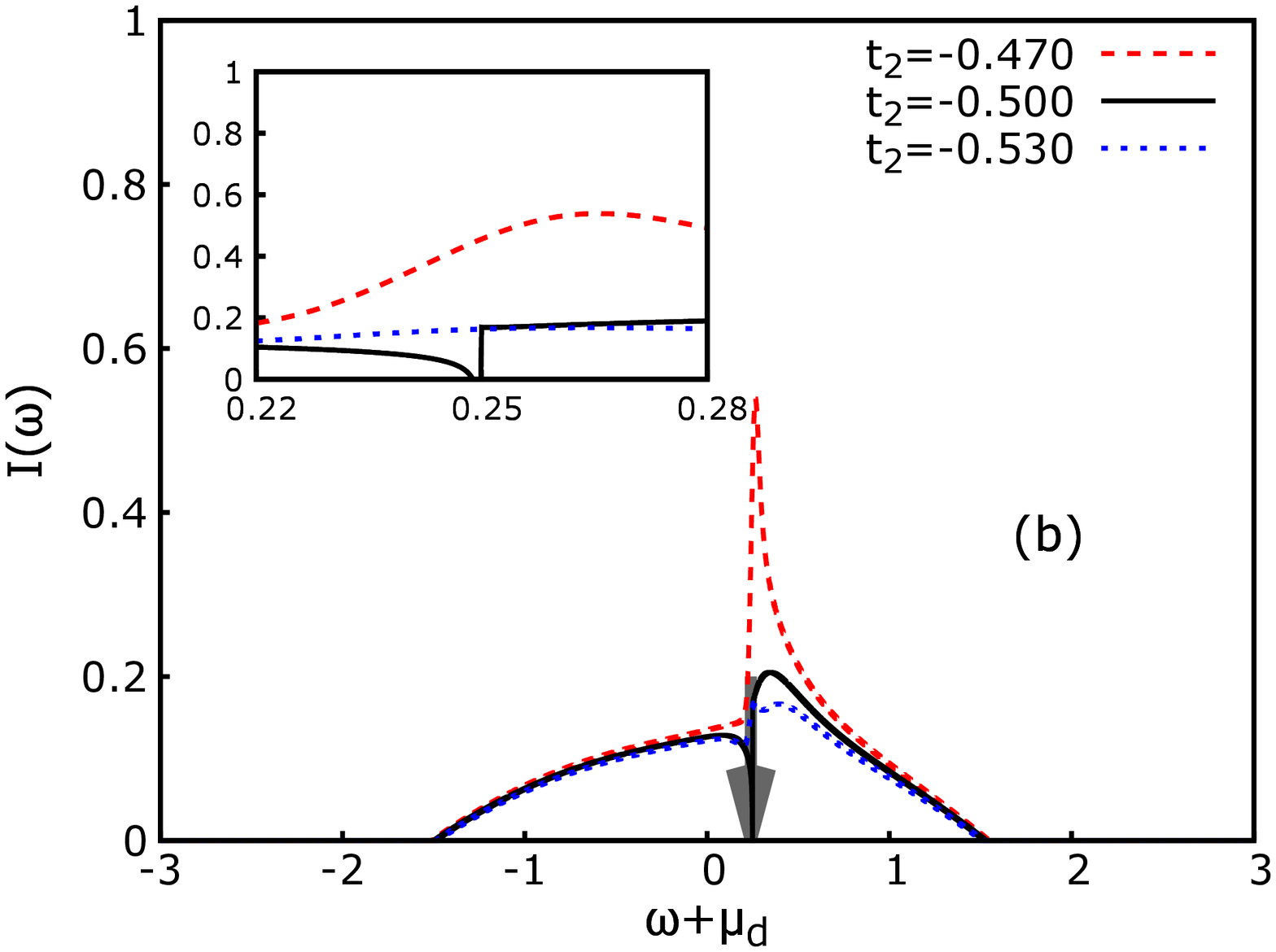}
	\caption{(Color online) The interacting DOS (panel a) and transport function (panel b) 
	plotted versus $\omega$ for $U=0.25$ 
	at half filling ($n_f=n_d=1/2$) 
	and for $t_2=-0.47$, $-0.5$, $-0.53$ ($t_3=0$). 
	The gray arrow indicates a narrow $\omega$-interval in which $E_{\textrm{F}}$  is  located for different values of $t_2$.  
	}
	\label{fig:dos_t2_-05U_025nf05}
\end{figure}

\begin{figure}    
	\centering
	\includegraphics[width=0.8\linewidth]{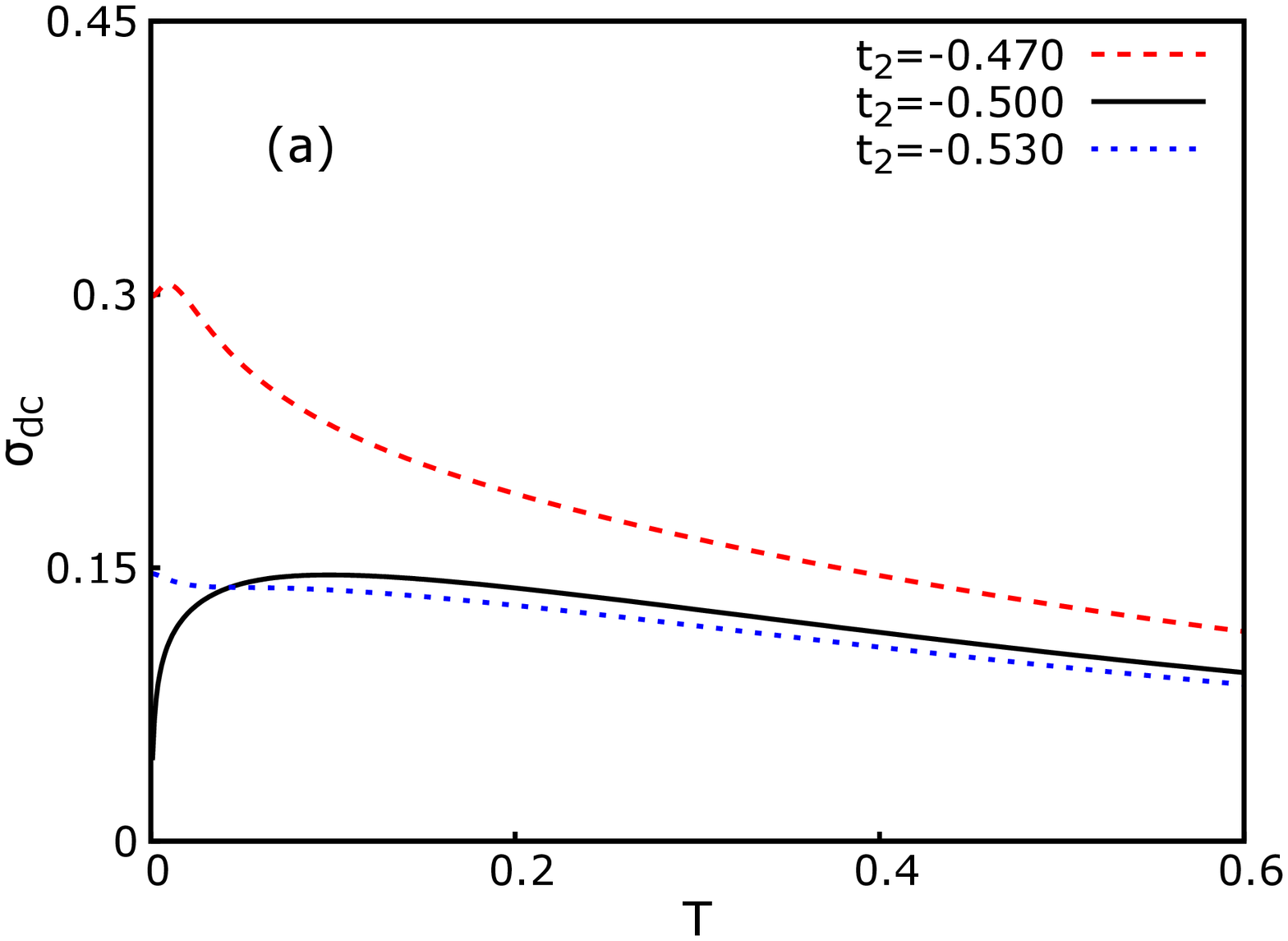}\\
	\includegraphics[width=0.8\linewidth]{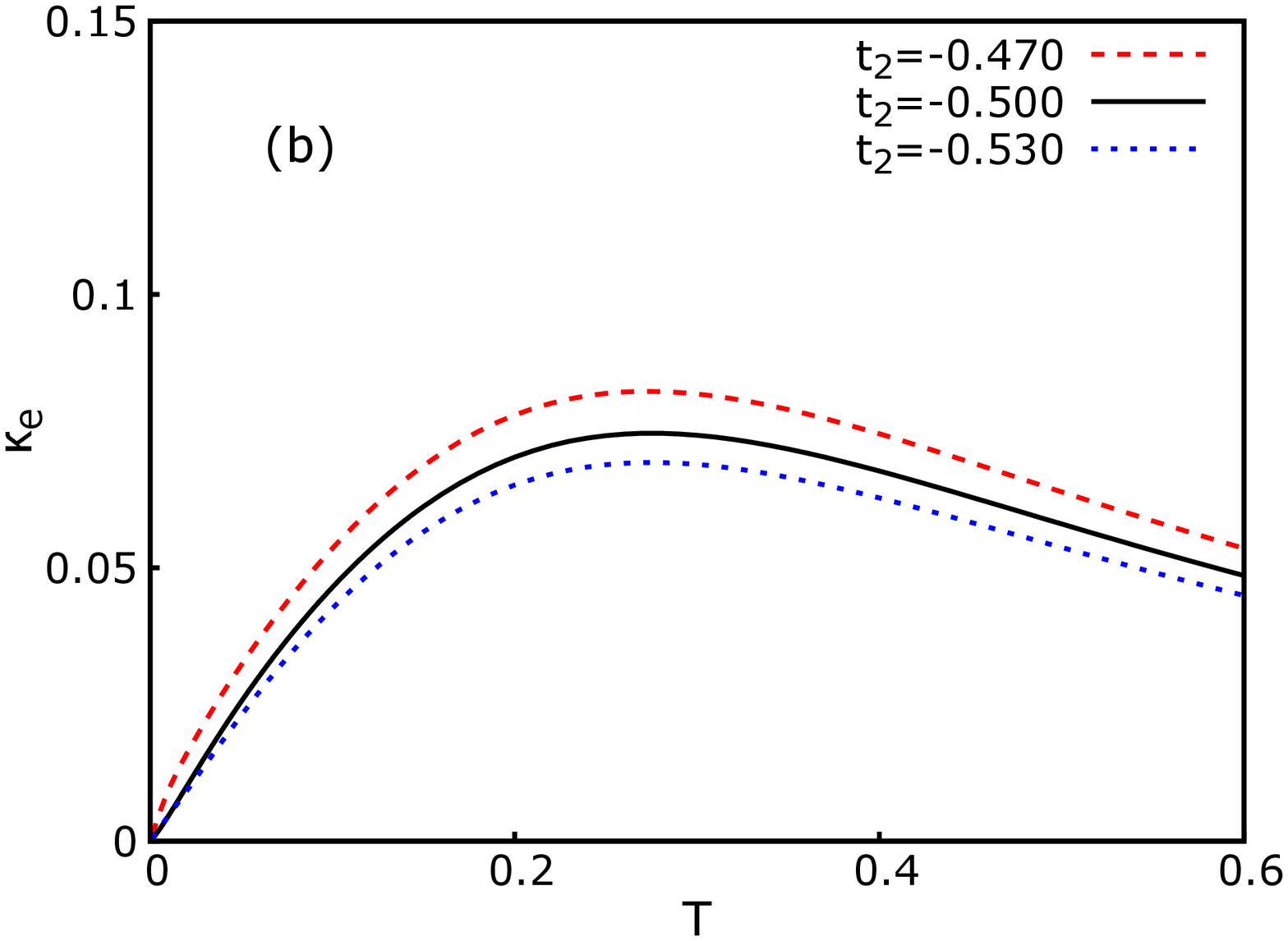}\\
	\includegraphics[width=0.8\linewidth]{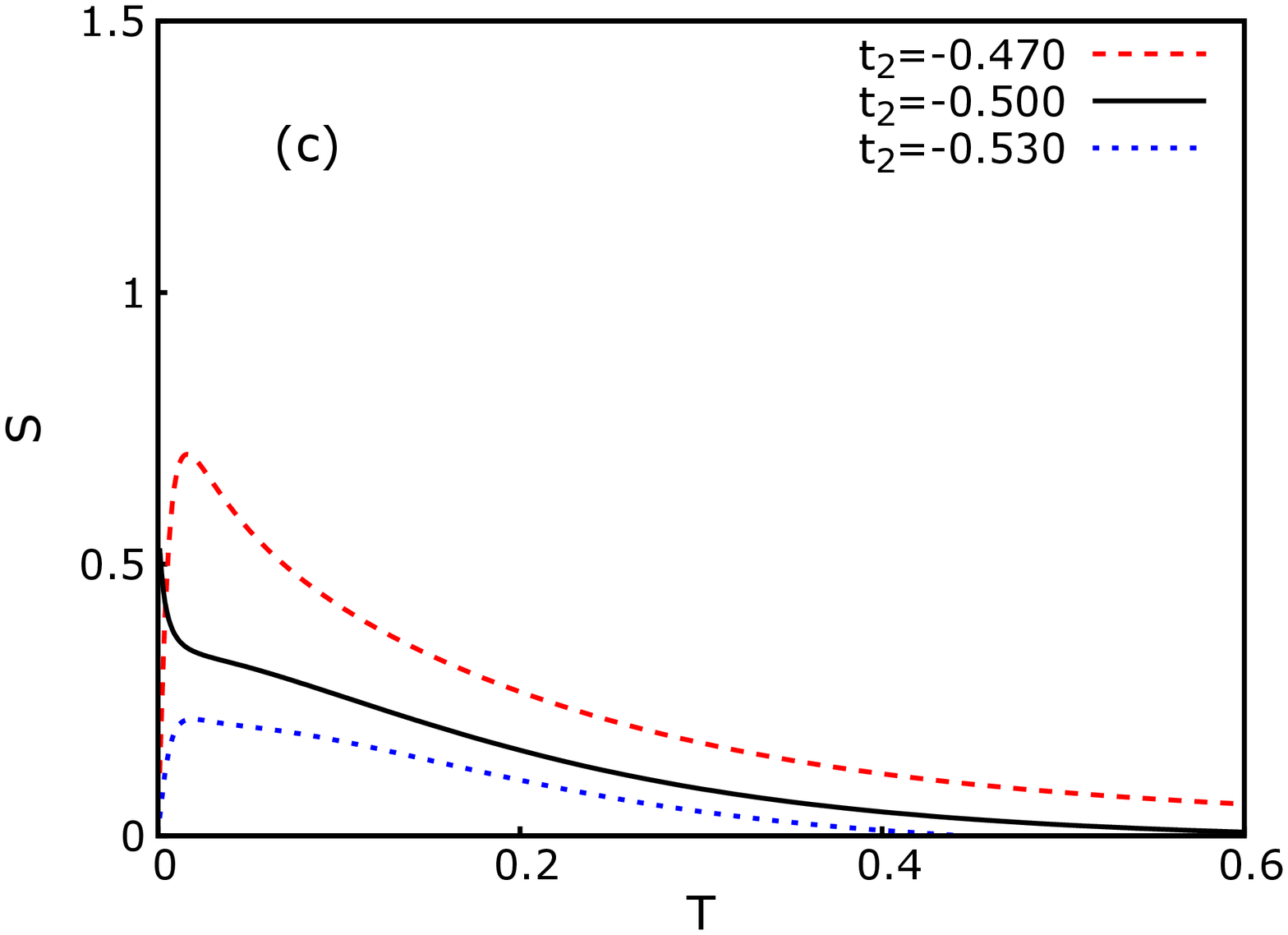}
	\caption{(Color online) Temperature dependences of the (a) dc conductivity $\sigma_{\textrm{dc}}$, 
		(b)  thermal conductivity $\kappa_{\textrm{e}}$, and (c) Seebeck coefficient $S$ is shown 
		for the same parameters as in Fig.~\ref{fig:dos_t2_-05U_025nf05}.
		}
	\label{fig:tr_t2_-05U_025nf05}
\end{figure}

\begin{figure}    
	\centering
	\includegraphics[width=0.8\linewidth]{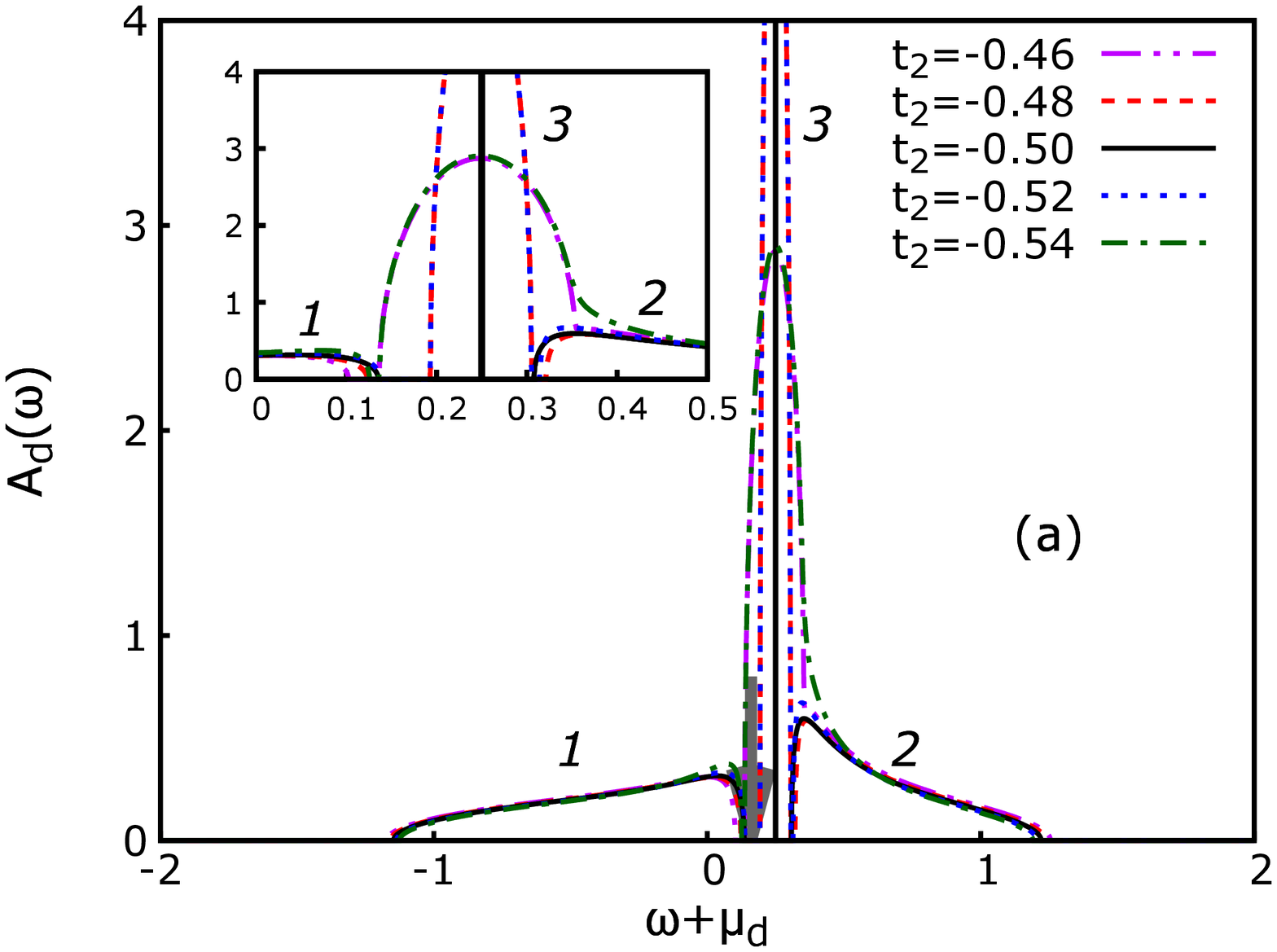}\\
	\includegraphics[width=0.8\linewidth]{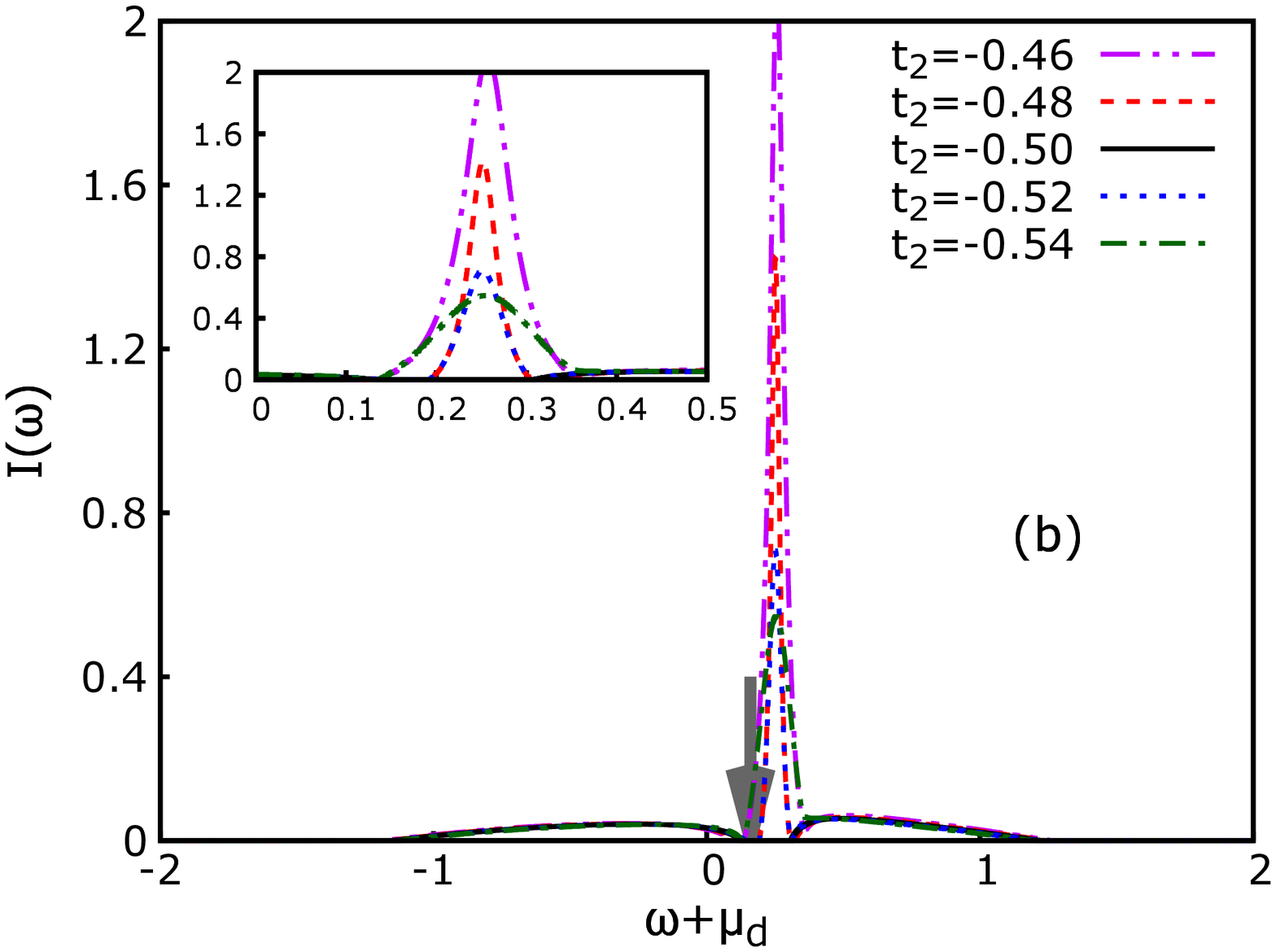}
	\caption{(Color online) 
	 The  interacting DOS  (panel a) and transport function  (panel b)  
	 plotted versus $\omega$ for $U=0.25$, $n_f=0.75$, $n_d=1-n_f=0.25$, 
	 and for $t_2=-0.46$, $-0.48$, $-0.5$, $-0.52$, $-0.54$. 
	Labels \textsl{1},  \textsl{2}, and \textsl{3} denote the lower Hubbard band, the upper Hubbard band, 
	and band of localized states, respectively.
         The gray arrow indicates a narrow $\omega$-interval in which $E_{\textrm{F}}$  is  located for different values of $t_2$.  
         }
	\label{fig:dos_t2_-05U_025nf075}
\end{figure}

\begin{figure}    
	\centering
	\includegraphics[width=0.8\linewidth]{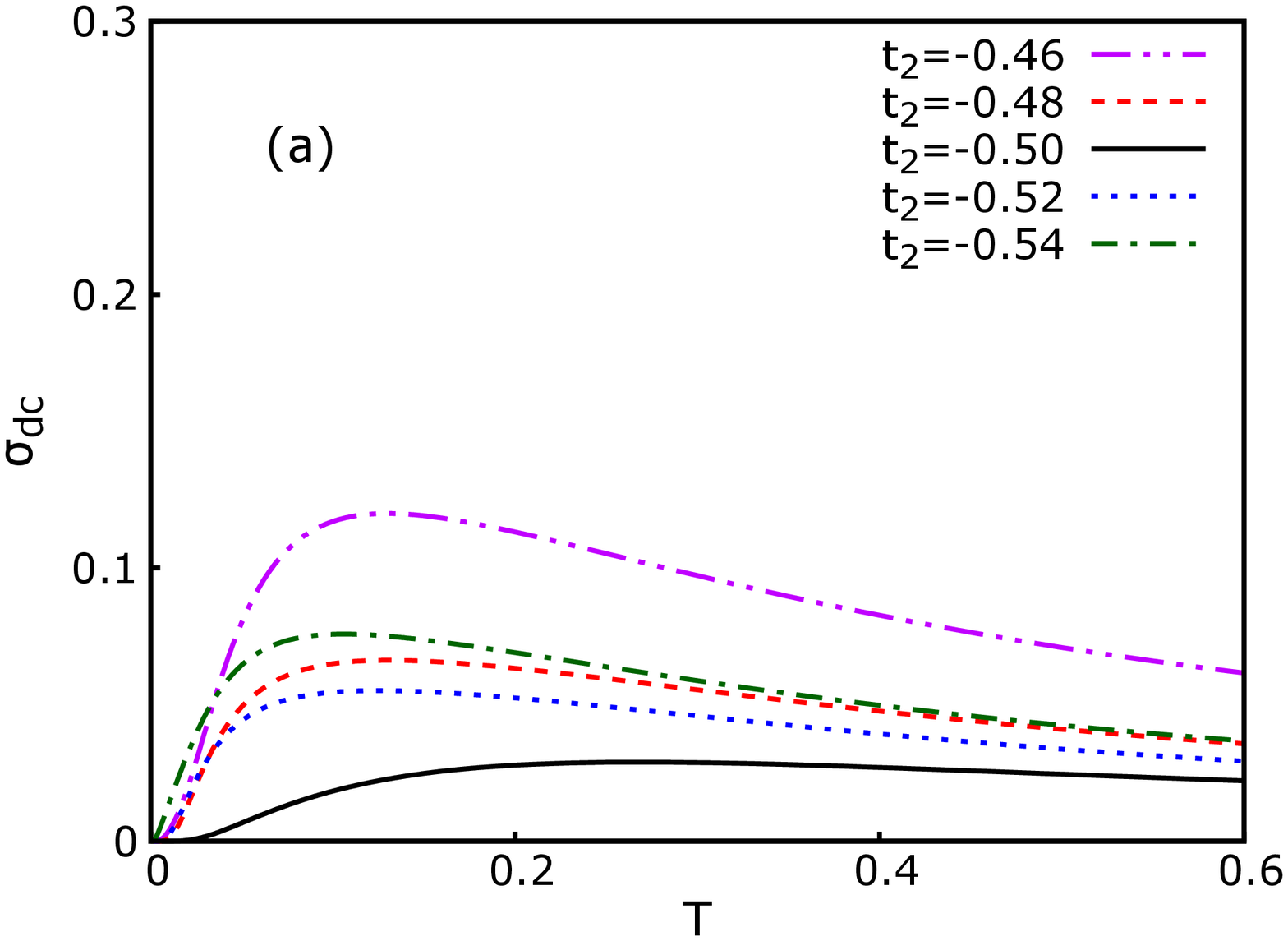}\\
	\includegraphics[width=0.8\linewidth]{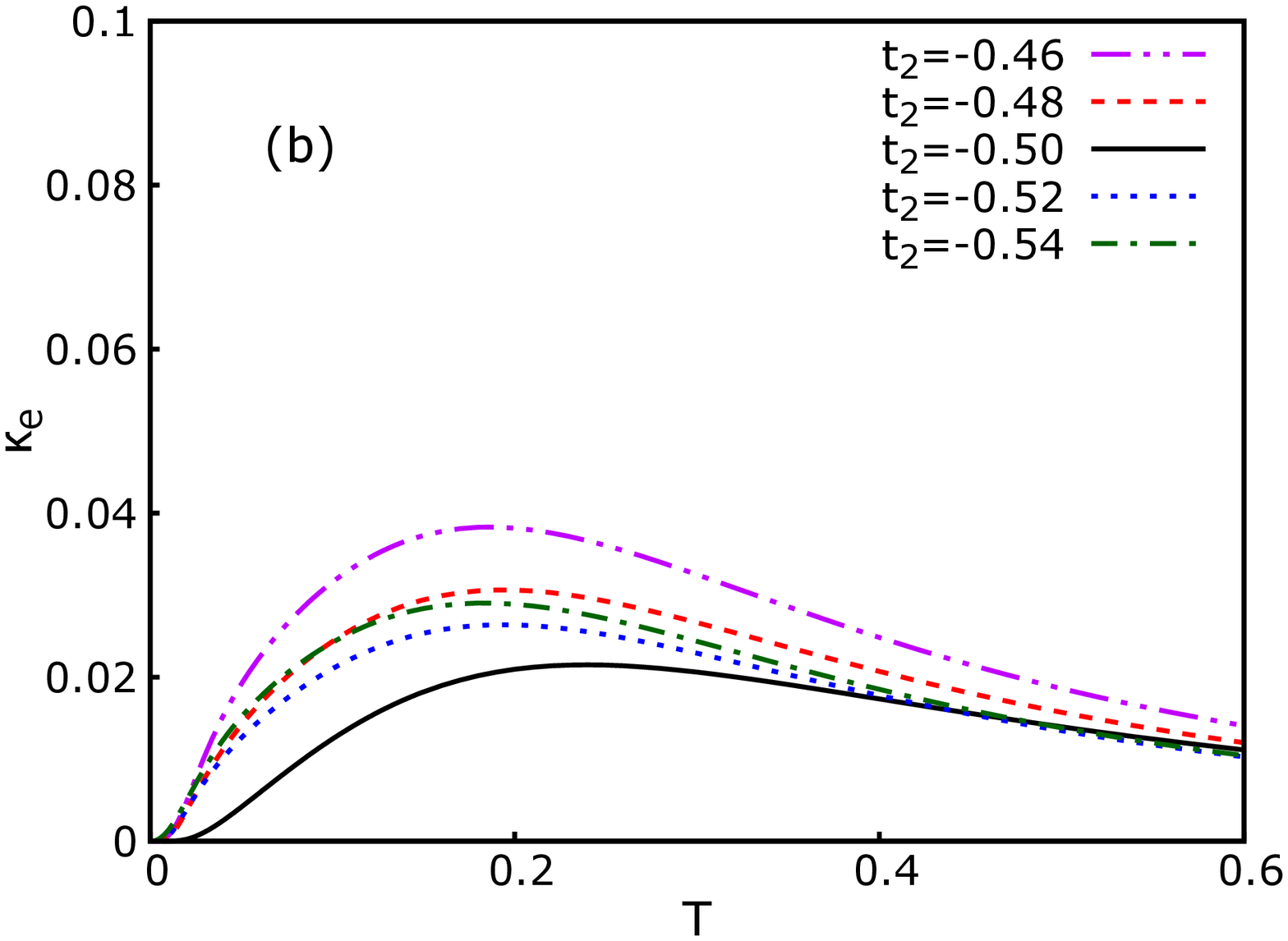}\\
	\includegraphics[width=0.8\linewidth]{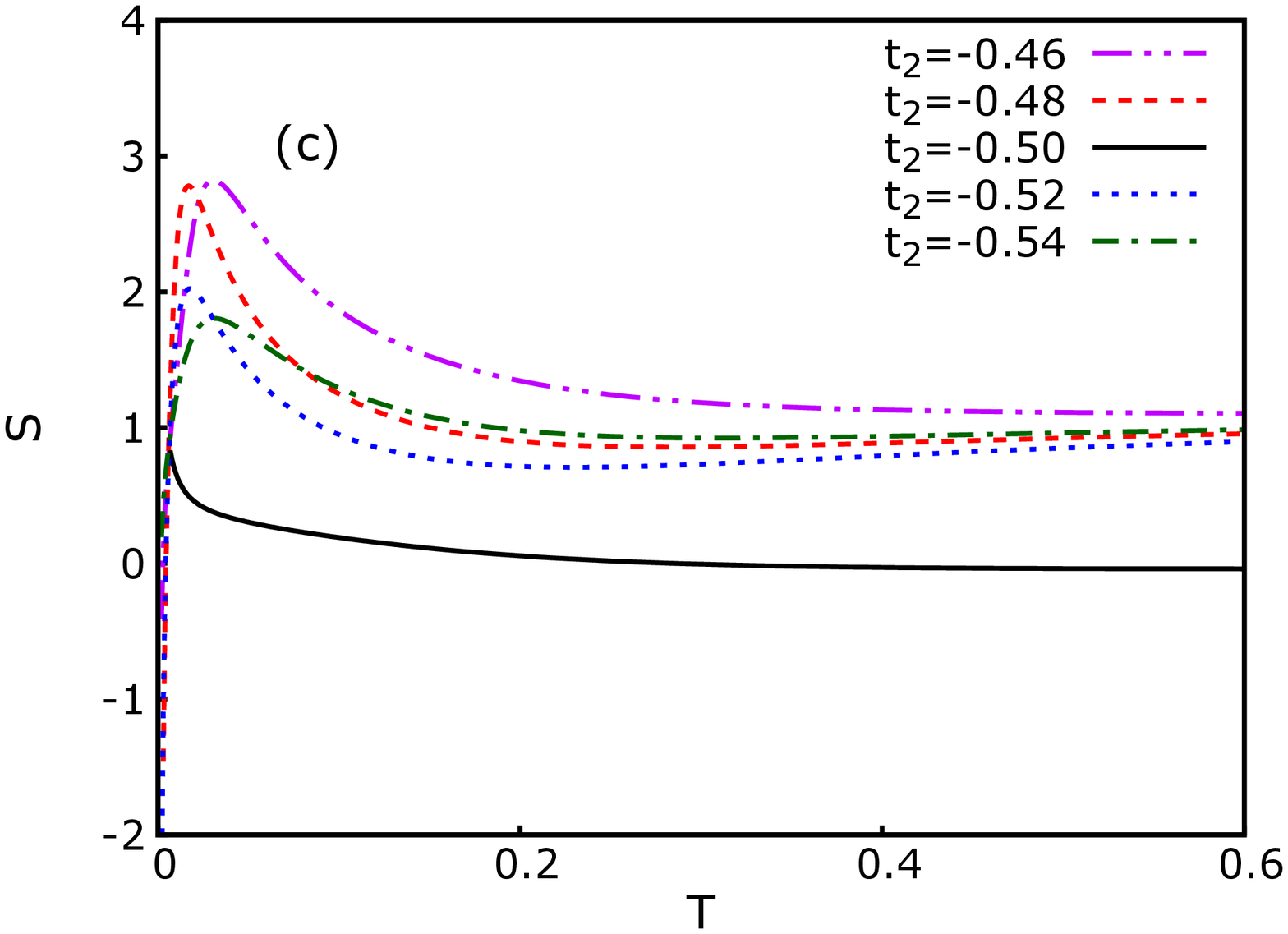}
	\caption{(Color online) 
	Temperature dependence of the (a) dc conductivity $\sigma_{\textrm{dc}}$, 
	(b)  thermal conductivity $\kappa_{\textrm{e}}$, and (c) Seebeck coefficient $S$ in the weak coupling regime 
	is shown for the same parameters as in Fig.~\ref{fig:dos_t2_-05U_025nf075}.
	}
	\label{fig:tr_t2_-05U_025nf075}
\end{figure}

In the weak coupling regime, the interaction constants are much smaller than the hopping integral: $|U|,|t_2|,|t_3|\ll |t_1|$. 
At half filling, $n_f=n_d=1/2$, the renormalized DOS  is slightly deformed with respect to the unperturbed 
semi elliptic one, i.e., the correlated hopping breaks the electron-hole symmetry 
and makes  $A_d(\omega)$ an asymmetric function [see Fig.~\ref{fig:dos_t2_00U_025nf05}(a)]. 
(At half filling and without the correlated hopping, $t_2=t_3=0$, the DOS is symmetric with respect to the frequency $\omega+\mu_d=U/2$.)
The width of the conduction band increases for $t_2>0$ and decreases for $t_2<0$. 

In contrast, the transport function $I(\omega)$, which is almost semi elliptic in the absence of the correlated hopping,  
becomes highly asymmetric as soon as $t_2\neq 0$. For $t_2=\pm0.05$, the resonant peak is outside the conduction band 
and the transport function is somewhat enhanced close to the band edges, but it is still smooth and almost flat around  
the chemical potential [see Fig.~\ref{fig:dos_t2_00U_025nf05}(b)]. 
Hence, the transport coefficients exhibit typical metallic behavior as functions of temperature (see Fig.~\ref{fig:tr_t2_00U_025nf05}). 
For $t_2=\pm0.1$, the resonant peak gives the main contribution to $I(\omega)$. At low temperatures, 
the resonant peak is outside the Fermi window,  $-{d f(\omega)}/{d \omega}\vert_{\omega\simeq\omega_{\text{res}}} \simeq 0$,  
and its contribution to the transport integrals is negligibly small. Hence, the transport coefficients exhibit a metallic behavior. 
However, as temperature increases,  the resonant peaks enter the Fermi window and we observe,  
first, an enhancement of the thermal conductivity, then, of the thermoelectric power, and, eventually, of the electric conductivity 
 (see Fig.~\ref{fig:tr_t2_00U_025nf05}).

Doping does not change much the shape of $A_d(\omega)$ and  $I(\omega)$ in the weak coupling regime, as shown 
by Fig.~\ref{fig:dos_t2_00U_025nf075} for  $n_f=0.75$ and $n_d=0.25$. 
The main effect  is the shift  of the  chemical potential towards the bottom of the conduction band, giving $\mu_d\sim -0.7$. 
For  $t_2=0.1$, the resonant peak overlaps the Fermi level, so that the electric conductivity, the thermal conductivity, and 
the thermopower are enhanced with respect to the half-filled case and when the resonant peak is absent (see Fig.~\ref{fig:tr_t2_00U_025nf075}). 
For $t_2=-0.1$, the resonant peak is too far away from the chemical potential  to contribute to the electric and 
thermal conductivity but a large asymmetry of $I(\omega)$, shown in Fig.~\ref{fig:dos_t2_00U_025nf075}(b), 
gives rise to an enhanced thermopower at high temperatures.

As shown in Fig.~\ref{fig:log_u=20}, an increase of local Coulomb interaction $U$ shifts the resonant frequency  
in the region of large positive and negative values of $t_2$, so that the resonant peaks do not affect much 
the transport properties for small values of correlated hopping $t_2$. The only effect of correlated hopping 
is to break the electron-hole symmetry and to make $A_d(\omega)$ and $I(\omega)$ asymmetric functions of $\omega$. 
Hence, in this part of the parameter space, the transport coefficients behave similarly as in the case of a doped Falicov-Kimball model 
without correlated hopping.\cite{joura:165105,zlatic:266601,zlatic:155101}

A different behavior emerges for small values of   $t^{++}$, when the direct hopping amplitude 
between the sites occupied by $f$ particles is much reduced, while the hopping between occupied and unoccupied, 
and between the unoccupied sites is large. For the Gaussian density of states, this case was analyzed in 
Ref.~[\onlinecite{shvaika:43704}] (see Figs.~6 and 7 of  Ref.~[\onlinecite{shvaika:43704}]). 
For the semi elliptic density of states, we find similar behavior but the data reveal more details and we discover new interesting regimes, 
like the three-band structure in the DOS and the clusters of sites occupied by $f$ electrons.

At half filling and $t^{++}=0$ ($t_2=-0.5$),  the DOS  acquires a tiny gap at the Fermi level,  
bounded by the singularity at the bottom of the upper Hubbard band [see Fig.~\ref{fig:dos_t2_-05U_025nf05}(a)]. 
The transport function also displays a gap  but the band-edge singularity is replaced by a step like function 
 [Fig.~\ref{fig:dos_t2_-05U_025nf05}(b)].
The presence of the gap at the Fermi level  reduces the dc and thermal conductivities at low temperatures, 
whereas it increases the thermopower due to the asymmetry of $I(\omega)$ 
around the chemical potential (see Fig.~\ref{fig:tr_t2_-05U_025nf05}). 
At high temperatures, when the width of the Fermi window exceeds the gap, 
the doped metallic or semiconductor behavior is restored.

At half filling and $t^{++}\neq 0$,  the gap closes rapidly and the DOS exhibits a smooth peak around the chemical potential. 
For negative  $t^{++}$ ($t_2<-0.5$), 
the transport function is smooth and we find a metallic behavior. 
For positive $t^{++}$ ($t_2>-0.5$), the transport function acquires a resonant peak, such that the transport coefficients are enhanced 
at low temperatures (see Fig.~\ref{fig:tr_t2_-05U_025nf05}). 

Away from half filling, the three band structure of $A_d(\omega)$  emerges for small values of $t^{++}$. 
As shown in Fig.~\ref{fig:dos_t2_-05U_025nf075}, for $t^{++}=0$ ($t_2=-0.5$) exactly,  
$A_d(\omega)$ exhibits two broad bands [the lower (\textsl{1}) and upper (\textsl{2}) Hubbard bands with spectral weights $1-n_f$], 
and a $\delta$-peak (\textsl{3}) at $\omega+\mu_d=U$ with spectral weight $2n_f-1$. 
The $\delta$-peak in $A_d(\omega)$ is due to the localized $d$ states in the clusters of lattice sites occupied by $f$ electrons. 
Because we have $n_d=1-n_f$, the Fermi level is in the gap between the lower Hubbard band and the localized band. 
Unlike the single particle DOS, the transport function $I(\omega)$ has only two contributions, due to the lower 
and upper Hubbard band. The absence of any feature at $\omega+\mu_d=U$ is obvious. 
Since the localized states cannot contribute much to the charge or heat transport, the shape of  $I(\omega)$ 
and temperature dependences of the  transport coefficients (see Fig.~\ref{fig:tr_t2_-05U_025nf075}) 
are similar to the one obtained for a doped Mott insulator.\cite{joura:165105,zlatic:266601,zlatic:155101}

Away from half filling and for $t^{++}\neq 0$	(Fig.~\ref{fig:dos_t2_-05U_025nf075}), 
the $\delta$-peak of localized $d$ states (\textsl{3}) 
broadens into a band which merges with the upper 
Hubbard band for larger values of $|t^{++}|$.
The contribution of the localized state to transport function depends on the sign of $t^{++}$. 
For negative values of $t^{++}<0$ ($t_2<-0.5$), this contribution 
increases at first and, then, as  $t^{++}$ becomes more negative and the localized band merges with the upper Hubbard band, 
it decreases. 
For  $t^{++}>0$ ($t_2>-0.5$), the resonant peak due to the localized states  dominants  the transport function. 
In both cases, the thermoelectric properties are enhanced, most prominently for positive  $t^{++}$, 
as the resonant peak affects strongly the transport properties (see Fig.~\ref{fig:tr_t2_-05U_025nf075}).

\subsubsection{Strong coupling regime}

A large enough Coulomb interaction $U$ leads to the reconstruction of the DOS and transport function. 
In the strong coupling regime, a large Mott-Hubbard gap governs the transport properties at low temperatures. 
Besides, the upper Hubbard band becomes narrow and acquires a high density of states. 

\begin{figure} 
	\centering
	\includegraphics[width=0.8\linewidth]{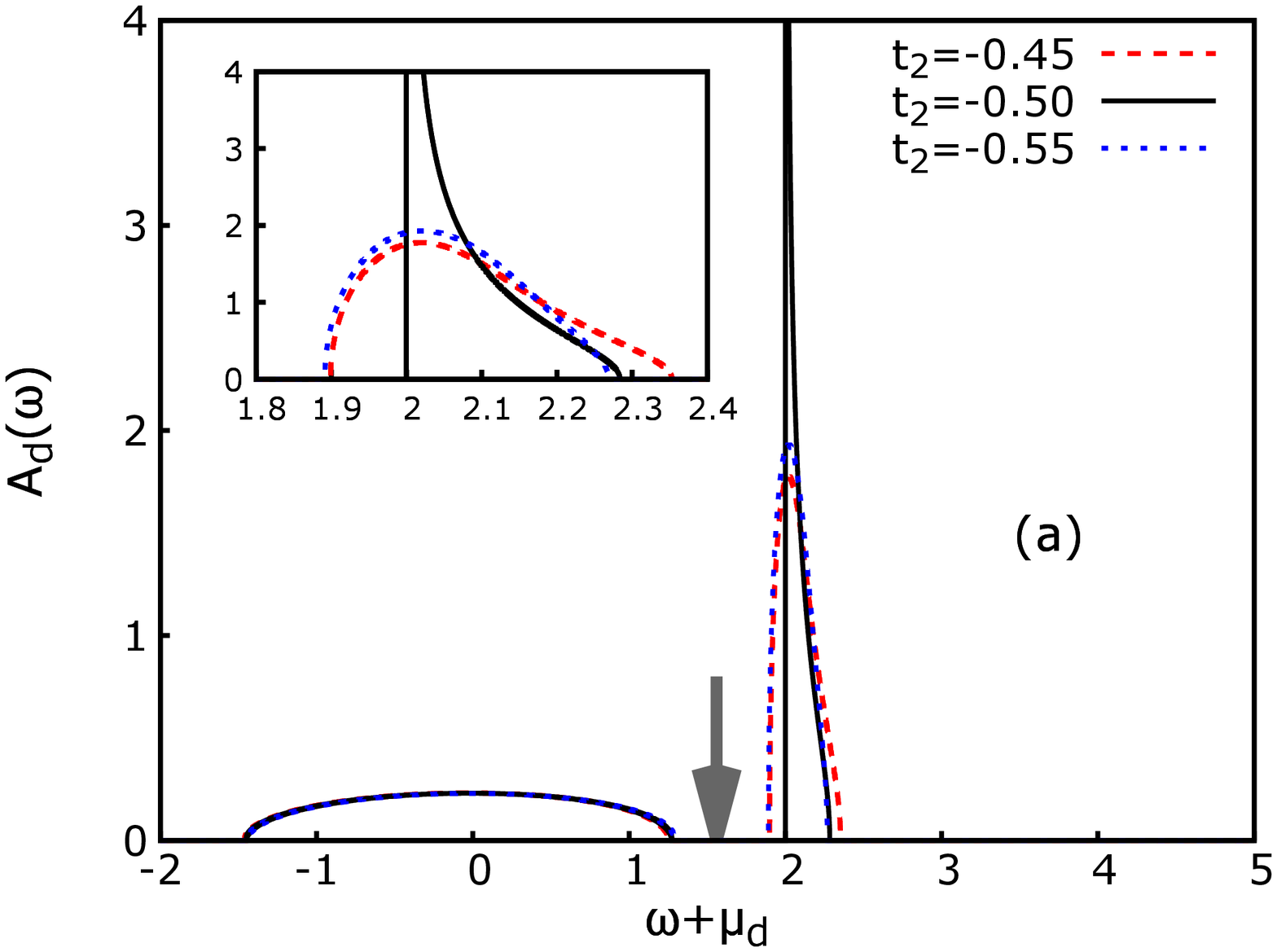}\\
	\includegraphics[width=0.8\linewidth]{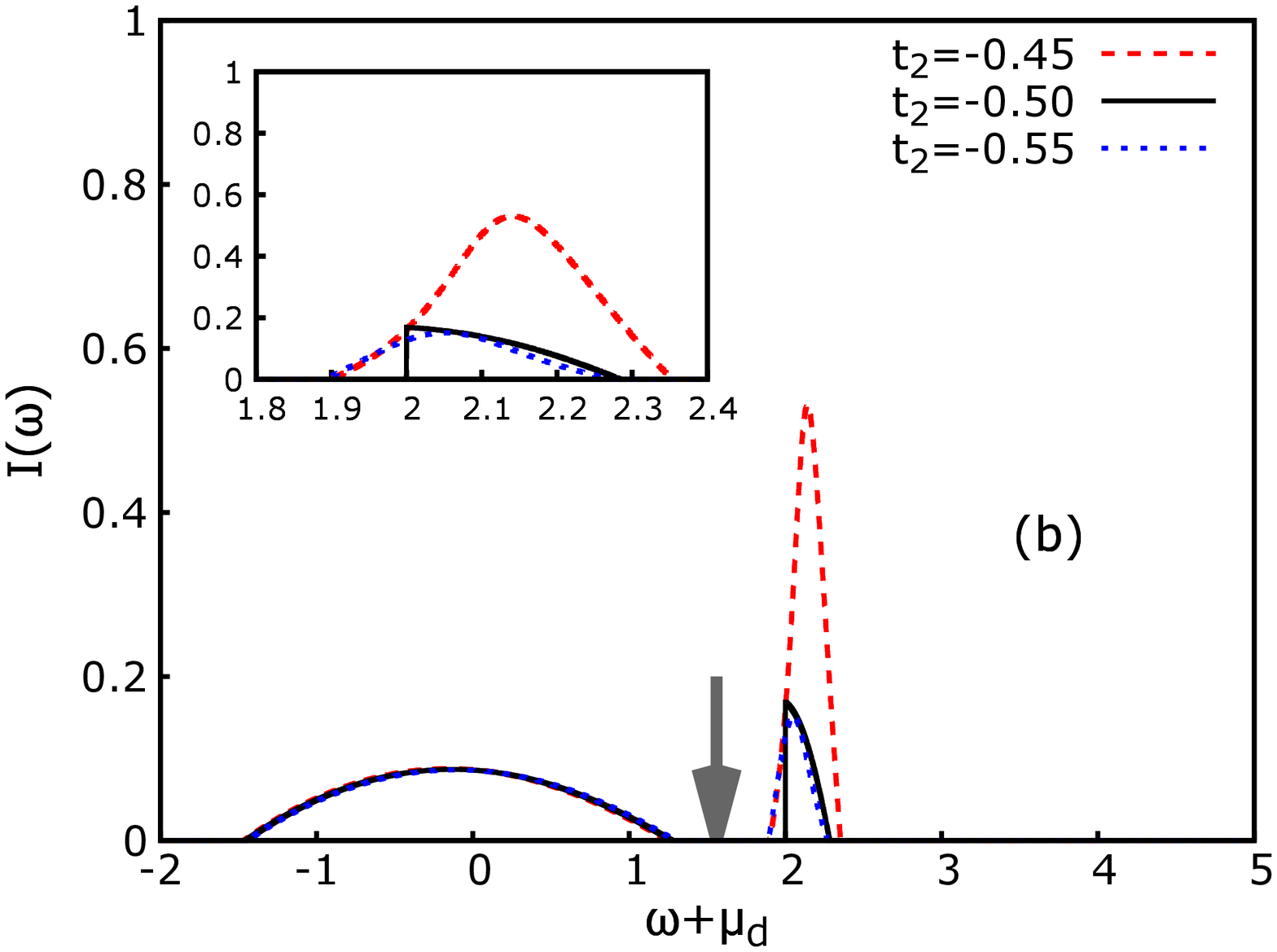}
	\caption{(Color online) The interacting DOS (panel a) and transport function (panel b) 
	is plotted versus $\omega$ for $U=2$ at half filling ($n_f=n_d=1/2$) 
	and for $t_2=-0.45$, $-0.5$, $-0.55$ ($t_3=0$).
         The gray arrow indicates a narrow $\omega$-interval in which $E_{\textrm{F}}$  is  located for different values of $t_2$.           
	     }
	\label{fig:dos_t2_-05U_20nf05}
\end{figure}

\begin{figure} 
	\centering
	\includegraphics[width=0.8\linewidth]{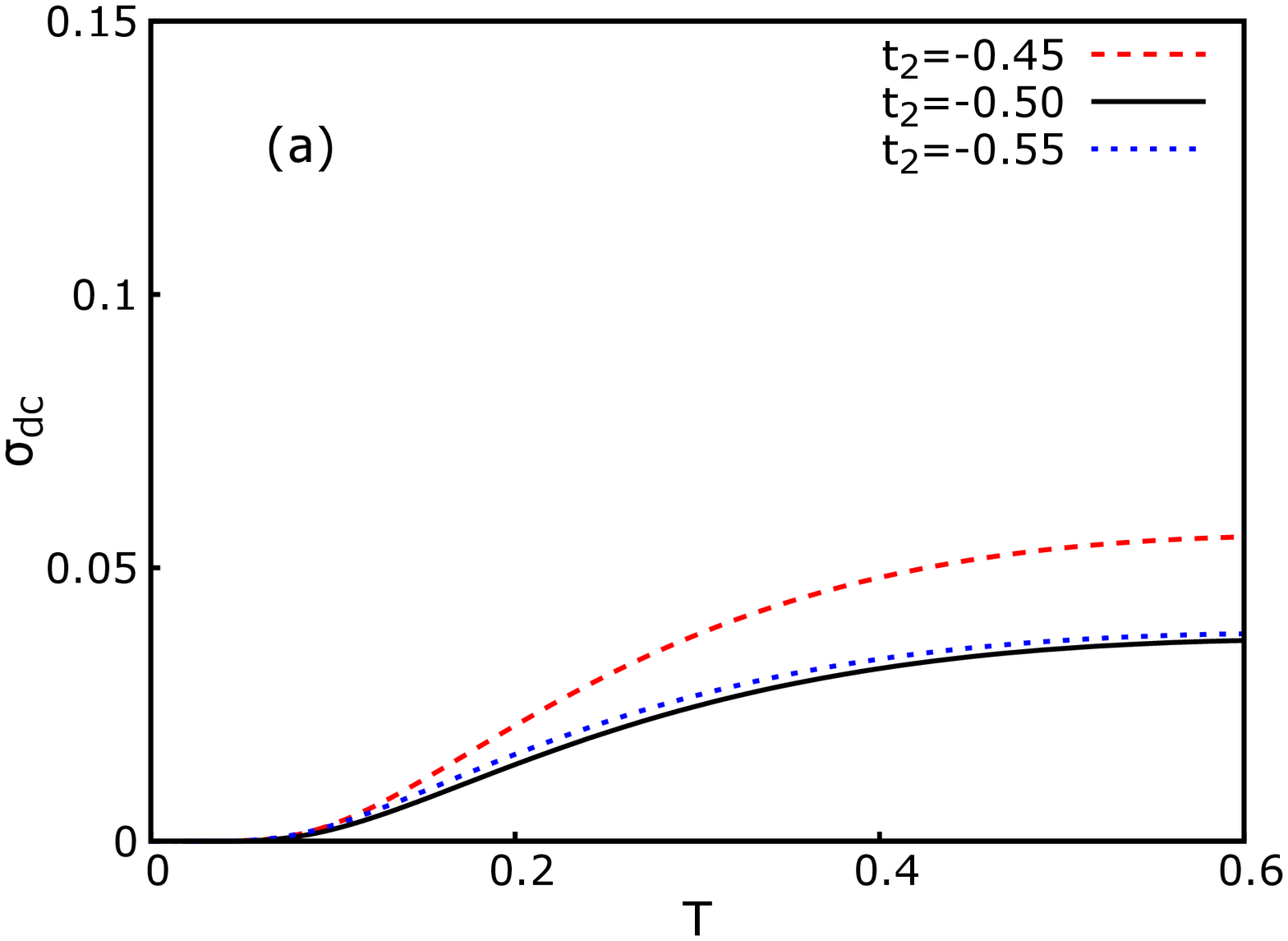}\\
	\includegraphics[width=0.8\linewidth]{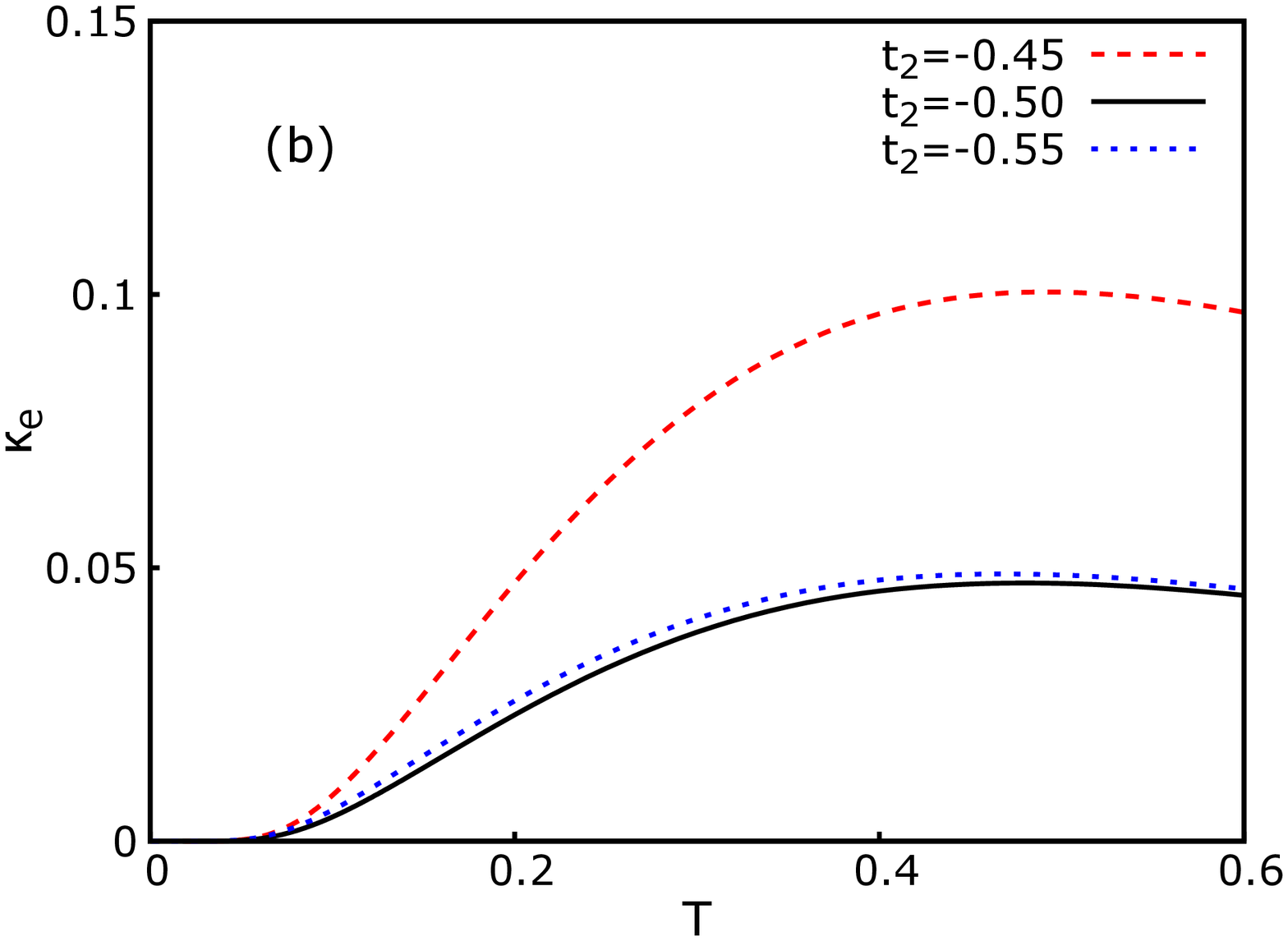}\\
	\includegraphics[width=0.8\linewidth]{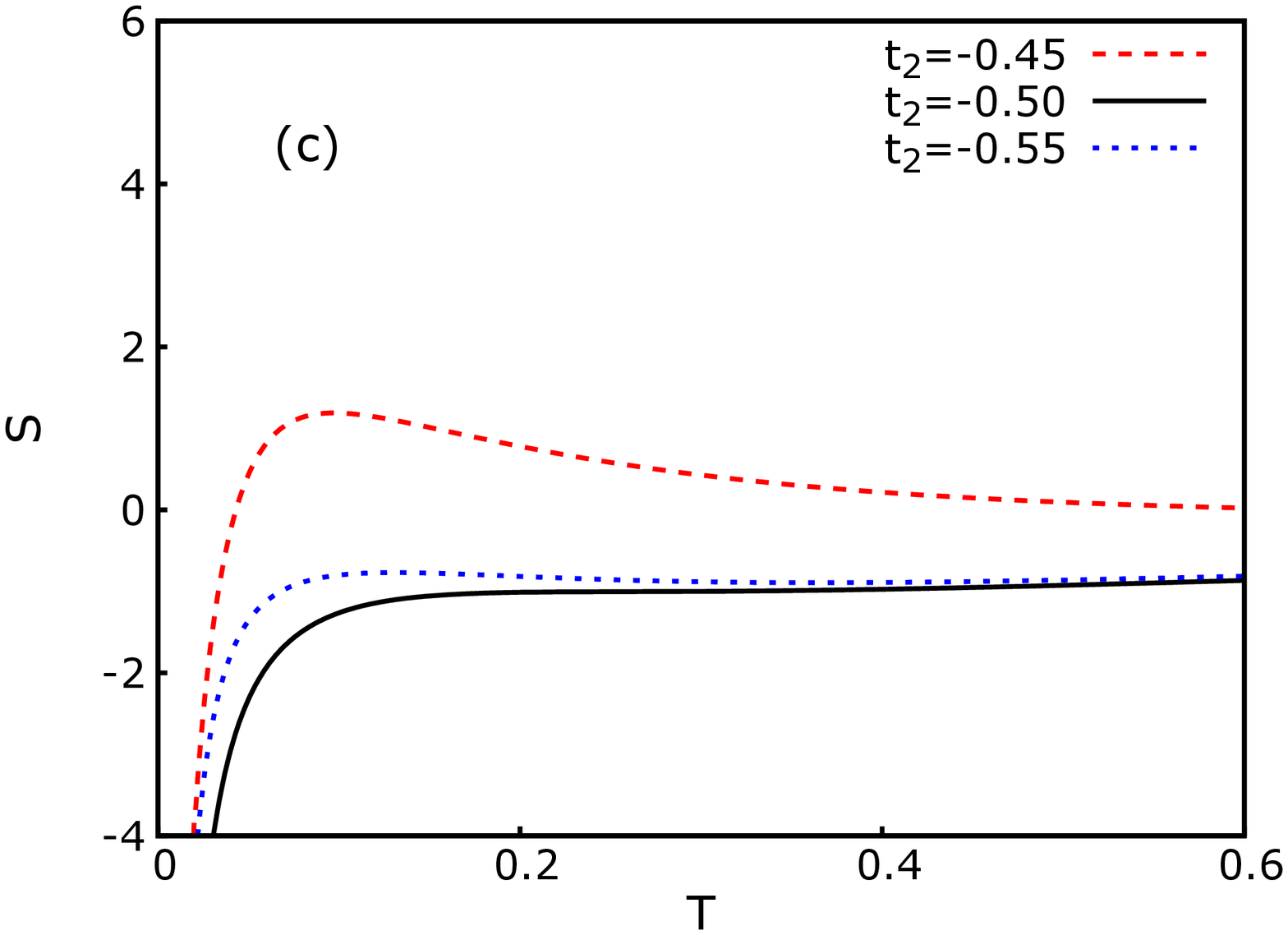}
	\caption{(Color online) Temperature dependence of the dc conductivity $\sigma_{\textrm{dc}}$,  
	the thermal conductivity $\kappa_{\textrm{e}}$, and Seebeck coefficient $S$ in the strong coupling regime is shown for the same parameters as in Fig.~\ref{fig:dos_t2_-05U_20nf05}.
        }
	\label{fig:tr_t2_-05U_20nf05}
\end{figure}

At half filling and $t^{++}=0$ ($t_2=-0.5$),  the upper Hubbard band has a singularity at the bottom edge, 
where the transport function shows only a step-like feature  (see Fig.~\ref{fig:dos_t2_-05U_20nf05}). 
The dc charge and thermal conductivities are similar to what one sees in doped Mott insulators,\cite{joura:165105,zlatic:266601,zlatic:155101} 
but the Seebeck coefficient is different: it is negative and displays anomalous behavior at low temperatures 
(see Fig.~\ref{fig:tr_t2_-05U_20nf05}).
Such a behavior is reflected in the anomalous temperature dependence of the chemical potential caused by the singularity in the DOS. 
For $t^{++}\neq 0$, the features in the upper Hubbard band are smoothed.
For $t^{++}>0$ ($t_2>-0.5$), the resonant peak dominates  the transport function, so that the dc charge and thermal conductivities increase, 
and the Seebeck coefficient is positive.   The anomalous behavior is shifted to lower temperatures.

\begin{figure} 
	\centering
	\includegraphics[width=0.8\linewidth]{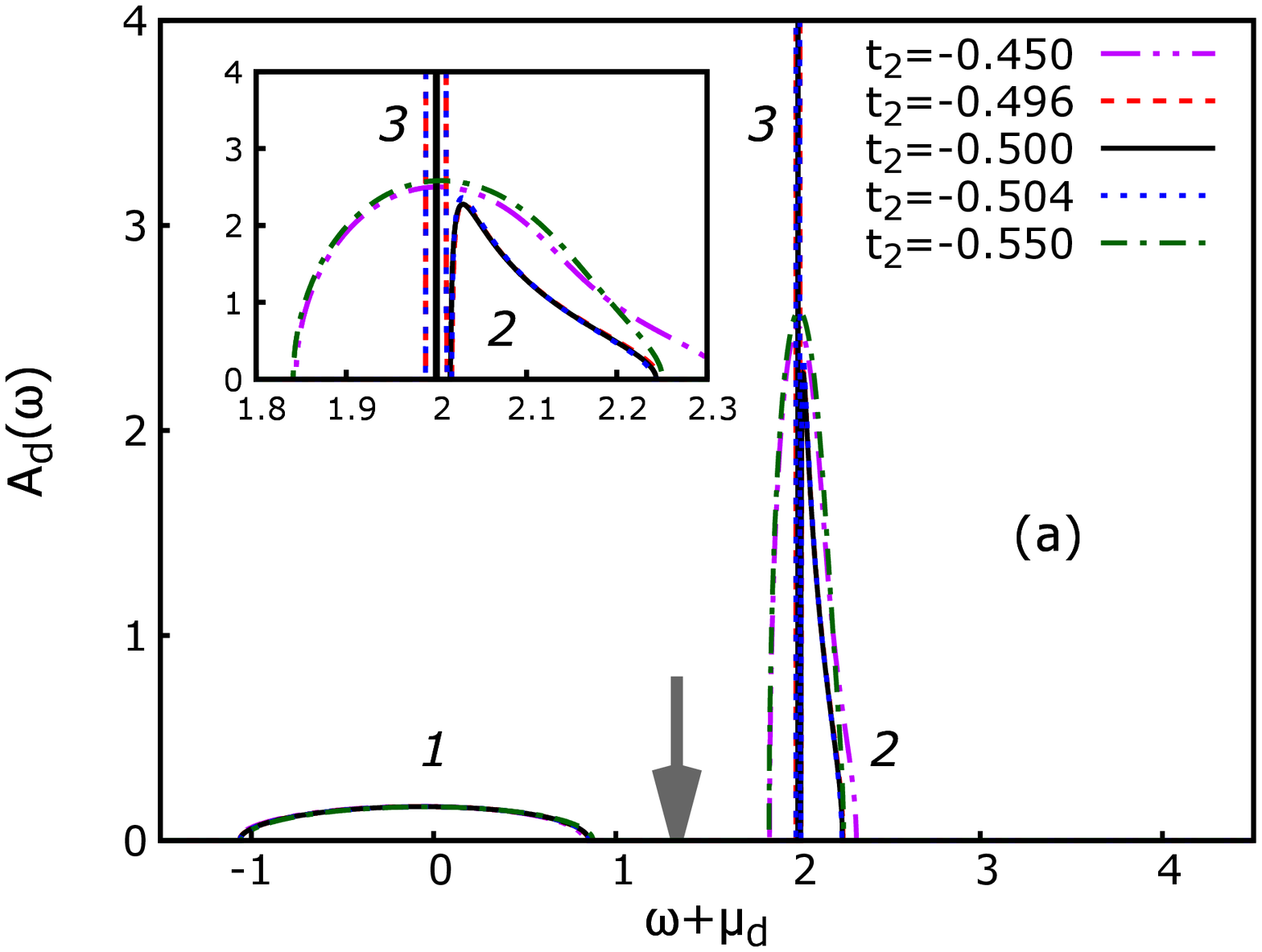}\\
	\includegraphics[width=0.8\linewidth]{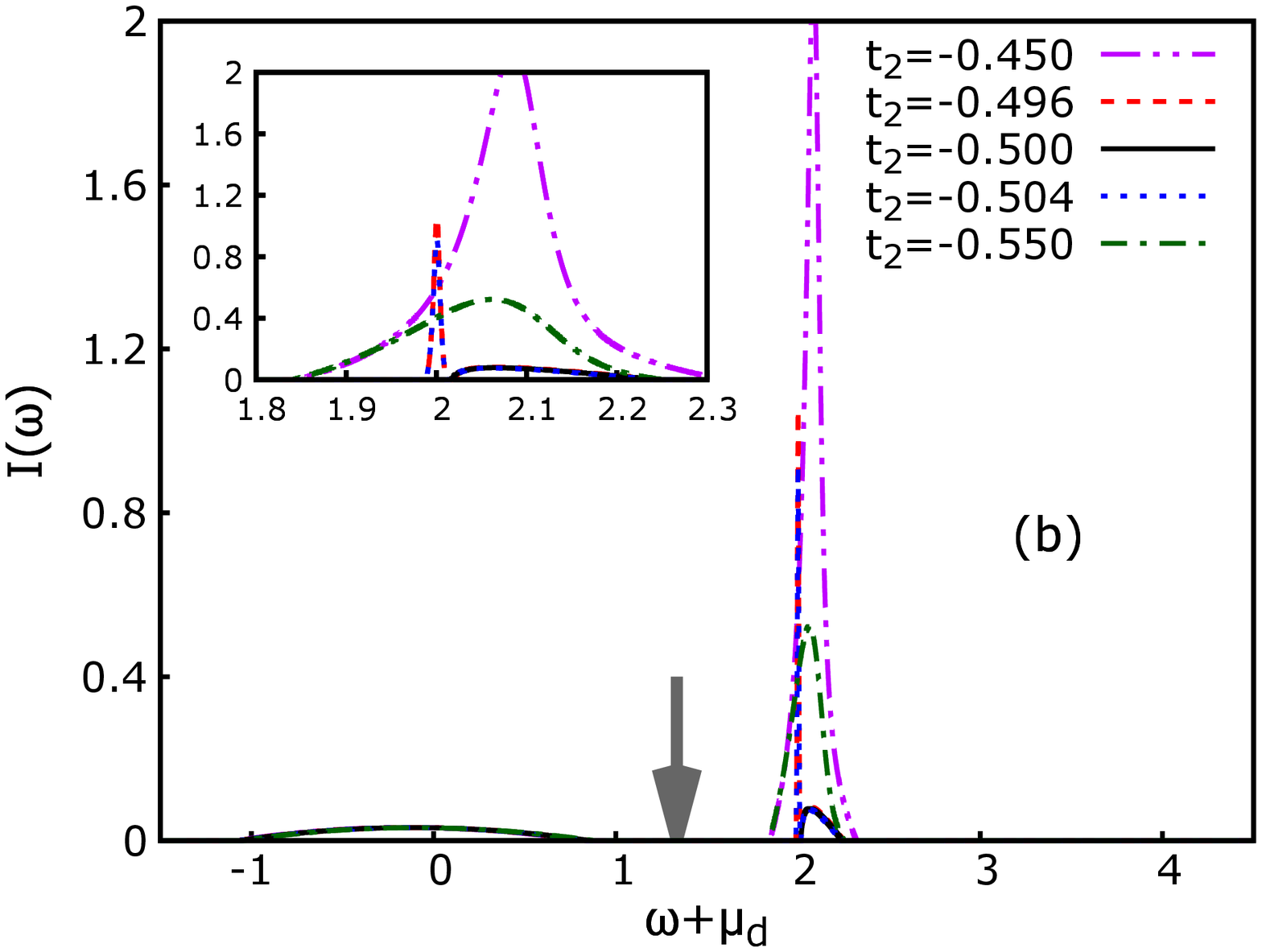}
	\caption{(Color online) The interacting DOS (panel a) and transport function (panel b)  
	for $U=2$ and $n_f=0.75$, and $n_d=1-n_f=0.25$.
	Here,  $t_2=-0.45$, $-0.496$, $-0.5$, $-0.504$, $-0.55$ and 
	labels \textsl{1}, \textsl{2}, and \textsl{3} denote the lower Hubbard band, the upper Hubbard band, 
	and the band of localized states, respectively.  
	 The gray arrow indicates a narrow $\omega$-interval in which $E_{\textrm{F}}$  is  located for different values of $t_2$. 
	 }
	\label{fig:dos_t2_-05U_20nf075}
\end{figure}

\begin{figure} 
	\centering
	\includegraphics[width=0.8\linewidth]{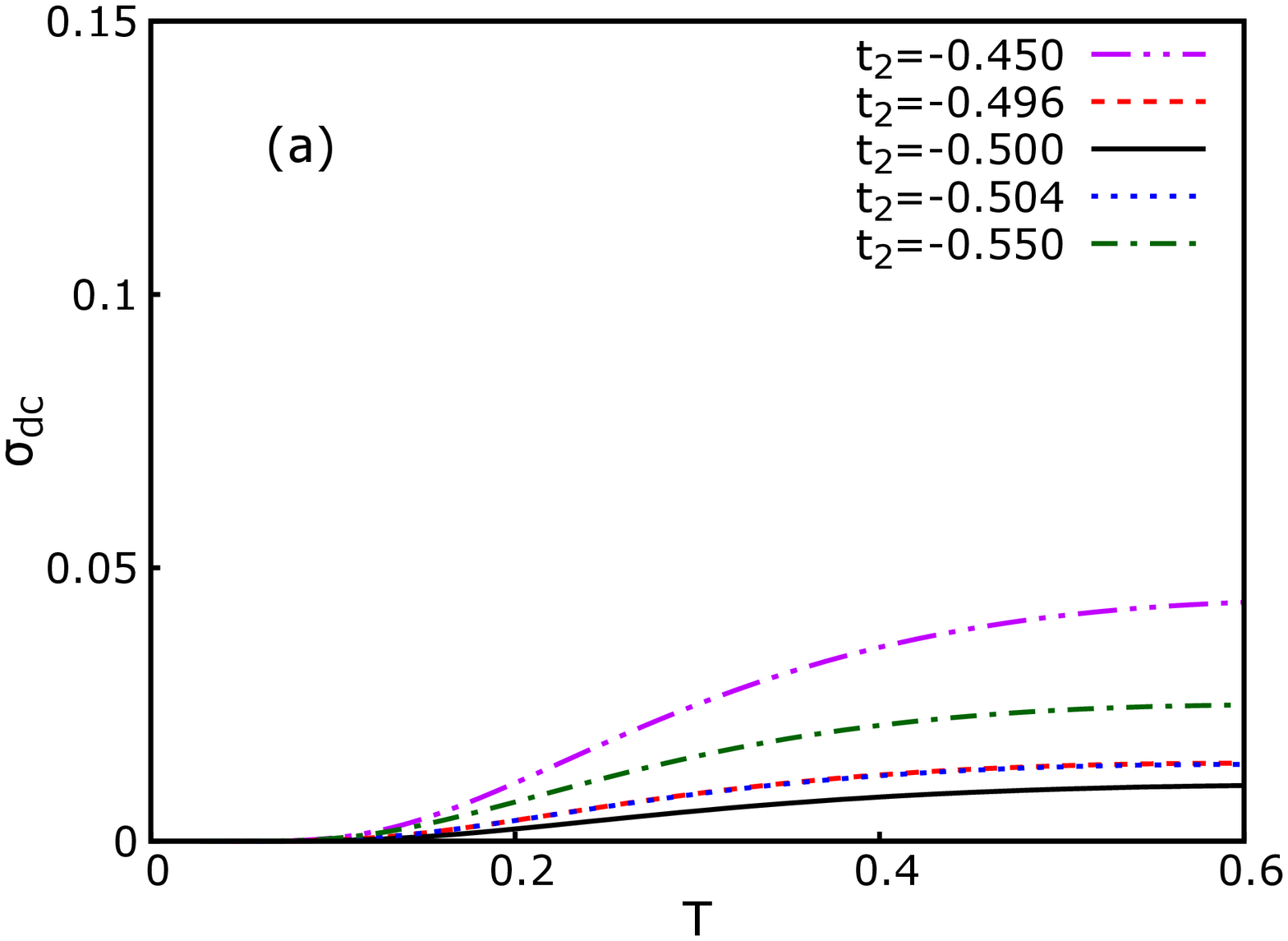}\\
	\includegraphics[width=0.8\linewidth]{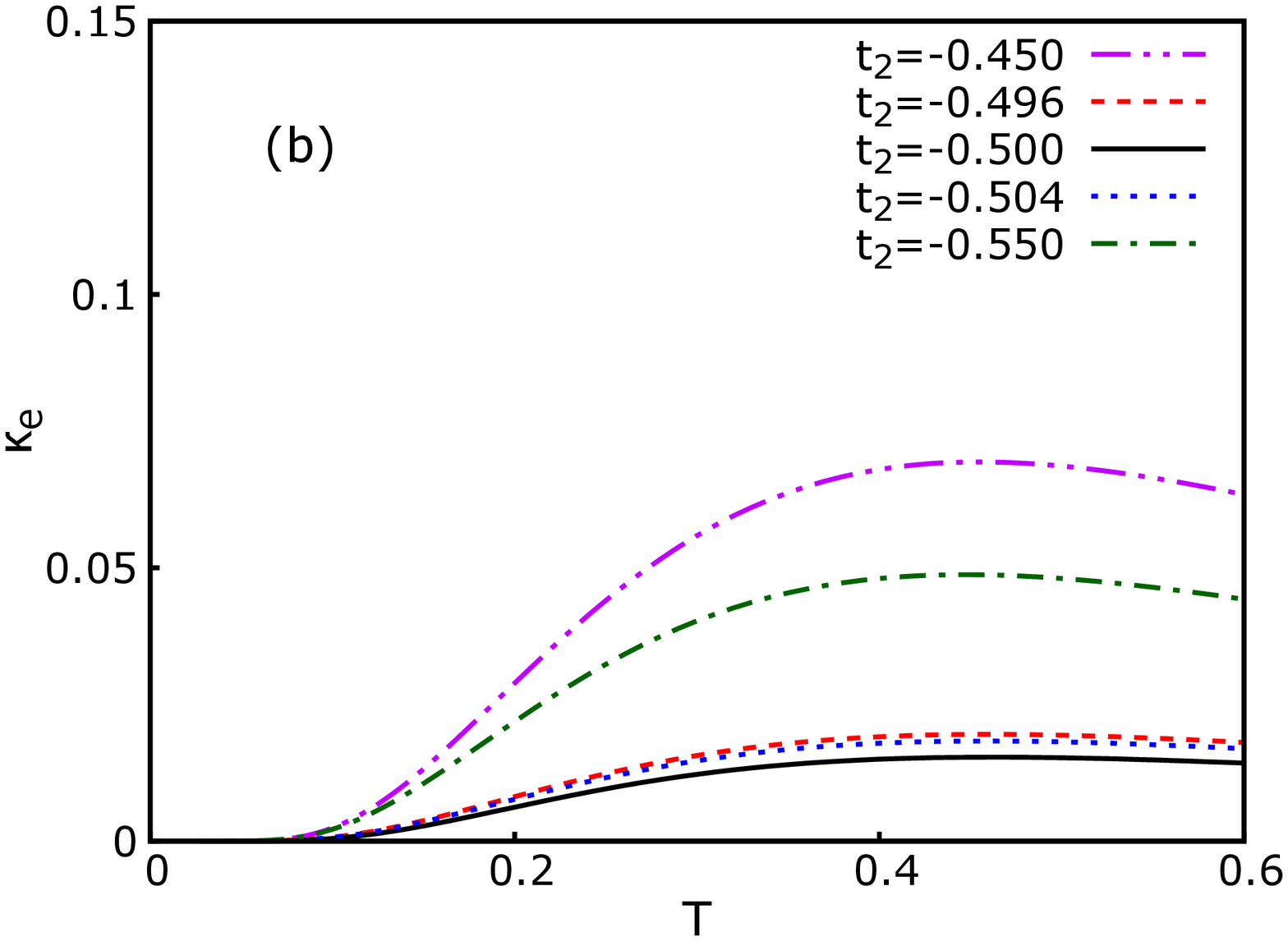}\\
	\includegraphics[width=0.8\linewidth]{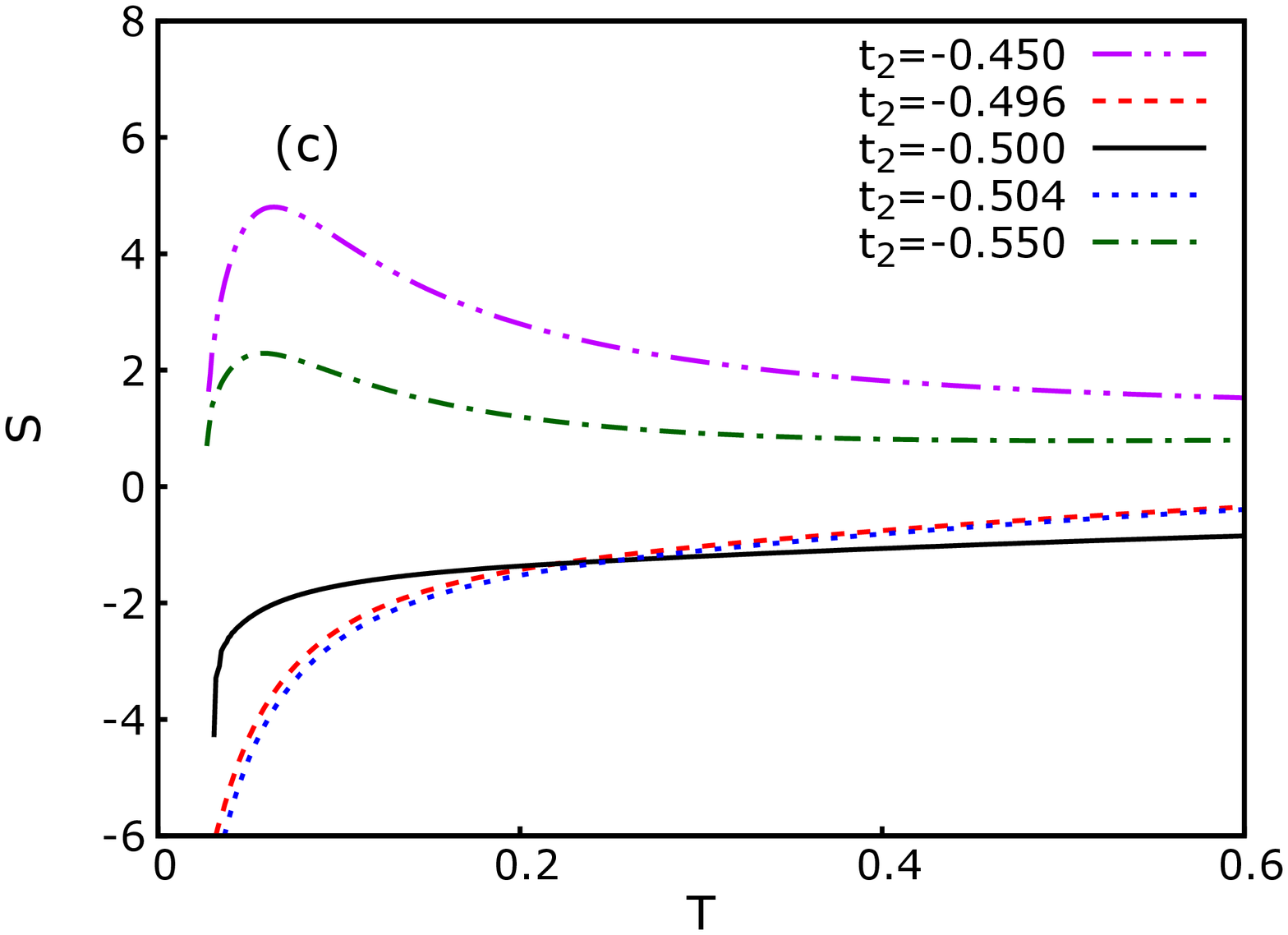}
	\caption{(Color online) 
	Temperature dependence of the dc conductivity $\sigma_{\textrm{dc}}$,  
	the thermal conductivity $\kappa_{\textrm{e}}$, and Seebeck coefficient $S$ 
	for the same parameters as in Fig.~\ref{fig:dos_t2_-05U_20nf075}.}
	\label{fig:tr_t2_-05U_20nf075}
\end{figure}

An increase of doping has a twofold effect. 
First, it expands the $\omega$--$t_2$ region in which the Mott gap is observed [see panels (c) and (d) in Fig.~\ref{fig:log_u=20}]. 
Second,  for $t^{++}$  in a narrow interval around $t^{++}=0$ ($t_2=-0.5$), it leads to the three-band structure in the DOS and transport function.
This is illustrated in Fig.~\ref{fig:dos_t2_-05U_20nf075}(a), 
where  the lines (\textsl{1}), (\textsl{2}), and (\textsl{3}) represent the lower Hubbard band, the upper Hubbard band, 
and the band of localized states, respectively.
The $\delta$-peak in the DOS, 
observed for  $t^{++}=0$ exactly,
is represented by the vertical line (\textsl{3}).  
This singular feature is coming from the localized $d$ states and the corresponding singularity in the transport function is absent. 
Any deviation of $t^{++}$ from zero, broadens  the $\delta$-peak into a narrow band of excitations which contribute to  the 
heat and charge transport (see Fig.~\ref{fig:tr_t2_-05U_20nf075}). 
For a sufficiently large value  of $|t^{++}|$, the band of localized states [line (\textsl{3})  in Fig.~\ref{fig:dos_t2_-05U_20nf075}] 
merges with the upper Hubbard band [line (\textsl{2})], so that the sharp features in the DOS are smoothed out.  In comparison with the half-filled case shown in Figs.~\ref{fig:dos_t2_-05U_20nf05} and \ref{fig:tr_t2_-05U_20nf05}, 
the Mott gap is now larger, so that we obtain smaller values of the dc charge and thermal conductivities,  
and Seebeck coefficient, for $t^{++}$ close to zero. 
For larger values of $|t^{++}|$, the resonant peak  dominates the transport function, 
which becomes large for positive $t^{++}$ ($t_2>-0.5$) and relatively small for negative $t^{++}$ ($t_2<-0.5$), 
so that the dc charge and thermal conductivities increase. 
The comparison with the undoped case shows that the temperature dependence of the Seebeck coefficient is modified  
and $S(T)$ is now larger and positive.

\section{Discussion and conclusions}\label{sec:conclusions}

In this article we studied the effects of correlated hopping on the density of states and transport coefficients of the Falicov-Kimball model 
on a Bethe lattice with a semielliptic DOS.

Using dynamical mean field theory, we derived an exact solution for the renormalized DOS $A_d(\omega)$ and 
the transport function $I(\omega)$, and computed the transport coefficients via the Boltzmann relations (Jonson-Mahan theorem). The band width of $A_d(\omega)$ and $I(\omega)$ is found to be the same but their functional form is completely different, showing that correlated hopping renormalizes the single particle and the two particle properties in a different way.
As regards $A_d(\omega)$,  the main effect of correlated hopping is an effective narrowing of the band width, accompanied 
by the opening of a Mott-Hubbard gap and, at finite doping, the emergence of a band of localized states. 
As regards $I(\omega)$, its behavior is dominated by the resonant peak which appears in the region 
of the parameter space (small values of $|t^{++}|$), where the single particle sates are localized. 
Our interpretation of the resonance is that it manifests a two-particle interference in random media, 
e.g., weak (anti)localization.\cite{bergmann:2914} 
The correlated hopping mimics the random media, with different values of the hopping integral corresponding to 
the random distances between the atoms. From this point of view, Eq.~\eqref{eq:res_cond} is the interference condition 
for $d$ particles following different trajectories over the lattice sites which are either occupied or unoccupied by the $f$ particles.

In our previous article,\cite{shvaika:43704} it was found that the charge and heat transport of particles 
described by the Falicov-Kimball model on a hypercubic lattice exhibits a number of surprising features. 
However, the anomalies were associated with the peculiarities of the Gaussian density of states, which does not have 
the clear-cut band edge 
and has a finite ``quasiparticle'' scattering time for frequencies outside the band.
To our surprise, similar behavior is obtained for a Bethe lattice with a semi elliptic density of states, so that the anomalous 
transport coefficients seem to be a common feature of systems with correlated hopping. 

The anomalous features of  $I(\omega)$,  due to the resonant two-particle contribution, 
are observed for a wide range of the correlated hopping parameter, $t_{2}$. 
In this parameter range, 
the renormalized one-particle DOS does not display any anomalies. 
The analytic expression for the resonant frequency, given by Eq.~\eqref{eq:w_res}, 
shows that by tuning the concentration of the itinerant electrons we can bring the Fermi level 
close to the resonance and increase substantially the conductivities and thermoelectric power.  

The reduction of the amplitude of correlated hopping between the sites occupied by the $f$ particles creates, 
for $t^{++}\to 0$, clusters of sites occupied by $f$ electrons and the ensuing band of localized $d$ states. 
When the $f$-electron concentration is above half filling, $n_f>0.5$, this band can be seen in the single particle DOS, 
in addition to the lower and upper Hubbard band. 
Depending on the concentration of $d$ electrons, the Fermi level is either in the lower ($n_d<1-n_f$) or  in the upper ($n_d>n_f$) Hubbard band 
or in the band of localized states ($1-n_f<n_d<n_f$). 
In each of these cases, the  thermoelectric coefficients exhibit completely different behaviors.

Finally, we  remark that a large enhancement of the conductivities by correlated hopping, 
driven by the emergence of two-particle resonant states, has been found here for the Falicov-Kimball model with static interaction. 
It would be interesting to check whether similar transport anomalies 
emerge in the models with dynamic interactions, e.g.\ in the Hubbard model with correlated hopping.

\bigskip

\begin{acknowledgments}
V.Z. acknowledges the support by the Ministry of Science of Croatia under the bilateral agreement 
with the USA on the scientific and technological cooperation, Project No. 1/2016. 
\end{acknowledgments}

\bibliography{thermoel}

\end{document}